\documentclass[oneside,english,italian]{book}
\usepackage[T1]{fontenc}
\usepackage{textcomp}
\usepackage[latin9]{luainputenc}
\usepackage{prettyref}
\usepackage{amsmath}
\usepackage{amssymb}
\usepackage{graphicx}

\makeatletter

\providecommand{\tabularnewline}{\\}
\newcommand{\lyxdot}{.}

\makeatother

\usepackage{babel}
\begin{document}
\title{\textsc{\includegraphics[scale=0.2]{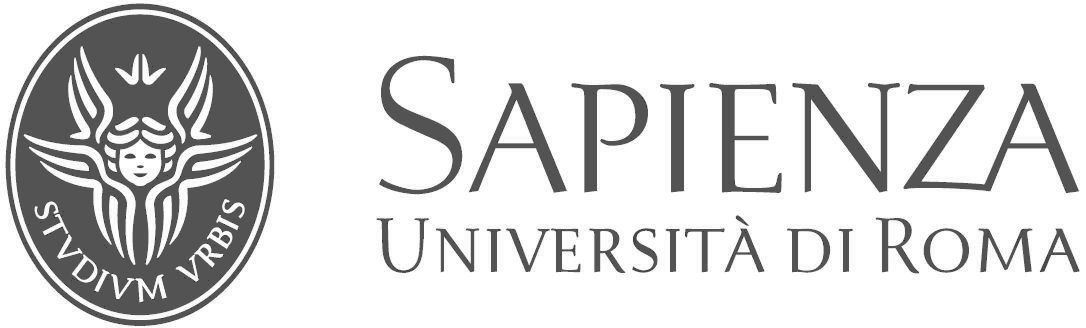}}}
\date{{\small\textsc{Facoltà di Scienze Matematiche Fisiche e Naturali }}\textsc{}\\
\textsc{Corso di Laurea Specialistica in Fisica }\\
\textsc{}\\
\textsc{}\\
\textsc{}\\
{\huge\textsc{Catene ideali con numero }}\\
{\huge\textsc{}}\\
{\huge\textsc{fissato di auto-intersezioni}}{\Large\textsc{}}\\
\textsc{}\\
\textsc{}\\
\textsc{}\\
\textsc{Simone Franchini}\\
{\normalsize\textsc{matr. n° 697666}}\textsc{}\\
\textsc{}\\
\textsc{}\\
{\normalsize\textsc{relatore}}\textsc{}\\
\textsc{Prof. Giorgio Parisi}{\Large\textsc{}}\\
\textsc{}\\
\textsc{}\\
\textsc{}\\
\textsc{A.A. 2010/2011}}

\maketitle
\newpage{}

\begin{center}
~
\par\end{center}

\newpage{}

\tableofcontents{}

\newpage{}

\chapter*{Introduzione}

Sin dalla sua introduzione, il Random Walk (\cite{Spitzer,Huges,Feller})
è stato uno dei modelli probabilistici di maggiore successo in molti
campi scientifici. La definizione più intuitiva che se ne può dare
è quella di una traccia lasciata da un punto che si muove a salti:
ognuno di questi salti ha direzione e modulo casuali, determinati
da una distribuzione di probabilità di qualche tipo. Questa definizione
può essere poi generalizzata, pensando al RW semplicemente come una
successione ordinata di punti nello spazio. 

Proprio questa interpretazione geometrica ha suggerito l'impiego del
Random Walk come primo modello di polimero, da cui sono nate una infinità
di varianti per includere alcune proprietà non banali dei sistemi
reali (\cite{de=000020Gennes,des=000020Cloiseaux-Jannink,Flory}).
Ad esempio, l'introduzione di vincoli statistici che limitassero il
numero di auto-intersezioni ha premesso di modellizzare l'effetto
di \emph{volume escluso} (il cui paradigma è rappresentato dal Self-Avoiding
Walk, \cite{Madras-Slade}).

In questo lavoro tratteremo in dettaglio le proprietà di auto-intersezione
nei Random Walks (\cite{Spitzer,Huges}). Benché relativamente poco
studiate (anche a causa della complessità matematica del problema),
queste proprietà rappresentano senza dubbio uno degli aspetti più
importanti dei cammini aleatori, essendo correlate in larga misura
a quasi tutti i comportamenti dei polimeri reali.

Il nostro scopo principale sarà quello di studiare catene ideali (Random
Walks) su reticolo $\mathbb{Z}^{d}$, $d\geq3$, in cui il rapporto
fra numero di auto-intersezioni (visite multiple di un dato sito reticolare)
e lunghezza totale della catena è fissato ad un numero reale $m\in\left[0,1\right]$.
Come vedremo, questo modello mostrerà interessanti proprietà in funzione
del parametro $m$, come ad esempio una transizione di fase per un
particolare valore $m_{c}\in\left(0,1\right)$. 

Il primo capitolo di questo lavoro sarà dedicato essenzialmente all'esposizione
dell'attuale teoria sulle proprietà di auto-intersezione dei Random
Walks, introducendo poi alcune quantità di interesse per lo studio
delle intersezioni. Il secondo presenterà i risultati di alcuni esperimenti
al computer, mirati all'indagine delle proprietà geometriche dei cammini
al variare del parametro $m$: in particolare, verranno esaminate
la distanza end-to-end, e il numero di occupazione dei siti intorno
all'origine (\emph{ambiente}).

Nel terzo capitolo verrà introdotto e studiato un processo stocastico
(esplicitamente dipendente dal tempo), che consentirà di derivare
alcune proprietà matematiche, utili nello studio della distribuzione
delle intersezioni.

Il quarto si concentrerà sulle proprietà della funzione \emph{ambiente},
introducendo una teoria microscopica approssimata: in particolare,
sarà possibile ottenere una approssimazione lineare, abbastanza accurata
per reticoli $d\geq4$.

In fine, nel quinto capitolo si utilizzeranno i risultati ottenuti
in precedenza per studiare un particolare modello termodinamico (il
modello di Stanley, \cite{Stanley}), che permetterà a sua volta di
approfondire ulteriormente le proprietà dei cammini con $m$ intersezioni
per monomero.

\chapter{Range e intersezioni di Random Walks}

Questo primo capitolo è dedicato all'esposizione delle proprietà di
intersezione dei random walks su reticolo. Dopo una breve introduzione
del formalismo utilizzato, verranno definite alcune quantità di interesse
per questo lavoro: in particolare, le nozioni di \emph{range} e \emph{intersezione}
per random walk semplici, e la distribuzione delle intersezioni sull'ensemble.
Per ognuna di queste quantità verranno riassunti i principali risultati
dell'attuale teoria, riguardanti essenzialmente valor medio e varianza:
per quest'ultima verrà riportato esplicitamente (e, a nostra conoscenza,
per la prima volta) un calcolo numerico esplicito degli andamenti
asintotici. Si procederà poi con l'introduzione della funzione \emph{ambiente},
e del suo utilizzo per esprimere la distribuzione delle intersezioni.

\section{Notazione e generalità}

Un Random Walk è un processo stocastico definito su un grafo connesso.
Detto $\Im$ il grafo di supporto, sia $X_{i}\in\Im$, $i\in\left[1,n\right]$
una sequenza di $n$ variabili casuali indipendenti, distribuite su
$\Im$ secondo le distribuzioni $\Pi_{i}\left(X_{i}\right)$. Chiameremo
Random Walk $\omega_{n}\left(A,B\right)$ il set di variabili $S_{i}\in\Im$,
$i\in\left[0,n\right]$ individuate dalla relazione
\begin{equation}
S_{k}=S_{0}+\sum_{i=1}^{k}X_{i},\label{eq:1.1.1}
\end{equation}
\\
dove $S_{0}=A$, $S_{n}=B$ (con $A,B\in\Im$). Da un punto di vista
geometrico, $\omega_{n}\left(A,B\right)$ è una \emph{catena} di punti
in $\Im$ (nodi), associati ad un set di coppie di punti (archi).
Sia inoltre $\Omega_{n}\left(A,B\right)$ l'insieme dei possibili
set $\omega_{n}\left(A,B\right)$.

I Random Walk che tratteremo, ove non specificato, sono quelli definiti
sul reticolo $d-$dimensionale $\mathbb{Z}^{d}$: i nodi di questo
grafo sono i vettori $x=\left(x_{1},\,x_{2},\,...,\,x_{d}\right)$
con $x_{j}\in\mathbb{Z}$, mentre gli archi sono le coppie di punti
$x,y\in\mathbb{Z}^{d}$. In particolare la nostra attenzione si rivolgerà
sul cosiddetto Simple Random Walk (SRW), per il quale tutte le $X_{i}$
hanno $\left|X_{i}\right|=1$, e $\Pi_{i}\left(X_{i}\right)=1/2d$
per ogni primo vicino di $S_{i-1}$: per i pattern ottenuti, i nodi
collegati dagli archi godranno dunque della proprietà $\left|S_{i}-S_{i+1}\right|=1$
(chiameremo queste coppie di punti \emph{primi vicini,} o semplicemente
NN). Dato il largo uso dei SRW nella modellizzazione dei polimeri
ideali, indicheremo i nodi del SRW con il nome di \emph{monomeri. }

Si consideri l'insieme $\Omega_{n}\equiv\bigcup_{y\in\mathbb{Z}^{d}}\Omega_{n}\left(x,y\right)$
dei SRW $\omega_{n}\left(x,y\right)$, $\forall y$, sul reticolo
$d-$dimensionale $\mathbb{Z}^{d}$ (si omette la dimensione nella
notazione). Dove non specificato altrimenti, la misura su di esso
sarà quella uniforme, considerando dunque un peso pari a $1/\left|\Omega_{n}\right|$
per elemento (per gli insiemi il modulo $\left|\cdot\right|$ rappresenterà
la cardinalità). Si noti che per un SRW sul reticolo ipercubico $\mathbb{Z}^{d}$
si ha semplicemente $\left|\Omega_{n}\right|=\left(2d\right)^{n}$.

Verranno ora elencate le definizioni di alcune quantità fondamentali.
Chiamiamo $P_{n}\left(x,y\right)$ la probabilità che all'$n-$esimo
passo un cammino partito da $x$ abbia raggiunto il sito $y$.
\begin{equation}
P_{n}\left(x,y\right)=\mbox{P}\left[S_{0}=x,\,S_{n}=y\right],\label{eq:1.1.2}
\end{equation}
La funzione gode di un invarianza traslazionale: introducendo $\delta x=y-x$
abbiamo $P_{n}\left(x,y\right)=P_{n}\left(y-x\right)=P_{n}\left(\delta x\right)$.
Gode inoltre della seguente rappresentazione integrale: 
\begin{equation}
P_{n}\left(\delta x\right)=\left(2\pi\right)^{-d}\int_{\left[-\pi,\pi\right]^{d}}dq\,e^{iq\delta x}\tilde{\lambda}_{d}\left(q\right)^{n}.\label{eq:1.1.4}
\end{equation}
La quantità $\tilde{\lambda}_{d}\left(q\right)$ è detta fattore di
struttura del reticolo, ed è legata al tipo di reticolo in esame;
nel caso del reticolo $\mathbb{Z}^{d}$ si ha 
\begin{equation}
\tilde{\lambda}_{d}\left(q\right)=d^{-1}\sum_{i=1}^{d}\cos\left(q_{i}\right).\label{eq:1.1.5}
\end{equation}
Data l'importanza che avrà nei prossimi sviluppi, è conveniente indicare
il caso particolare $p_{n}=P_{n}\left(0,0\right)$, che rappresenta
la probabilità di ritrovarsi nel punto d'origine al passo $n-$esimo.
La sua rappresentazione integrale segue dalle \prettyref{eq:1.1.4},
\prettyref{eq:1.1.5}, mentre è possibile ottenere un andamento asintotico
per grandi $n$ sviluppando il fattore di struttura: si avrà $p_{n}\sim\left(d/2\pi n\right)^{d/2}$. 

Il valore atteso del numero di visite al sito $y$ dopo $n$ passi
è detto $G_{n}\left(x,y\right)$:
\begin{equation}
G_{n}\left(x,y\right)=\sum_{i=0}^{n}P_{i}\left(x,y\right),\label{eq:1.1.8}
\end{equation}
definiamo il limite $n\rightarrow\infty$ con $G\left(x,y\right)$.
In generale i limiti $n\rightarrow\infty$ di funzioni relative all'insieme
$\Omega_{n}\left(x,y\right)$ si indicheranno omettendo l'indice basso
$n$. La rappresentazione integrale segue dalla \prettyref{eq:1.1.4}.
Anche in questo caso prestiamo particolare attenzione al caso $G_{n}\left(0,0\right)=G_{n}$,
che rappresenta il numero medio di visite all'origine di un cammino
di $n$ passi. In particolare è necessario discutere il limite 
\begin{equation}
\rho_{d}=\lim_{n\rightarrow\infty}G_{n}.\label{eq:1.1.9}
\end{equation}
L'esistenza del limite \prettyref{eq:1.1.9} è infatti legata alla
dimensionalità, piuttosto che al tipo di reticolo. Dalla sua rappresentazione
integrale 
\begin{equation}
\rho_{d}=\left(2\pi\right)^{-d}\int_{\left[-\pi,\pi\right]^{d}}dq\left(1-\tilde{\lambda}_{d}\left(q\right)\right)^{-1}\label{eq:1.1.10}
\end{equation}
si nota (\cite{Polya}) che $\rho_{d}<\infty$ solo se $d\geq3$.
Per $d\leq2$ il numero di ritorni nell'origine è infinito, e il cammino
tende ad occupare $\mathbb{Z}^{d}$ uniformemente, Questo tipo di
cammini viene definito \emph{ricorrente}, mentre un $\rho_{d}$ finito
indica una occupazione parziale dello spazio, e i cammini vengono
classificati come \emph{transienti}. Il valore di $\rho_{d}$ è valutabile
con l'ausilio del calcolatore, ed è noto per un gran numero di valori
di $d$ (anche frazionari).

Detto $H$ un sottospazio di $\mathbb{Z}^{d}$, sia $P_{n}^{H}\left(x,y\right)$
la probabilità che un cammino di lunghezza $n\geq1$ (con estremi
$x$, $y$) non abbia incontrato il sottospazio $H$ per i primi $n-1$
passi.
\begin{equation}
P_{n}^{H}\left(x,y\right)=\mbox{P}\left[S_{i}\notin H,\,i\in\left(0,n\right)\right].\label{eq:1.1.11}
\end{equation}
Da qui passiamo al limite $n\rightarrow\infty$, introducendo la frazione
$F^{H}\left(x,y\right)$ di cammini che terminano in $y\in H$ senza
aver incontrato $H$ nei primi $n-1$ passi: 
\begin{equation}
F^{H}\left(x,y\right)=\sum_{i=1}^{\infty}P_{i}^{H}\left(x,y\right).\label{eq:1.1.12}
\end{equation}
Consideriamo infine il caso particolare $H=y$, che rappresenta la
frazione di cammini con estremi in $x$, $y$ che includono l'estremo
$y$ una sola volta.
\begin{equation}
F\left(x,y\right)=\sum_{i=1}^{\infty}P_{i}^{y}\left(x,y\right).\label{eq:1.1.12.1}
\end{equation}
come prima, per $n<\infty$ indicheremo con $F_{n}\left(x,y\right)$.
Chiameremo $f_{n}=P_{n}^{0}$ la probabilità che l'origine sia stata
raggiunta per la seconda volta esattamente all'$n-$esimo passo. Questo
ci consente di introdurre un altro valore fondamentale: la probabilità
$C_{d}$ di ritorno nell'origine per $n\rightarrow\infty$. 
\begin{equation}
C_{d}=\sum_{i=1}^{\infty}f_{i}=\mbox{P}\left[\exists i\,|\,S_{i}=S_{0},\,i\in\left(0,\infty\right)\right].\label{eq:1.1.13}
\end{equation}
Chiaramente la probabilità complementare $\bar{C}_{d}=1-C_{d}$ rappresenterà
la frazione di cammini che contengono l'origine una sola volta. Quest'ultima
quantità è legata alla $\rho_{d}$ dalla relazione $\bar{C}_{d}=\rho_{d}^{-1}$
(\cite{Feller}). Si noti come nei cammini ricorrenti il ritorno all'origine
sia un evento certo ($C_{1}=C_{2}=1$), mentre per $d\geq3$ si abbia
una frazione finita che riesce a sfuggire. In ultimo, si riporta da
\cite{Spitzer} che $F\left(x,y\right)=\left[1-F\left(0,0\right)\right]G\left(x,y\right)$
per $x\neq y$, con $F\left(0,0\right)=C_{d}$.

\section{Range di un SRW e distribuzione delle intersezioni}

Sia $R\left[\omega_{n}\left(x,y\right)\right]$ il numero di siti
di $\mathbb{Z}^{d}$ occupati almeno una volta dal cammino; questa
quantità è nota come \emph{range} del SRW $\omega_{n}\left(x,y\right)$.
Sia inoltre $M\left[\omega_{n}\left(x,y\right)\right]$ il numero
di volte in cui percorrendo $\omega_{n}\left(x,y\right)$ da un estremo
all'altro si incontra un sito già visitato (numero di \emph{intersezioni}).
Le due quantità sono legate dalla relazione 
\begin{equation}
R\left[\omega_{n}\left(x,y\right)\right]+M\left[\omega_{n}\left(x,y\right)\right]=n+1.\label{eq:1.2.1}
\end{equation}
Per evitare di appesantire la notazione, ove non sia necessario indicare
gli estremi dei cammini chiameremo queste due quantità con $R_{n}$
e $M_{n}$. Dalla \prettyref{eq:1.2.1} è evidente come tutti i risultati
validi per il range, siano immediatamente applicabili al numero $M_{n}$
delle intersezioni. Di conseguenza per tutta questa sezione ci occuperemo
della sola variabile \emph{range}, così da presentare direttamente
i notevoli risultati disponibili in letteratura. Introduciamo le notazioni
per aspettazione e varianza della variabile $s$, distribuita secondo
$P\left(s\right)$: il valore atteso è $\langle s\rangle=\sum sP\left(s\right)$,
la varianza invece verrà indicata con $\langle\Delta s^{2}\rangle=\langle\left(s-\langle s\rangle\right)^{2}\rangle$.

Riportiamo ora un risultato fondamentale su distribuzione, aspettazione
e varianza del range : dette $\xi_{n}$, $\xi$ le variabili 
\begin{equation}
\xi_{n}=\frac{R_{n}-\langle R_{n}\rangle}{\sqrt{\langle\Delta R_{n}^{2}\rangle}},\label{eq:1.2.2}
\end{equation}
\begin{equation}
\xi=\lim_{n\rightarrow\infty}\xi_{n},\label{eq:1.2.3}
\end{equation}
Jain, Orey e Pruitt (\cite{Jain-Pruitt,Jain-Orey}) hanno rigorosamente
dimostrato che, per $d\geq2$, la variabile $\xi$ ammette una distribuzione
\begin{equation}
P\left(\xi\right)=\frac{1}{\sqrt{2\pi}}\exp\left(-\frac{1}{2}\xi^{2}\right)\label{eq:1.2.3.1}
\end{equation}
 ovvero, nel limite di $n\rightarrow\infty$, la variabile $\xi_{n}$
tende ad avere una distribuzione normale con $\langle\xi_{n}\rangle=0$,
$\langle\Delta\xi_{n}^{2}\rangle=1$. 

In particolare Erdos, Dvoretzky, Jain, Orey, Pruitt e Torney (\cite{Jain-Pruitt,Jain-Orey,Dvoretzky-Erdos,Torney})
hanno calcolato gli andamenti di varianza e aspettazione per $d=2$
(cammini ricorrenti): 
\begin{equation}
\langle R_{n}\rangle=\frac{\pi n}{\log\left(n\right)}+\mathcal{O}\left(\log\log\left(n\right)\,\log^{-2}\left(n\right)\right)\label{eq:1.2.6}
\end{equation}
\begin{equation}
\langle\Delta R_{n}^{2}\rangle=\frac{\sigma_{2}^{2}n^{2}}{\log\left(8n\right)^{4}}+\mathcal{O}\left(n^{2}\log\left(n\right)^{-5}\right)\label{eq:1.2.7}
\end{equation}
La costante $\sigma_{2}$ può essere calcolata analiticamente in forma
chiusa. Si noti come il raggiungimento dell'andamento asintotico sia
particolarmente lento per la varianza in $d=2$; è in realtà possibile
fornire una stima migliore di ordine $\mathcal{O}\left(\sqrt{n}\log^{-2}\left(n\right)\right)$
attraverso una sommatoria. 

In effetti i risultati sul caso $d=2$ sono molteplici, dato l'enorme
interesse accademico suscitato sin dai primi anni '50. Ci troviamo
tuttavia costretti a omettere una esposizione sistematica, a causa
della loro scarsa rilevanza nell'ambito di questo lavoro: verranno
brevemente introdotti quando necessario durante l'esposizione. 

Come già detto in precedenza, la nostra trattazione si concentra sui
cammini transienti, dunque $d\geq3$. Per questa categoria valgono
i risultati di Jain e Pruitt (\cite{Jain-Pruitt,Jain-Orey}): 
\begin{equation}
\langle R_{n}\rangle=\left\{ \begin{array}{l}
\left(1-C_{d}\right)n+\Delta_{d}\left(2-\frac{d}{2}\right)^{-1}n^{2-\frac{d}{2}}+\mathcal{O}\left(1\right)\\
\left(1-C_{4}\right)n+\Delta_{4}\log\left(n\right)+\mathcal{O}\left(1\right)
\end{array}\,\,\,\begin{array}{r}
d\neq4\\
d=4
\end{array}\right..\label{eq:1.2.4}
\end{equation}
Le costanti $C_{d}$ sono calcolabili attraverso una integrazione
numerica (per $d=3$ anche in forma analitica). Le $\Delta_{d}$ sono
direttamente dipendenti dalle $C_{d}$: ne forniremo più avanti i
valori espliciti. Per quanto riguarda la varianza, sempre gli stessi
autori (\cite{Jain-Pruitt}) concludono che 
\begin{equation}
\langle\Delta R_{n}^{2}\rangle=\left\{ \begin{array}{l}
\sigma_{3}^{2}\,n\log\left(n\right)+\mathcal{O}\left(n\right)\\
\sigma_{d}^{2}\,n+\mathcal{O}\left(\sqrt{n}\right)
\end{array}\,\,\,\begin{array}{r}
d=3\\
d\geq4
\end{array}\right..\label{eq:1.2.5}
\end{equation}
Solo la costante $\sigma_{3}$ è effettivamente calcolata in forma
analitica. Ora, benché in \cite{Jain-Pruitt} questo non sia esplicitamente
indicato, è possibile derivare una forma integrale delle costanti
$\sigma_{d}$ anche per $d\geq4$, che le rende direttamente calcolabili
attraverso una integrazione numerica. La dimostrazione di questi risultati
è riportata in appendice nel caso $d\geq4$. 

In particolare, dal teorema di Jain-Pruitt si ottiene la seguente
forma esplicita della costante $\sigma_{d}$ nel limite (calcoli dettagliati
sono riportati in appendice): 
\begin{equation}
\sigma_{d}^{2}=C_{d}\bar{C}_{d}+2a\label{eq:1.2.8}
\end{equation}
\begin{equation}
a=\bar{C}_{d}\underset{x\neq0}{\sum}F\left(x\right)^{3}\left(1+F\left(x\right)\right)^{-1}\label{eq:1.2.8.1.1}
\end{equation}

Ricordiamo che $F\left(x\right)=F\left(0,x\right)$ rappresenta la
probabilità totale che un cammino, con estremi in $0$, $x$, e nel
limite di lunghezza infinita, abbia occupato l'origine esattamente
una volta, ed è uguale a $\bar{C}_{d}G\left(0,x\right)=\bar{C}_{d}G\left(x\right)$
per $x\neq0$.

Notiamo dalla \prettyref{eq:1.2.8.1.1} che $a<\infty$ se $\sum_{x}G\left(x\right)^{3}<\infty$:
questo è vero per $d\geq4$, mentre in $d=3$ si avrà una divergenza
logaritmica in $n$, che porta al risultato precedentemente enunciato
in (1.19). 

Per calcolare esplicitamente le $\sigma_{d}$ dobbiamo esprimere $G\left(x\right)$
in una forma che ne permetta una agevole integrazione: partiamo dalle
\prettyref{eq:1.1.4}, \prettyref{eq:1.1.8}, che danno per \foreignlanguage{english}{$G\left(x\right)$}
la seguente forma integrale 
\begin{equation}
G\left(x\right)=\left(2\pi\right)^{-d}\int_{\left[-\pi,\pi\right]^{d}}dq\,e^{iqx}\left(1-\tilde{\lambda}_{d}\left(q\right)\right)^{-1}.\label{eq:1.2.14}
\end{equation}
Quest'ultima espressione può essere scritta attraverso un integrale
di funzioni di Bessel modificate, secondo la relazione 
\begin{equation}
G\left(x\right)=\int_{0}^{\infty}d\tau\,e^{-\tau}\prod_{i=1}^{d}I_{x_{i}}\left(\frac{\tau}{d}\right),\label{eq:1.2.15}
\end{equation}
dove $x_{i}$ è la componente $i-$esima del vettore di posizione
$x$, e
\begin{equation}
I_{k}\left(u\right)=\int_{0}^{\pi}\frac{d\theta}{\pi}\,e^{u\cos\theta}\cos\left(k\theta\right)
\end{equation}
($k\in\mathbb{Z}$) è la funzione di Bessel modificata di ordine $k$.
Da questa espressione è possibile calcolare numericamente una stima
inferiore del valore di $a$, eseguendo la somma in \prettyref{eq:1.2.8.1.1}
per $\left|x\right|\leq Q$. Dato il rapido decadimento in $x$ di
$G\left(x\right)^{3}$, le stime così impostate sono già ottime per
valori di $Q<12$. Si noti inoltre che la proprietà $I_{m}\left(x\right)=I_{-m}\left(x\right)$
permette di suddividere il dominio di integrazione in sottoinsiemi
di integrali equivalenti, da risommare con le opportune molteplicità:
questo riduce notevolmente il numero calcoli da eseguire. 

Nella tabella \prettyref{Flo:tab1} sono riportati alcuni valori di
$a$, $\sigma_{d}$, calcolati con il metodo descritto (tranne i valori
di $\sigma_{2}$, $\sigma_{3}$ che sono tratti da \cite{Torney}),
assieme ai corrispondenti valori di $C_{d}$; l'incertezza è sulle
cifre in parentesi.

\begin{table}
\begin{centering}
\caption{}
\label{Flo:tab1}
\par\end{centering}
~
\centering{}%
\begin{tabular}{|c|c|c|c|}
\hline 
\noalign{\vskip\doublerulesep}
$d$ & $C_{d}$ & $a$ & $\sigma_{d}$\tabularnewline[\doublerulesep]
\hline 
\hline 
$2$ & $1$ & $\infty$ & $4.09489\left(8\right)$\tabularnewline
\hline 
$3$ & $0.3405373\left(3\right)$ & $\infty$ & $0.21514\left(5\right)$\tabularnewline
\hline 
$4$ & $0.1932016\left(7\right)$ & $0.06165\left(3\right)$ & $0.27918\left(1\right)$\tabularnewline
\hline 
$5$ & $0.1351786\left(1\right)$ & $0.02221\left(5\right)$ & $0.16133\left(5\right)$\tabularnewline
\hline 
$6$ & $0.1047155\left(0\right)$ & $0.01211\left(2\right)$ & $0.11797\left(4\right)$\tabularnewline
\hline 
$7$ & $0.0858449\left(3\right)$ & $0.00781\left(9\right)$ & $0.09411\left(4\right)$\tabularnewline
\hline 
$8$ & $0.0729126\left(5\right)$ & $0.00553\left(1\right)$ & $0.07865\left(8\right)$\tabularnewline
\hline 
\end{tabular}
\end{table}

\section{Probabilità di intersezione}

Vogliamo ora prendere in considerazione la probabilità, da parte di
un camminatore, di finire in un sito già visitato in precedenza nell'esecuzione
dell'$n-$esimo passo. Da un punto di vista statico, si può vedere
questa quantità come la probabilità che all'$n-$esimo passo si verifichi
una configurazione tale che $S_{n}=S_{n-k}$ per un qualche $k<n$.
In quest'ultima definizione è già possibile individuare la stretta
connessione con i risultati della sezione precedente, ottenuti per
il numero di intersezioni in situazione statica.

Definiamo dunque il funzionale $\pi\left[\omega\right]$, secondo
la relazione

\begin{equation}
\pi\left[\omega\right]=\left(2d\right)^{-1}\sum_{\left|x\right|=1}R_{x}\left[\omega\right].\label{eq:1.3.1}
\end{equation}
dove $R_{x}\left[\omega\right]$ è la funzione caratteristica del
set $\omega$ (con l'estremo di partenza nell'origine), il cui valore
è $1$ se $x\in\omega$, e $0$ altrimenti: è facile verificare come
una integrazione sullo spazio restituisca il range $R\left[\omega\right]$
della realizzazione $\omega$. Nella moderna letteratura $R_{x}\left[\omega\right]$
viene indicata come \emph{tempo locale di auto-intersezione} di ordine
0.

Ora, la $\pi\left[\omega\right]$, che chiameremo \emph{ambiente},
rappresenta la frazione di siti occupati intorno all'origine, dunque,
se proseguiamo il cammino aggiungendo un monomero all'estremo nell'origine,
$\pi\left[\omega\right]$ rappresenterà proprio la probabilità che
questo passo finisca in un sito già occupato dal resto del cammino. 

Consideriamo $\pi_{n}=\langle\pi\rangle$, valore atteso su $\Omega_{n}$:
questo valore rappresenta il numero medio di siti occupati intorno
all'origine in un random walk di $n$ passi. Data la simmetria geometrica
di un SRW per inversione dell'ordinamento temporale dei siti visitati,
questa quantità indica anche la probabilità, per un generico cammino,
di saltare in un sito occupato quando viene aggiunto un monomero.

Per calcolare $\pi_{n}$ prendiamo un cammino $\omega$ che all'$n-$esimo
passo si trovi ad occupare il sito $S_{k}$, già precedentemente occupato
al tempo $k<n$. Possiamo dividere il cammino in due sottoinsiemi:
\begin{equation}
\begin{array}{l}
\omega_{I}=\left\{ S_{j}\,|\,j\in\left[0,k\right)\right\} \\
\\\omega_{L}=\left\{ S_{j}\,|\,j\in\left[k,n\right]\right\} 
\end{array}\label{eq:1.3.2}
\end{equation}
 dove il sottoinsieme $\omega_{L}$ è un cammino chiuso di $n-k$
passi, mentre $\omega_{I}$ è un qualunque cammino di $k-1$ passi.
Siamo interessati a calcolare la probabilità di una intersezione di
qualunque ordine nel passo $n-1\rightarrow n$; possiamo ignorare
l'insieme $\omega_{I}$, e calcolare la probabilità che si verifichi
una qualunque configurazione chiusa $\omega_{L}$. Se ridefiniamo
l'origine in $S_{k}$, notiamo immediatamente l'equivalenza con la
probabilità $f_{n-k}$ di ritorno nell'origine esattamente al passo
$n-k$. Siccome a noi interessa che l'intersezione avvenga, indipendentemente
dal valore di $k$, sommiamo sulle possibili realizzazioni di $\omega_{L}$
con lunghezza di al massimo $n$, ottenendo 
\begin{equation}
\pi_{n}=\sum_{k=0}^{n}f_{n-k}=F_{n}.\label{eq:1.3.4}
\end{equation}
Dunque $\pi_{n}$ è uguale alla probabilità di ritorno all'origine
per un cammino di $n$ passi.

Dalla \prettyref{eq:1.3.4} si può immediatamente osservare che nel
limite di catena infinita si ha $\pi_{n}\rightarrow C_{d}$. Possiamo
ottenere anche una stima delle fluttuazioni come segue. Si consideri
il numero medio di visite all'origine nella forma $G_{n}=\bar{C}_{d}^{-1}-\Sigma_{n}$:
da uno sviluppo al primo ordine in analogia con la relazione $\bar{C}_{d}=\rho_{d}^{-1}$
(\cite{Feller}) si ottiene $\pi_{n}\simeq C_{d}-\bar{C}_{d}^{2}\Sigma_{n}$,
con $\Sigma_{n}$ data dalla relazione (per \prettyref{eq:1.1.4},
\prettyref{eq:1.1.8}) 
\begin{equation}
\Sigma_{n}=\left(2\pi\right)^{-d}\int_{\left[-\pi,\pi\right]^{d}}dq\,\tilde{\lambda}_{d}\left(q\right)^{n}\left(1-\tilde{\lambda}_{d}\left(q\right)\right)^{-1}.\label{eq:1.3.5}
\end{equation}
L'integrale \prettyref{eq:1.3.5} non è esattamente risolubile, tuttavia
è stimabile con buona precisione attraverso le seguenti approssimazioni.
Per prima cosa sostituiamo $\cos\left(q_{i}\right)$ nel fattore di
struttura con una espansione al secondo ordine in $q=0$; questo ci
porta alla forma $\tilde{\lambda}_{d}\left(q\right)\simeq1-\left(2d\right)^{-1}\left|q\right|^{2}$.
Dato che la $\tilde{\lambda}_{d}\left(q\right)$ così approssimata
può assumere valori negativi nel dominio di integrazione della \prettyref{eq:1.3.5},
dobbiamo cambiare anche il dominio $\left[-\pi,\pi\right]^{d}$ nella
ipersfera $q\in\left\{ \left|q\right|^{2}\leq2d\right\} $. Si noti
che il nuovo dominio esclude dall'integrazione le singolarità dell'integrando
ai vertici di $\left[-\pi,\pi\right]^{d}$, dando solo metà del valore
totale: è quindi necessario introdurre ``a mano'' un fattore correttivo
di $1/2$. Passando alle coordinate polari $dq\rightarrow dV_{q}=2\Gamma\left(\frac{d}{2}\right)^{-1}\pi^{-\frac{d}{2}}\left|q\right|^{d-1}d\left|q\right|$
si ha: 
\begin{equation}
\Sigma_{n}\simeq\frac{2^{2-d}d}{\Gamma\left(\frac{d}{2}\right)\pi^{\frac{d}{2}}}\int_{0}^{\sqrt{2d}}d\left|q\right|\left|q\right|^{d-3}\left(1-\frac{\left|q\right|}{2d}\right)^{n+1}.\label{eq:1.3.6}
\end{equation}
Cambiando la variabile di integrazione in $\tau=\left|q\right|^{2}/2d$,
otteniamo un integrale di Eulero del primo tipo
\begin{equation}
\Sigma_{n}\simeq\frac{1}{\Gamma\left(\frac{d}{2}\right)}\left(\frac{d}{2\pi}\right)^{\frac{d}{2}}\int_{0}^{1}d\tau\,\left(1-\tau\right)^{\frac{d}{2}-2}\tau^{n+1}.\label{eq:1.3.7}
\end{equation}
Ora, usando la nota relazione $\int_{0}^{1}d\tau\left(1-\tau\right)^{\alpha}\tau^{\beta}=\Gamma\left(\alpha\right)\Gamma\left(\beta\right)\Gamma\left(\alpha+\beta\right)^{-1}$,
e l'approssimazione di Stirling nel limite $n\rightarrow\infty$,
otterremo l'andamento 
\begin{equation}
\frac{\Gamma\left(\frac{d}{2}-2\right)\Gamma\left(n+1\right)}{\Gamma\left(\frac{d}{2}+n-1\right)}\simeq\Gamma\left(\frac{d}{2}\right)\frac{n^{1-\frac{d}{2}}}{\left(\frac{d}{2}-1\right)},
\end{equation}
da cui le relazioni per $\pi_{n}$: 
\begin{equation}
\pi_{n}=C_{d}-\Delta_{d}n^{1-\frac{d}{2}}+o\left(1\right),\label{eq:1.3.8}
\end{equation}
con l'ampiezza di fluttuazione $\Delta_{d}$ data dall'espressione
\begin{equation}
\Delta_{d}\simeq\frac{4\bar{C}_{d}^{2}}{\left(d^{2}-3d+2\right)}\left(\frac{d}{2\pi}\right)^{\frac{d}{2}}.\label{eq:1.3.9}
\end{equation}
Come ultima osservazione, notiamo che la somma $\sum_{t=0}^{n}\pi_{t}$
corrisponde al numero medio $\langle M_{n}\rangle$ di intersezioni
in un cammino di $n$ passi: dunque la $\Delta_{d}$ che compare nella
\prettyref{eq:1.3.8} corrisponde a quella che troviamo nelle relazioni
\prettyref{eq:1.2.4}, valide per il range $\langle R_{n}\rangle=n+1-\langle M_{n}\rangle$. 

Chiaramente le espressioni appena ottenute sono valide solo per cammini
transienti (nel nostro caso $d\geq3$), in quanto per quelli ricorrenti
l'integrale \prettyref{eq:1.3.5} diverge; per $d=2$ è comunque possibile
ricavare l'andamento di $\pi_{n}$ differenziando in $n$ la relazione
\prettyref{eq:1.2.6}.

\section{Funzione \emph{ambiente} e distribuzione delle intersezioni}

E' possibile calcolare la distribuzione delle intersezioni $P_{n}\left(M\right)$
a partire dalla media di $\pi\left[\omega\right]$ sull'insieme dei
cammini $\Omega_{n}\left(M\right)$, definito come
\begin{equation}
\Omega_{n}\left(M\right)\equiv\left\{ \left|\omega\right|=n,\,M\left[\omega\right]=M\right\} 
\end{equation}
Detta dunque $\left|\Omega_{n}\left(M\right)\right|$ la cardinalità
dell'insieme dei cammini con esattamente $M$ intersezioni, definiamo
la funzione $\pi_{n}\left(M\right)$ attraverso la relazione:

\begin{equation}
\pi_{n}\left(M\right)=\left|\Omega_{n}\left(M\right)\right|^{-1}\sum_{\omega\in\Omega_{n}\left(M\right)}\pi\left[\omega\right]\label{eq:3.0.1}
\end{equation}
La $\pi_{n}\left(M\right)\in\left[0,\,1\right]$ rappresenta la probabilità
di incontrare un sito già visitato quando aggiungiamo un monomero
ad un cammino di $n$ passi, con esattamente $M$ intersezioni.

Si consideri ora il processo stocastico nella variabile $M_{t}$,
definito dalle seguenti equazioni:\\
\begin{equation}
M_{t+1}=M_{t}+\delta_{t}M_{t},\label{eq:3.1.1-1}
\end{equation}

\begin{equation}
\delta_{t}M_{t}=\theta\left(w_{t}-\pi_{t}\left(M_{t}\right)\right),\label{eq:3.1.2-1}
\end{equation}
\\
dove $\theta\left(\cdot\right)$ è la funzione di Heaviside, e $w_{t}\in\left[0,\,1\right]$
è una variabile aleatoria di supporto, distribuita in modo uniforme
nell'intervallo secondo la distribuzione $\partial_{w}P\left(w\right)=0$.
L'eguaglianza tra $P_{n}\left(M\right)$ del SRW e la distribuzione
derivante dalle \prettyref{eq:3.1.1-1}, \prettyref{eq:3.1.2-1} segue
immediatamente dalla definizione di $\pi_{t}\left(M\right)$.

Una diretta relazione funzionale fra $\pi_{n}\left(M\right)$ e la
distribuzione $P_{n}\left(M\right)$ può essere fornita attraverso
l'introduzione di una distribuzione ausiliaria $\varphi_{n}\left(t\right)$,
che assume esclusivamente i valori $0$ o $1$ al variare del parametro
intero $t\in\left[0,\,n\right]$. Introduciamo il funzionale 
\begin{equation}
W_{n}\left[\varphi\right]=\prod_{t=1}^{n}\left[1-\varphi_{n}\left(t\right)-\pi_{t}\left(M_{t}\left[\varphi\right]\right)+2\,\varphi_{n}\left(t\right)\,\pi_{t}\left(M_{t}\left[\varphi\right]\right)\right],\label{eq:3.3.1}
\end{equation}
\begin{equation}
M_{t}\left[\varphi\right]=\sum_{\tau=1}^{t}\varphi_{n}\left(\tau\right):
\end{equation}
questa quantità corrisponde al peso statistico di una configurazione
$\omega$ le cui intersezioni sono avvenute esattamente ai tempi $t'\,|\,\varphi_{n}\left(t'\right)=1$.
Le configurazioni impossibili sono comprese, e danno un peso $W_{n}\left[\varphi\right]$
nullo. 

Considerando che la somma $\pi\left[\omega\right]+\bar{\pi}\left[\omega\right]=1$
dei pesi su tutte le possibili realizzazioni di $\varphi_{n}\left(t\right)$
deve valere 
\begin{equation}
\sum_{\forall\varphi}W_{n}\left[\varphi\right]=\prod_{t=1}^{n}\sum_{\varphi=1,0}\left[1-\varphi_{n}\left(t\right)-\pi_{t}\left(M_{t}\left[\varphi\right]\right)+2\,\varphi_{n}\left(t\right)\,\pi_{t}\left(M_{t}\left[\varphi\right]\right)\right]=1,\label{eq:3.3.2}
\end{equation}
possiamo esprimere la distribuzione di $M$ attraverso la somma
\begin{equation}
P_{n}\left(M\right)=\sum_{\varphi\in\Phi_{n}\left(M\right)}W\left[\varphi\right],\label{eq:3.3.3}
\end{equation}
definita sui $\varphi$ appartenenti all'insieme 
\begin{equation}
\Phi_{n}\left(M\right)=\left\{ \left|\varphi_{n}\left(t\right)\right|=n,\,M_{n}\left[\varphi\right]=M\right\} ,
\end{equation}
che comprende costruzioni di $\varphi$ con uguale $M_{n}\left[\varphi\right]=M$;
si noti come la somma su $M$ della \prettyref{eq:3.3.3} dia immediatamente
la \prettyref{eq:3.3.2}.

Attraverso semplici passaggi algebrici, è possibile legare $\pi_{n}\left(M\right)$
a probabilità di intersezione $\pi_{n}$ e numero medio di intersezioni
$\langle M_{n}\rangle$ del SRW:
\begin{equation}
\pi_{n}=\sum_{M=0}^{n-1}\pi_{n}\left(M\right)\cdot P_{n}\left(M\right)\label{eq:3.0.2}
\end{equation}
\begin{equation}
\langle M_{n}\rangle=\sum_{M=0}^{n-1}M\cdot P_{n}\left(M\right)=\sum_{t=0}^{n}\sum_{M=0}^{t-1}\pi_{t}\left(M\right)\cdot P_{t}\left(M\right)\label{eq:3.0.3}
\end{equation}
Si individuano subito due valori particolari di $\pi_{n}\left(M\right)$:
il primo si ha per $M=0$, che rappresenta la media dell'ambiente
$\pi\left[\omega\right]$ sull'insieme $\Omega_{n}\left(0\right)$
dei \emph{Self-Avoiding Walks} (SAW, in appendice è presente un breve
riassunto dei risultati noti per questo modello). Questo valore è
$\pi_{n}\left(0\right)\rightarrow\eta_{d}$, che può essere calcolato
a partire dalla costante connettiva del SAW $\mu_{d}$ secondo la
relazione 
\begin{equation}
\eta_{d}=1-\frac{\mu_{d}}{2d}.
\end{equation}
Il secondo può essere individuato a partire dalle \prettyref{eq:1.2.2},
\prettyref{eq:1.2.3}, \prettyref{eq:1.2.3.1}: la distribuzione $P_{n}\left(M\right)$
nel limite di $n\rightarrow\infty$ tende ad una delta in $\langle M_{n}\rangle$,
da cui segue che $\lim_{n\rightarrow\infty}\pi_{n}\left(\langle M_{n}\rangle\right)=\lim_{n\rightarrow\infty}\pi_{n}=C_{d}$.
Come vedremo più avanti, è possibile imporre una terza condizione
legata alla varianza del SRW.

\chapter{Intersezioni per monomero vincolate}

Fino ad ora abbiamo considerato catene di lunghezza $n$, con $M\in\left[0,\,n-1\right]$
intersezioni; passeremo ora allo studio di cammini caratterizzati
da un rapporto fissato $\tilde{m}=n^{-1}M$ tra numero di intersezioni
e numero di monomeri. A questo scopo si introducono il parametro reale
$m\in\left[0,\,1\right]$ e l'insieme associato $\Omega_{m}$, definito
come\\
\begin{equation}
\Omega_{m}\equiv\left\{ \left|\omega\right|=n,\,M\left[\omega\right]=\left\lfloor m\left(n-1\right)\right\rfloor \right\} ,
\end{equation}
\\
dove $\left\lfloor a\right\rfloor $ indica la parte intera di $a\in\mathbb{R}$.
Nel limite di $n$ grandi si ha $m\left(n-1\right)\simeq m\,n$, dunque
possiamo considerare l'insieme $\Omega_{m}$ come l'insieme dei cammini
che hanno $m\simeq\tilde{m}$ intersezioni per monomero. Da qui in
poi, dove non specificato diversamente, lavoreremo in approssimazione
$M=m\,n$.

In questo capitolo studieremo l'ensemble $\Omega_{m}$ al variare
del parametro: in particolare ci si concentrerà sull'interpretazione
di simulazioni numeriche, consistenti in cammini di lunghezza $\mathcal{O}\left(10^{3}\right)$
per reticoli con $2\leq d\leq6$, realizzati utilizzando una variante
a un parametro del Pruned-Enriched Rosenbluth Method (PERM, si veda
la sezione relativa in Appendice).

\section{Proprietà geometriche dei cammini al variare di $m$}

Come primo approccio vogliamo osservare il comportamento della distanza
quadratica end-to-end mediata su $\Omega_{m}$: 
\begin{equation}
\Re_{n}^{2}\left(m\right)=\langle\,\,\left|S_{n}-S_{0}\right|^{2}\rangle_{\Omega_{m}}
\end{equation}
Introduciamo la distribuzione $P_{n}\left(m\right)$, definita attraverso
il cambio di variabile $M\rightarrow n^{-1}M$ su $P_{n}\left(M\right)$
: dalle \prettyref{eq:1.2.2}, \prettyref{eq:1.2.3} si osserva che
$P_{n}\left(m\right)$ tende ad una delta in $C_{d}$ per $n\rightarrow\infty$.
Ne segue che media e massimo della distribuzione coincidono nel limite
considerato, e dunque vale la relazione 
\begin{equation}
\Re_{n}^{2}\left(C_{d}\right)\simeq\langle\Re_{n}^{2}\left(C_{d}\right)\rangle_{m},
\end{equation}
con $\langle\Re_{n}^{2}\left(C_{d}\right)\rangle_{m}=n$ raggio quadratico
medio del SRW. Nel caso $m=0$ , $\Re_{n}^{2}\left(0\right)$ corrisponderà
alla distanza quadratica media del SAW: 
\begin{equation}
\Re_{n}^{2}\left(0\right)\simeq D\,n^{2\nu_{d}}:
\end{equation}
gli esponenti caratteristici per il SAW saranno indicati con $\nu_{d}$
(per una trattazione più dettagliata si veda in appendice). 

Dalle simulazioni si osserva che per tutte le dimensioni vale la relazione
$\Re_{n}^{2}\left(m\right)\propto n^{2\nu\left(m\right)}$, con $\nu\left(m\right)$
esponente critico dipendente da $m$. Di seguito discuteremo i comportamenti
a seconda della dimensionalità $d$ del reticolo.

Il caso $d=1$ è piuttosto semplice: in una catena unidimensionale
il range è proporzionale alla distanza end-to-end, dunque fissare
il parametro ad un valore $m<1$ significa imporre una distanza end-to-end
con andamento balistico (proporzionale ad $n$). In particolare per
grandi $n$ si avrà \foreignlanguage{english}{$\Re_{n}^{2}\left(m\right)\sim\left(1-m\right)^{2}n^{2}$},
andamento simile a quello del SAW unidimensionale ($\nu_{1}=1$),
che viene recuperato esattamente imponendo $m=0$. 

Un SRW in $d=1$ ha $\nu=1/2$, dunque il range medio avrà un andamento
$\langle R_{n}\rangle\propto n^{1/2}$: ne deriva che, per $m\sim1-\mathcal{O}\left(n^{-1/2}\right)$,
l'esponente dovrà cadere da $\nu=1$ a $\nu=1/2$, determinando una
discontinuità di $\nu\left(m\right)$ in $m=1$. Come vedremo a breve,
questo comportamento sarà comune a tutte le dimensioni reticolari
per un valore critico $m_{c}$ del parametro.

\begin{figure}
\begin{centering}
\caption{}
\includegraphics[scale=0.5]{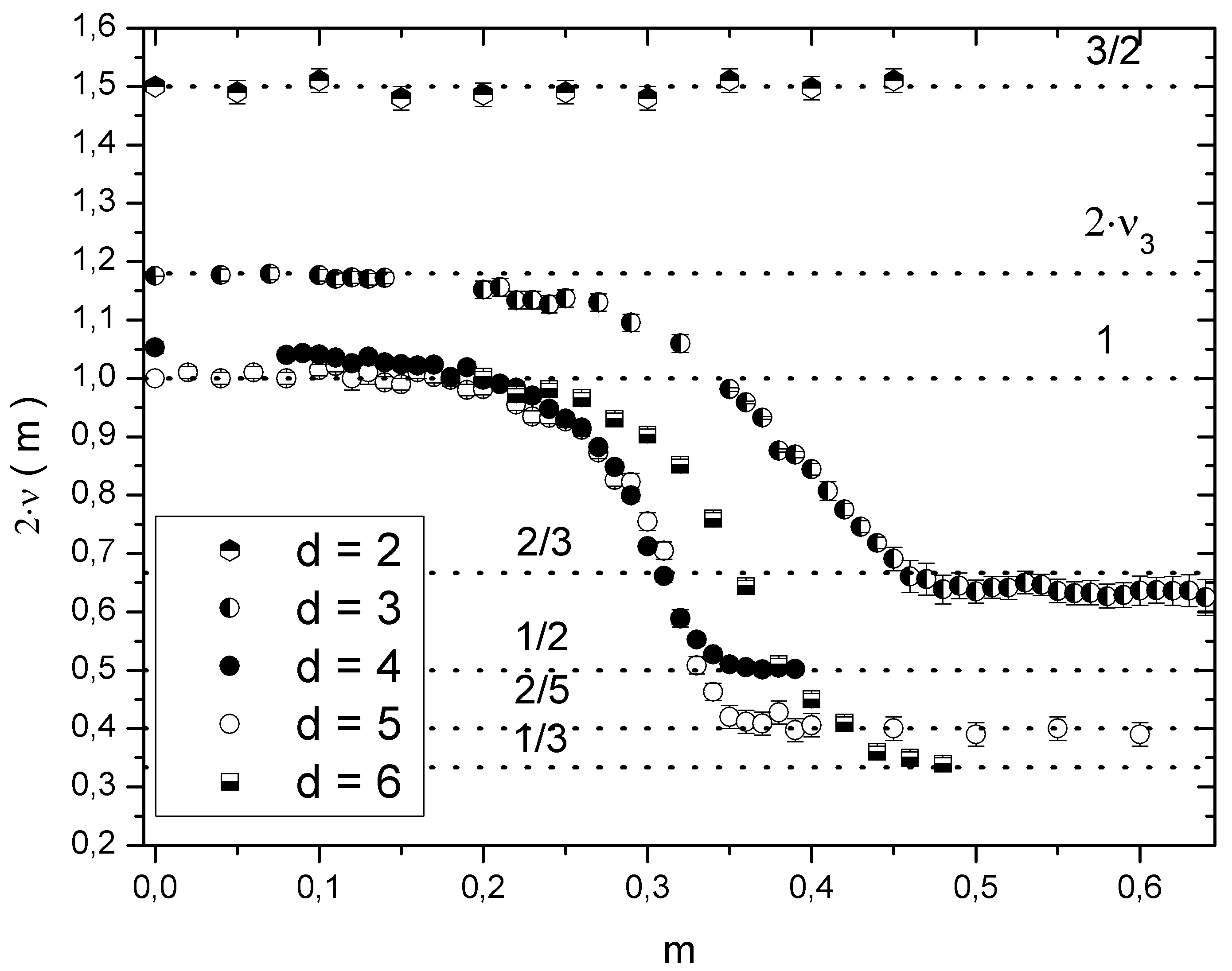}\label{Flo:FIGURA1}
\par\end{centering}
\begin{centering}
\label{Flo:ESPONENTI}
\par\end{centering}
\emph{Nel grafici sono riportati i valori di $\nu\left(m\right)$,
ottenuti da fit di $\log\left(\Re_{n}^{2}\left(m\right)\right)$ vs
$\log\left(n\right)$: il range per ogni fit è $n\in\left[0.8\,n_{M},\,n_{M}\right]$.
La lunghezza massima dei cammini simulati è stata: per $d=2$, $n_{M}=2\cdot10^{3}$,
per $d=3$, $n_{M}=10^{3}$, per $d=4$, $n_{M}=2\cdot10^{3}$, per
$d=5$, $n_{M}=10^{3}$, per $d=6$, $n_{M}=0,5\cdot10^{3}$. Nel
caso $d=3$, inoltre, la $\log\left(\Re_{n}^{2}\left(m\right)\right)$
è stata realizzata imponendo un parametro $m=C_{3}M/\langle M_{n}\rangle$
dipendente da $n$, dove $C_{3}/\langle M_{n}\rangle$ è una correzione
inserita per tenere conto che $n<\infty$ (si veda la Figura 2.3).
Per $d>3$ gli effetti di taglia finita sono molto più deboli, e questo
accorgimento non è necessario.}

\emph{Le simulazioni per $d=4$ mostrano un esponente leggermente
maggiore di $1$ di un ordine $\mathcal{O}\left(10^{-2}\right)$:
questo è compatibile con la correzione logaritmica prevista, che per
catene di $2\cdot10^{3}$ dà un esponente apparente di $\nu\sim1.067$.}
\end{figure}

\begin{figure}
\begin{centering}
\caption{}
\par\end{centering}
\begin{centering}
\includegraphics[scale=0.16]{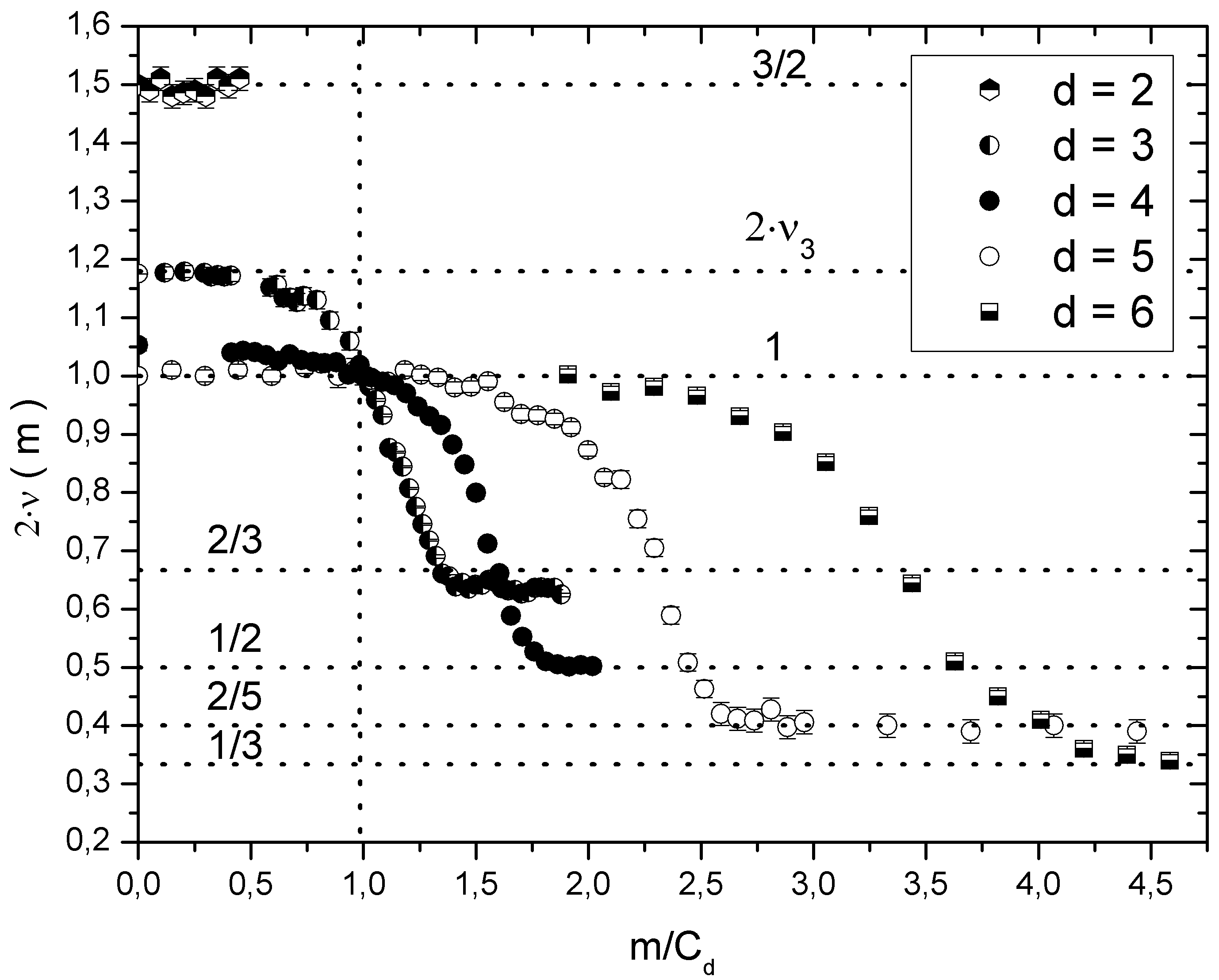}\label{Flo:FIGURA2}
\par\end{centering}
\begin{centering}
\label{Flo:ESPONENTI=000020NORMAL}
\par\end{centering}
\emph{In questo grafico gli stessi valori di $\nu\left(m\right)$
in Figura \prettyref{Flo:FIGURA1} sono riportati in funzione del
rapporto $m/C_{d}$ (con $C_{2}=1$). Questo rende più agevole il
confronto tra reticoli a dimensionalità differente: in particolare
la figura sembra confermare la relazione $m_{c}>C_{d}$ per $d=5,\,6$.
Si noti che il rapporto $m_{c}/C_{d}$ è apparentemente crescente
in $d$.}
\end{figure}

Per $d=2$ la situazione è concettualmente simile. Se $m<1$ si ritrova
$\nu\left(m\right)=\nu_{2}$. I cammini bidimensionali sono ancora
ricorrenti, dunque si avrà $C_{2}=1$, e la conseguente discontinuità
in $m=1$ (da $\nu_{2}$ ad $1/2$) per $\nu\left(m\right)$. In particolare
la \prettyref{eq:1.2.6} prevede il salto per $m=1-\mathcal{O}\left(\log\left(n\right)^{-1}\right)$.
A parte una verifica di $\nu\left(m<1\right)=\nu_{2}$ (si vedano
le Figure \prettyref{Flo:FIGURA1}, \prettyref{Flo:FIGURA2}), il
caso bidimensionale non è stato trattato, preferendo concentrare l'interesse
sui reticoli $d\geq3$.

Il caso $d\geq3$ presenta infatti un fenomeno interessante: le Figure
\prettyref{Flo:FIGURA1}, \prettyref{Flo:FIGURA2} indicano l'esistenza
di un valore critico $0<m_{c}<1$, oltre il quale le catene collassano
in una configurazione semi-compatta con $\nu\left(m\right)=1/d$ (range
proporzionale al volume). Per $m<m_{c}$ si trova invece $\nu\left(m\right)=\nu_{d}$:\\
\begin{equation}
\nu\left(m\right)=\left\{ \begin{array}{l}
\nu_{d}\\
\nu_{c}\\
\frac{1}{d}
\end{array}\begin{array}{l}
m<m_{c}\\
m=m_{c}\\
m>m_{c}
\end{array}\right.,\label{eq:2.1.E}
\end{equation}
\begin{table}
\begin{centering}
\caption{}
\label{Flo:tab1-1}
\par\end{centering}
~
\begin{centering}
\begin{tabular}{|c|c|c|c||c|}
\hline 
\noalign{\vskip\doublerulesep}
$d$ & $\nu\left(m<m_{c}\right)$ & $\nu\left(m=m_{c}\right)$ & $\nu\left(m>m_{c}\right)$ & $m_{c}/C_{d}$\tabularnewline[\doublerulesep]
\hline 
\hline 
$3$ & $\nu_{3}$ & $1/2$ & $1/3$ & $1$\tabularnewline
\hline 
$4$ & $\,\,\,\,\,\,\,\,1/2\,_{l.c.}$ & $1/2$ & $1/4$ & $1$\tabularnewline
\hline 
$5$ & $1/2$ & $-$ & $1/5$ & $1.5\backsim2.7$\tabularnewline
\hline 
$6$ & $1/2$ & $-$ & $1/6$ & $2.1\backsim4.4$\tabularnewline
\hline 
\end{tabular}\\
$\,$
\par\end{centering}
\centering{}{\small\emph{l.c. indica la correzione logaritmica \prettyref{eq:logcorr4d}}}{\small\par}
\end{table}
\\
dove abbiamo indicato con $\nu_{c}$ l'esponente relativo all'andamento
critico. Le simulazioni indicano dunque una transizione nel parametro
$m$, dove lo stato esteso scala come un SAW, e quello collassato
come un cluster compatto.

Questi dati sono consistenti con un lavoro di Van den Berg, Bolthausen,
den Hollander (\cite{Den=000020Hollander}), in cui viene affrontato
il problema di una Wiener Sausage $W^{a}\left(t\right)$ vincolata
ad avere un volume $\left|W^{a}\left(t\right)\right|\leq\bar{m}t$:
si tratta dell'analogo continuo del problema da noi trattato (i risultati
di questo lavoro sono brevemente riassunti in Appendice). 

Il lavoro \cite{Den=000020Hollander} mostra rigorosamente l'esistenza
di una $m_{c}$, giungendo alle medesime conclusioni per quanto riguarda
la configurazione geometrica delle catene sotto il valore critico
di $\bar{m}_{c}$: in particolare si ipotizza per il caso collassato
una configurazione \emph{a gruviera}, con i buchi taglia $\mathcal{O}(1)$,
sempre più densi quando ci si allontana dall'origine.

Per le simulazioni in $d=3$ possiamo fare alcune considerazioni.
Dai risultati noti in letteratura per il modello di Stanley repulsivo,
quando $m<C_{3}$ si deve avere $\nu\left(m\right)=\nu_{3}$ (il motivo
verrà chiarito più avanti nel capitolo relativo ai modelli termodinamici),
mentre, assumendo il medesimo comportamento del modello continuo,
da \cite{Den=000020Hollander} si deduce che $\nu\left(m>C_{3}\right)=1/3$:
sotto queste ipotesi si ha quindi $m_{c}=C_{3}$. Si può inoltre concludere
che l'andamento di $\Re_{n}^{2}\left(m\right)$ al punto critico deve
avere $\nu_{c}=1/2$, ovvero, per $d=3$ si ha una transizione di
tipo \emph{Coil-globule} nel parametro $m$ (questo rappresenta un
risultato sorprendente, di cui tratteremo diffusamente nel capitolo
conclusivo).

Si noti tuttavia che la configurazione collassata $\nu\left(m>C_{3}\right)=1/3$
viene raggiunta solo dopo un lungo crossover. La Figura 2.3 mostra
$\nu$ contro $n^{-1}M$ e $C_{3}M/\langle M_{n}\rangle$: il primo
caso non tiene conto della taglia finita del sistema simulato, e mostra
un esponente $\nu>1/3$ (cluster non compatto, con buchi percolanti
di taglia divergente $\mathcal{O}(n^{\epsilon})$), mentre il secondo
considera sommariamente le fluttuazioni di $\langle M_{n}\rangle$,
dando un risultato compatibile con l'ipotesi $\nu\left(m>C_{3}\right)=1/3$.

Allo stesso modo possiamo discutere il caso $d=4$: sempre dal modello
di Stanley, che per $d=4$ , $\beta<\infty$ prevede lo stesso scaling
del SAW, quando $m<C_{4}$ si avrà $\nu\left(m\right)=1/2$ con una
correzione logaritmica all'andamento a potenza:
\begin{equation}
\Re_{n}^{2}\left(m<C_{4}\right)\propto n\,\log\left(n\right)^{\frac{1}{4}}.\label{eq:logcorr4d}
\end{equation}
\\
Questo risultato è confermato dalle simulazioni (Figure \prettyref{Flo:FIGURA1},
\prettyref{Flo:FIGURA2}), dove troviamo un esponente più grande dell'unità
di un ordine $\mathcal{O}\left(\log\left(\log\left(n\right)\right)/\log\left(n\right)\right)$. 

Dai dati si osserva anche che $\nu\left(m>C_{4}\right)=1/4$, in accordo
con il modello continuo: dunque, anche per $d=4$ la transizione è
prevista per $m=C_{4}$, con $\nu_{c}=1/2$ (transizione di tipo \emph{Coil-globule}). 

Per quanto riguarda $d\geq5$, i dati delle simulazioni con $d=5,\,6$
sembrano confermare la \prettyref{eq:2.1.E}, con $m_{c}>C_{d}$ (in
accordo con \cite{Den=000020Hollander}): in Tabella \prettyref{Flo:tab1-1}
sono riassunte le considerazioni fatte fino a questo punto per $3\leq d\leq6$. 

Un ultimo commento va ai grafici che mostrano la quantità $n^{-2\nu_{d}}\Re_{n}^{2}\left(m\right)$
in funzione del parametro $m$. Il limite 
\begin{equation}
\varrho_{d}\left(m\right)=\lim_{n\rightarrow\infty}\frac{\Re_{n}^{2}\left(m\right)}{\Re_{n}^{2}\left(0\right)}\label{eq:VARRO}
\end{equation}
si annulla nel punto critico, e potrebbe essere considerato un buon
parametro d'ordine per caratterizzare la transizione: purtroppo i
dati nelle Figure 2.4, 2.5, 2.6, 2.7, 2.8, pur rilevando chiaramente
il passaggio dalla fase estesa a quella collassata, non hanno una
taglia sufficiente per mostrare l'andamento asintotico in prossimità
di $m_{c}$, e fornire informazioni sull'ordine della transizione.

Per $d=3,4$ è comunque possibile stabilire una connessione con il
modello di Stanley, che ci consente di fare alcune ipotesi circa il
tipo di transizione. In particolare, si anticipa che per $d=3$ la
transizione è supposta continua. La questione sarà trattata più avanti,
nel capitolo sui modelli termodinamici.

\section{La funzione \emph{ambiente}}

Passiamo ora all'ambiente $\pi\left[\omega\right]$, mediato su $\Omega_{m}$:
definiamo $\pi_{n}\left(m\right)=\langle\pi\left[\omega\right]\rangle_{\Omega_{m}}$,
e il suo limite 
\begin{equation}
\pi\left(m\right)=\lim_{n\rightarrow\infty}\pi_{n}\left(m\right).
\end{equation}
Con un ragionamento identico a quello fatto per $\Re_{n}^{2}\left(m\right)$,
anche in questo caso individuiamo due valori particolari: per $m=C_{d}$
si deve avere $\pi\left(C_{d}\right)=C_{d}$, mentre $\pi\left(0\right)=\eta_{d}$.
Per quanto riguarda $m<m_{c},$ le simulazioni indicano l'esistenza
di $\pi\left(m\right)$: dalla Figura 2.11 (il caso mostrato è $d=5$,
ma si osserva lo stesso anche nelle altre dimensioni) si nota chiaramente
come $\pi_{n}\left(m\right)$ converga ad una funzione $\pi\left(m\right)$
per $m<m_{c}$.

Dalle Figure 2.9, 2.10, 2.11, 2.12, 2.13 è anche evidente come $\pi_{n}\left(m\right)$
abbia una discontinuità in $m_{c}$, oltre la quale, per problemi
di efficienza computazionale, non è stato possibile ottenere dati
conclusivi su una eventuale forma di $\pi\left(m\right)$. 

Possiamo comunque fare alcune osservazioni: i dati suggeriscono infatti
$\pi_{n}\left(m\right)<1$ per $m>m_{c}$, con $\pi\left(m\right)$
continua per $m\rightarrow m_{c}^{+}$, $m\rightarrow m_{c}^{-}$.
Se per $m_{c}<m<1$ l'ambiente è ancora $\pi\left(m\right)<1$, ne
deriva che il range continua ad avere una frazione di spazio vuoto,
di taglia complessiva comparabile, distribuita al suo interno. Dato
che se $\nu\left(m\right)=1/d$, si deve avere un volume occupato
dalla catena di ordine $\mathcal{O}\left(\langle R\rangle_{\Omega_{m}}\right)=\mathcal{O}\left(n\right)$,
allora anche il volume complessivo occupato dai buchi dovrà essere
di ordine $\mathcal{O}\left(n\right)$. Questo è consistente con il
comportamento descritto in \cite{Den=000020Hollander} per la WS collassata,
dove si ipotizzano buchi di taglia $\mathcal{O}\left(1\right)$ uniformemente
distribuiti nel volume del cluster (dando un vuoto complessivo di
ordine $\mathcal{O}\left(n\right)$).

E' intuitivamente chiaro che 
\begin{equation}
\lim_{m\rightarrow1}\pi\left(m\right)=1:
\end{equation}
nel limite $m\rightarrow1$ infatti il range equivale al volume occupato
dalla catena, dunque, se l'origine si trova strettamente contenuta
in esso, non avrà siti liberi nel suo intorno. Dai dati per $n$ finito
si osserva tuttavia un crollo di $\pi_{n}\left(m\right)$ verso il
valore $1/2d$ (quando si ha $m=1$, la catena è ridotta ad un singolo
arco tra i siti reticolari, dando necessariamente $\pi_{n}\left(1\right)=1/2d$):
questo avviene quando il volume del cluster diventa molto piccolo,
tanto da non consentirci di trascurare la probabilità di avere l'origine
sulla frontiera (a titolo d'esempio, in Figura 2.11 è mostrato il
caso $d=5$). 

Da queste due relazioni deriva l'idea che, per $n\rightarrow\infty$,
sia possibile scomporre $\pi_{n}\left(m\right)$ in due funzioni:
una $\pi\left(m\right)$ dipendente solo dal rapporto $m=M/n$, e
una $\bar{\psi}_{n}\left(m\right)=1-\psi_{n}\left(m\right)$, indicante
la probabilità di trovarsi sulla frontiera del volume occupato da
una catena collassata: avremo dunque

\begin{equation}
\pi_{n}\left(m\right)\simeq\pi\left(m\right)\psi_{n}\left(m\right)+\lambda\left(m\right)\bar{\psi}_{n}\left(m\right),
\end{equation}
\\
dove solo $\psi_{n}\left(m\right)$ contiene gli effetti di taglia
finito del sistema, e $\lambda\left(m\right)$ è il valore dell'\emph{ambiente}
condizionato al fatto di trovarsi sulla frontiera del cluster ($\lambda\left(1\right)=1/2d$).

Abbiamo detto che gli effetti di bordo insorgono per $m$ prossimo
all'unità, quando la catena è un cluster così compresso da avere una
frontiera comparabile con il volume: è logico concludere che nel limite
$n\rightarrow\infty$ questi effetti siano trascurabili per ogni $m\in\left[0,\,1\right)$,
e che la funzione $\psi_{n}\left(m\right)$ sia apprezzabilmente minore
di $1$ solo in una zona $m\in\left[1-\mathcal{O}\left(n^{-s}\right),\,1\right]$,
$s<1$. Come vedremo a breve, l'ordine di decadimento nella zona rilevante
è supposto di tipo $n^{-s}$. Si sottolinea che un effetto di bordo
dell'ordine di una potenza può insorgere solo nel caso collassato,
perché i decadimenti per effetti di bordo in catene estese sono tipicamente
di ordine esponenziale. 

Non è stato possibile indagare numericamente $\bar{\psi}_{n}\left(m\right)$
per valori significativi di $n$; tuttavia la seguente argomentazione
euristica fornisce una stima che, come vedremo più avanti, risulterà
in accordo con quanto si osserva sulla WS. 

Consideriamo un moto browniano al tempo $t$, e supponiamo che sia
contenuto in una sfera di raggio $L$: da una analisi agli autovalori
dell'equazione di diffusione per la distribuzione $p\left(t,x\right)$

\begin{equation}
\partial_{t}p\left(t,x\right)=\mathfrak{D}\,\nabla^{2}p\left(t,x\right),\label{eq:EQUVALOR}
\end{equation}
\\
condizionata \foreignlanguage{english}{$p_{t}\left(0\right)=\delta\left(0\right)$},
$p_{t}\left(L\right)=0$, si può mostrare (\cite{Huges}) che la probabilità
$\bar{p}_{t}$ che il confinamento avvenga è di ordine $\exp\left(-c\,tL^{-2}\right)$,
con $c$ costante positiva. 

Ora, supponendo che lo stesso valga anche per cammini discreti, la
probabilità che un cammino di $n$ passi, interamente contenuto in
una sfera di raggio $L$, vi resti anche al passo $\left(n+1\right)-$esimo
sarà

\begin{equation}
\psi_{n}\simeq\frac{\bar{p}_{n+1}}{\bar{p}_{n}}\propto\exp\left(-cL^{-2}\right).
\end{equation}
\\
 Approssimiamo la distribuzione dei monomeri a una distribuzione uniforme
nell'interno della sfera, così da avere range e numero di intersezioni
proporzionale al volume; questo ci darà $L\sim R^{1/d}$, da cui segue
$\psi_{n}\propto\exp\left(-cR^{-2/d}\right)$. Infine, espandendo
al primo ordine per $R\rightarrow\infty$, e imponendo $\psi_{n}+\bar{\psi}_{n}=1$,
si arriva alla relazione $\bar{\psi}_{n}\propto R^{-2/d}$. 

La $\bar{\psi}_{n}$ trovata è una approssimazione per $\bar{\psi}_{n}\left(m\right)$
nel limite di grandi $n$, con $m\rightarrow1$: sostituendo $R=\left(1-m\right)n$
otterremo
\begin{equation}
\bar{\psi}_{n}\left(m\right)\propto\left[\left(1-m\right)n\right]^{-\frac{2}{d}}.\label{eq:FROV}
\end{equation}

Per il momento non discuteremo $\pi\left(m\right)$: verrà trattata
diffusamente in un capitolo dedicato, al termine del quale vedremo
anche come la stima di $\bar{\psi}_{n}\left(m\right)$ fornita in
\prettyref{eq:FROV} sia in accordo con la distribuzione di volume
della WS (\cite{Den=000020Hollander}) per $m\rightarrow1$, $n\rightarrow\infty$.

\pagebreak{}

\begin{figure}
\begin{centering}
\caption{}
\includegraphics[scale=0.5]{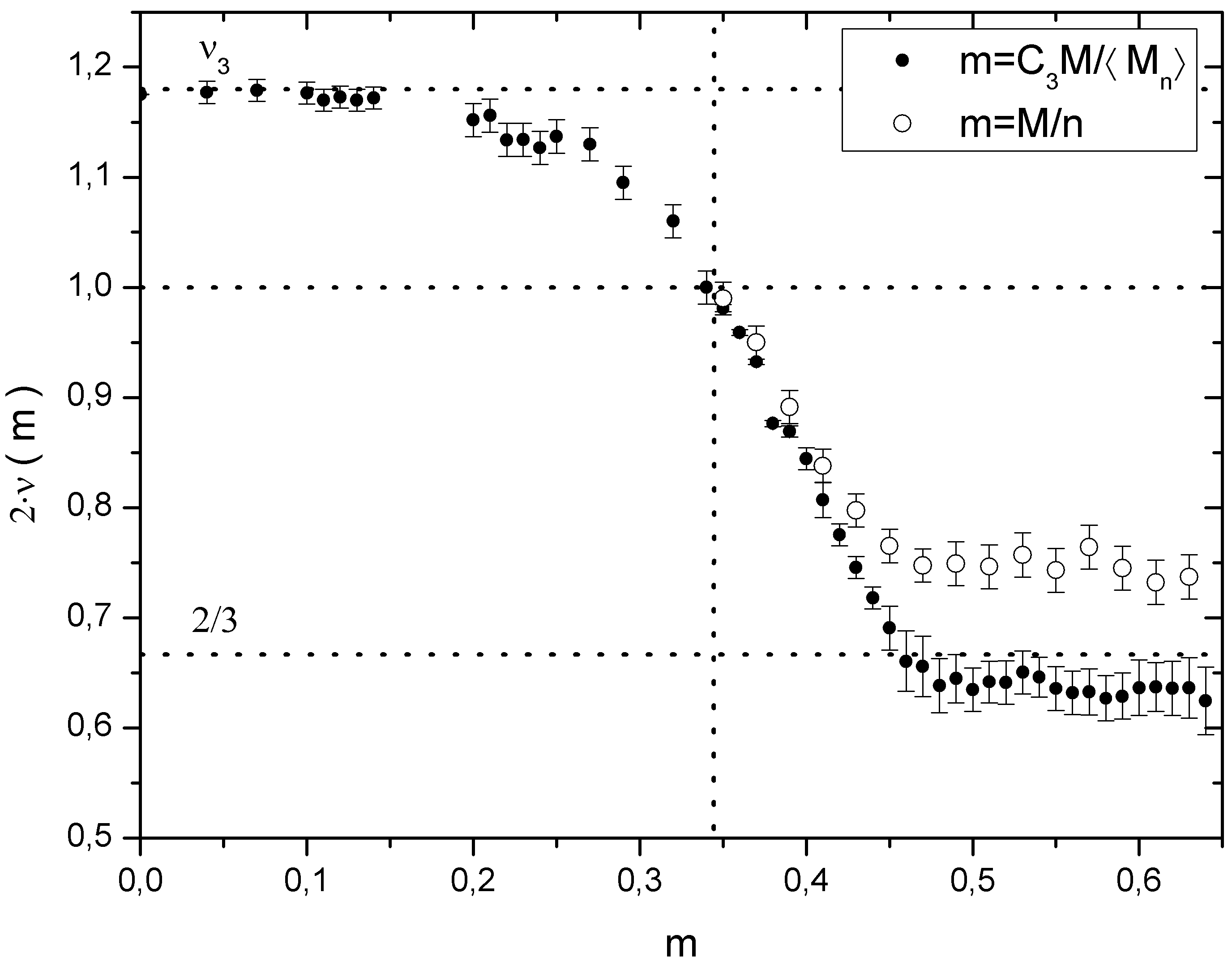}
\par\end{centering}
\begin{centering}
\label{Flo:3D=000020doppiohh}\label{Flo:FIGURA3}
\par\end{centering}
\emph{Valori di $\nu\left(m\right)$ per $d=3$, $n_{M}=10^{3}$:
nel grafico sono confrontati i dati ottenuti dai fit di $\log\left(\Re_{n}^{2}\left(m\right)\right)$
vs $\log\left(n\right)$ (punti vuoti) e $\log\left(\Re_{n}^{2}\left(C_{3}n\,m/\langle M_{n}\rangle\right)\right)$
vs $\log\left(n\right)$ (punti pieni), con $\langle M_{n}\rangle=n-\langle R_{n}\rangle$
dato dalla \prettyref{eq:1.2.4}. }

\emph{In pratica, $m$ viene riscalato con il valore del supporto
(range) previsto per un SRW con $n<\infty$. Questo procedimento,
esatto per il punto $m=C_{d}$, costituisce un notevole miglioramento
delle simulazioni, almeno per $m$ non troppo lontano da $C_{d}$. }

\emph{Dai dati con $m$ non corretto appare evidente come, per $n_{M}=10^{3}$,
si sia ancora lontani dall'andamento asintotico (le due previsioni
devono coincidere): il sistema mostra dunque un comportamento transiente,
di cui (a titolo speculativo) si presenta la seguente interpretazione.}

\emph{L'andamento transiente osservato ha un esponente prossimo a
$\nu_{\theta}=11/30$, previsto in \cite{Douglas=000020Sponge=0000201996}
per uno }``Sponge Polymer''\emph{ al punto $\theta$: a livello
speculativo si potrebbe ipotizzare un valore $n^{*}$ critico tale
che, per $n<n^{*}$, la taglia dei buchi all'interno dei cluster sia
$\mathcal{O}\left(n^{11/10}\right)$ (buchi percolanti). Superato
$n^{*}$, la crescita smette e si arriva alla configurazione compatta,
con $\nu\left(m\right)=1/3$ e buchi di taglia $\mathcal{O}\left(1\right)$.}

\end{figure}

\begin{figure}
\begin{centering}
\caption{}
\includegraphics[scale=0.5]{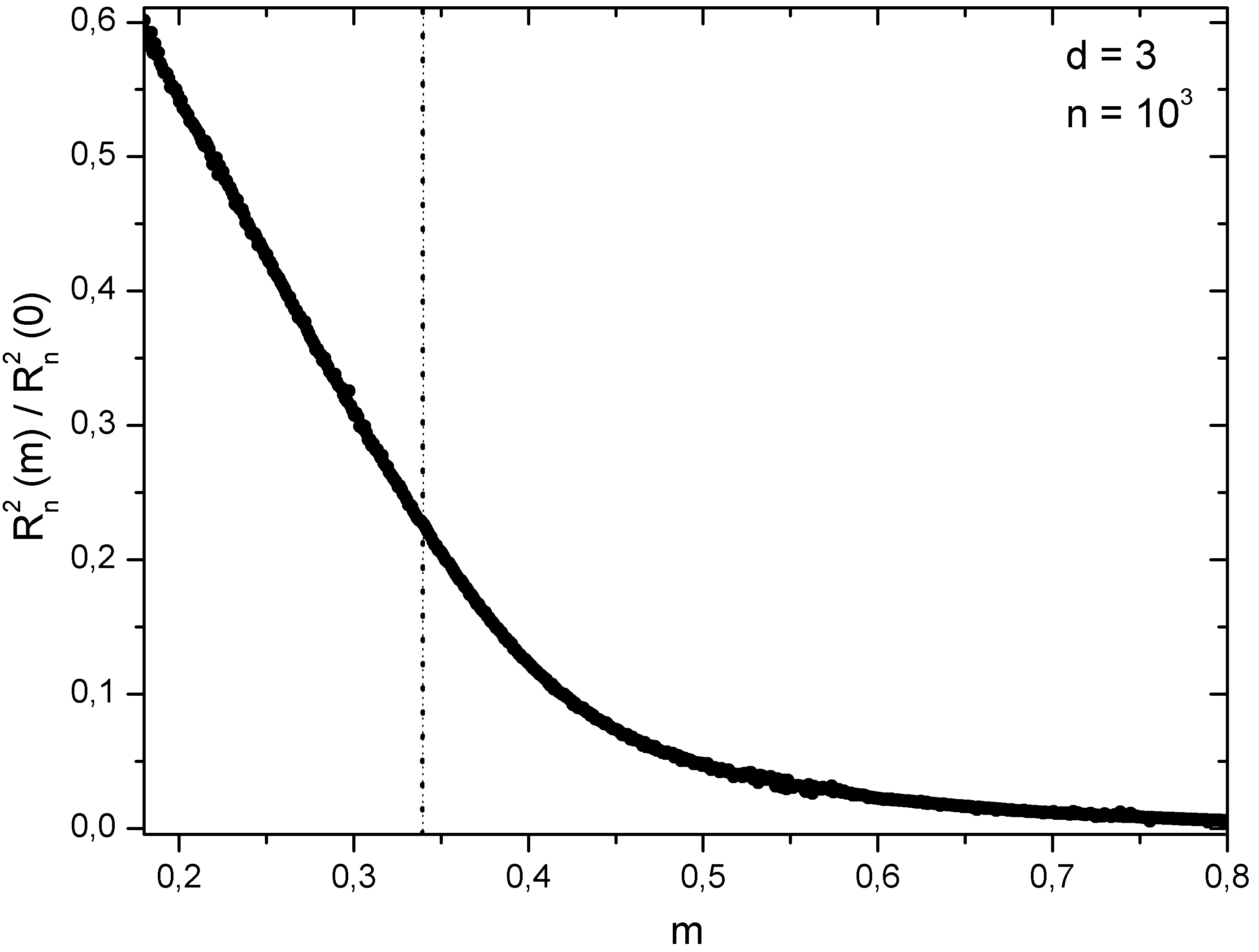}
\par\end{centering}
\begin{centering}
\label{Flo:R^2/n-3D}
\par\end{centering}
\emph{Andamento di $\Re_{n}^{2}\left(m\right)/\Re_{n}^{2}\left(0\right)$
per $d=3$, $n=10^{3}$: la linea verticale rappresenta il punto $m=C_{3}$,
dove è previsto l'annullamento di $\varrho_{3}\left(m\right)$. Dal
grafico non è tuttavia possibile chiarire la natura della transizione:
in particolare se il limite $\varrho_{3}\left(m\right)$ è discontinuo
in $m_{c}$.}
\end{figure}

\begin{figure}
\begin{centering}
\caption{}
\includegraphics[scale=0.5]{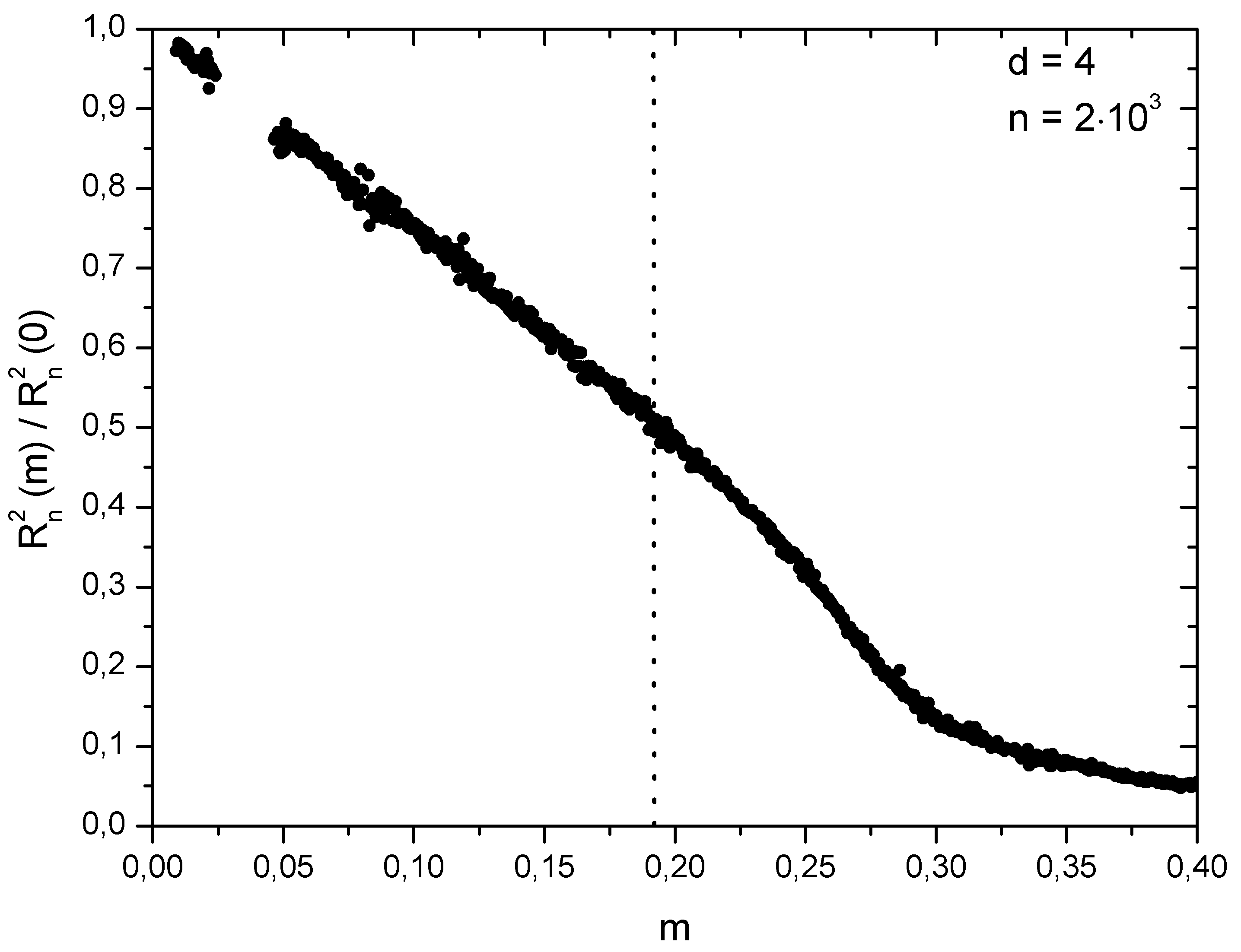}
\par\end{centering}
\begin{centering}
\label{Flo:R^2/n=000020-=0000204D}
\par\end{centering}
\emph{Il grafico mostra $\Re_{n}^{2}\left(m\right)/n$ per $d=4$,
$n=2\cdot10^{3}$: la linea verticale è $m=C_{4}$, dove si prevede
l'annullamento di $\varrho_{4}\left(m\right)$.}

\emph{Dallo scaling del SRW si deve avere $\Re_{n}^{2}\left(C_{4}\right)/n\rightarrow1$,
mentre dal modello di Stanley si desume che il limite del rapporto
$\Re_{n}^{2}\left(m\right)/n$ con $m<C_{4}$ diverga per $n\rightarrow\infty$
come }\foreignlanguage{english}{$\log\left(n\right)^{1/4}$}\emph{.
Intuitivamente, queste assunzioni portano ad una situazione in cui
$\varrho_{4}\left(C_{d}^{-}\right)>0$, $\varrho_{4}\left(C_{d}^{+}\right)=0$,
come previsto per la WS. }

\emph{Benché i dati riportati siano consistenti con questa descrizione,
anche in questo caso la taglia delle catene è insufficiente per chiarire
l'ordine della transizione: si tratta del caso più difficile, in quanto
la discrepanza con il SRW è logaritmica, e potrebbe necessitare di
catene così lunghe da rendere di fatto impossibile un osservazione
sperimentale del fenomeno.}
\end{figure}

\begin{figure}
\begin{centering}
\caption{}
\includegraphics[scale=0.5]{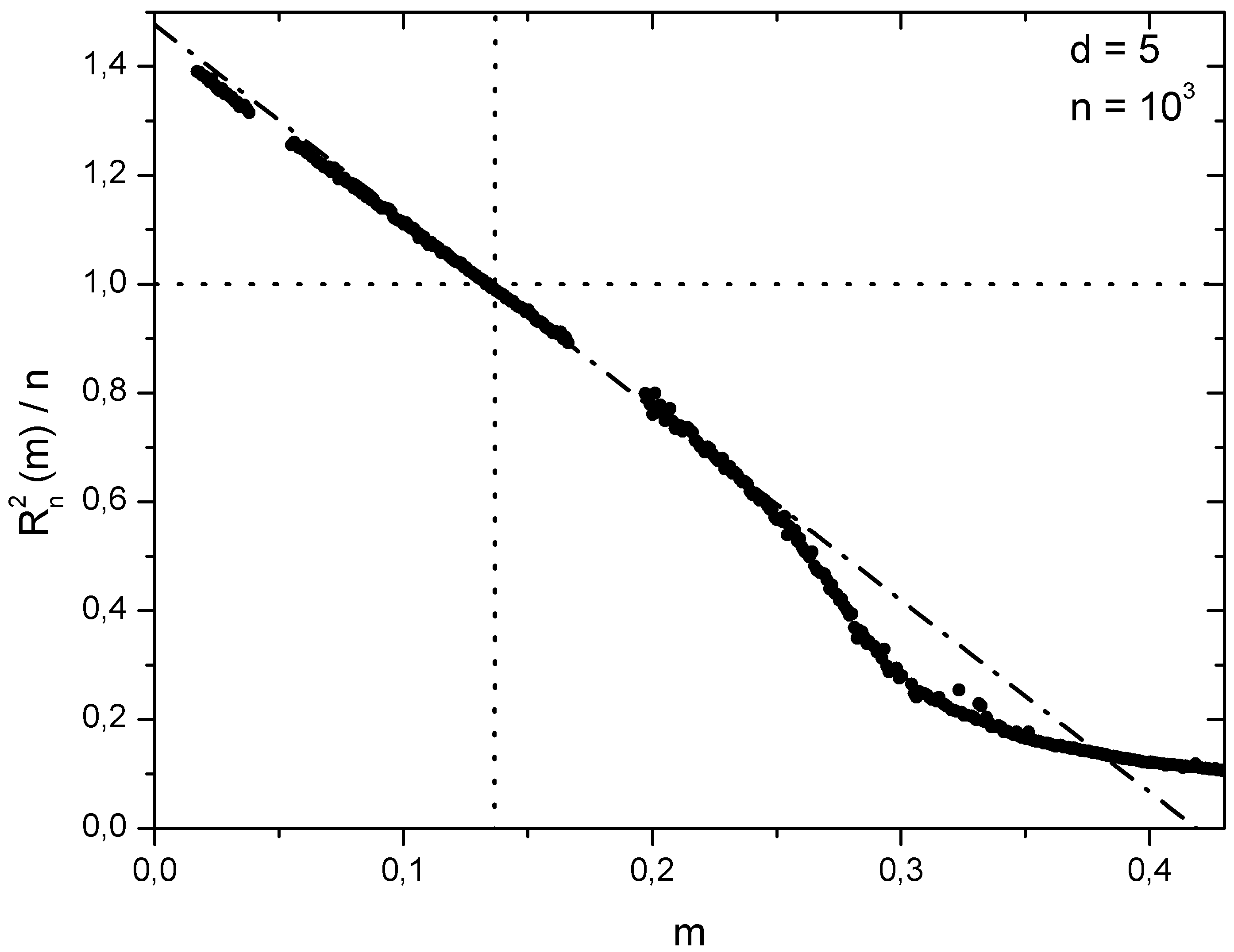}
\par\end{centering}
\begin{centering}
\label{Flo:R^2/n-5D}
\par\end{centering}
\emph{Il grafico mostra $\Re_{n}^{2}\left(m\right)/n$ per $d=5$,
$n=10^{3}$: le linee di punti rappresentano $m=C_{5}$ (verticale),
$\Re_{n}^{2}\left(m\right)/n=1$ (orizzontale). La retta tratteggiata
è $D_{5}+\left(1-D_{5}/C_{5}\right)m$, con $D_{5}$ costante diffusiva
del SAW (}\foreignlanguage{english}{$\Re_{n}^{2}\left(0\right)/n\rightarrow D_{5}$}\emph{).}
\end{figure}

\begin{figure}
\begin{centering}
\caption{}
\includegraphics[scale=0.5]{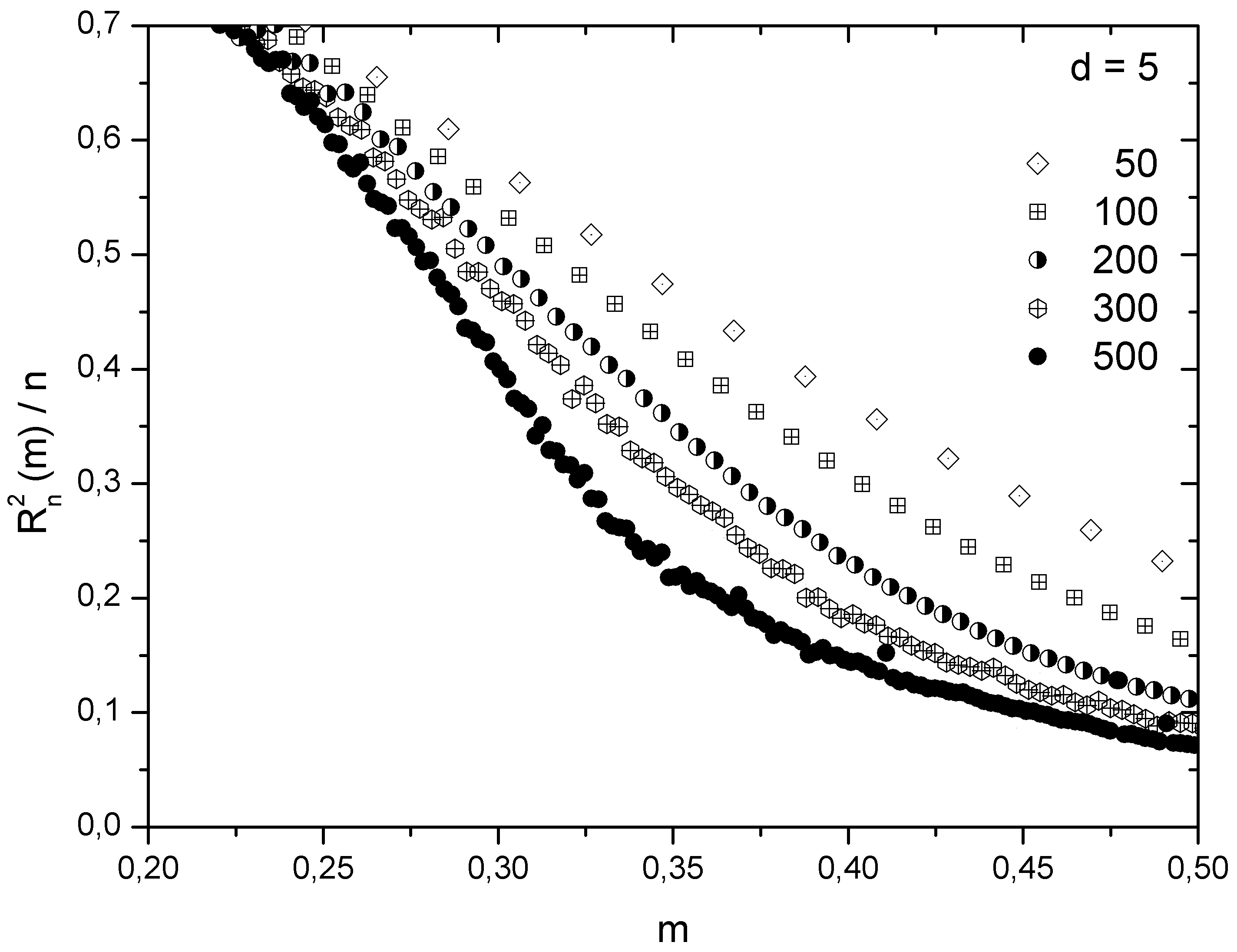}
\par\end{centering}
\begin{centering}
\label{Flo:R^2/n-5D-2}
\par\end{centering}
\emph{In figura $\Re_{n}^{2}\left(m\right)/n$, $d=5$ per alcuni
valori di $n$. Il grafico evidenzia l'insorgere della transizione
in $m$, ipotizzata per $\varrho_{5}\left(m\right)$: al crescere
di $n$, si osserva il progressivo abbassamento di $\Re_{n}^{2}\left(m\right)/n$
nel range mostrato ($m\in\left[0.2,\,0.5\right]>C_{5}$). L'ipotesi
è che, nel limite $n\rightarrow\infty$, $\Re_{n}^{2}\left(m\right)/n$
si annulli per un particolare valore $m_{c}>C_{5}.$}
\end{figure}

\begin{figure}
\begin{centering}
\caption{}
\includegraphics[scale=0.5]{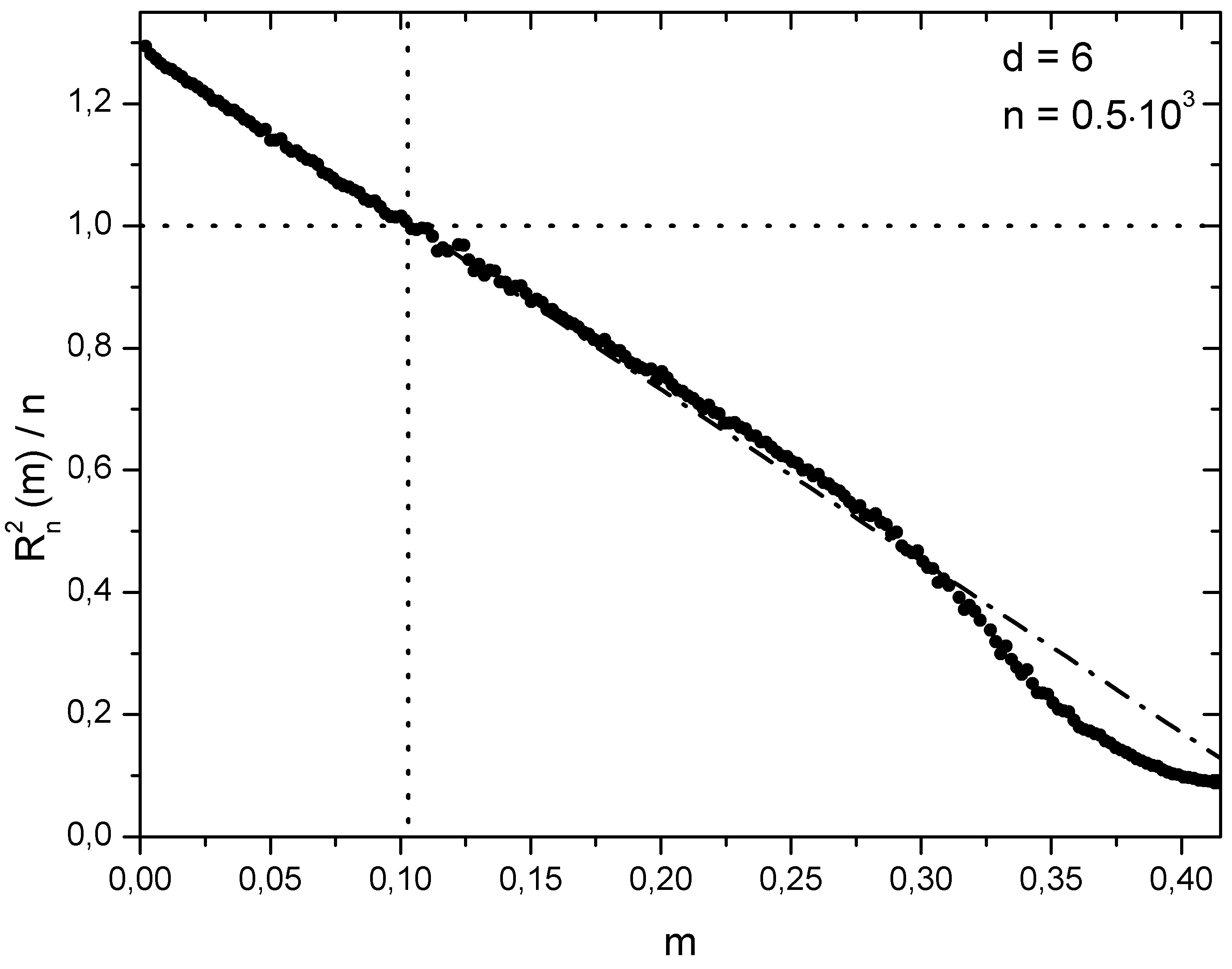}
\par\end{centering}
\begin{centering}
\label{Flo:R^2/n-6D}
\par\end{centering}
\emph{Il grafico mostra $\Re_{n}^{2}\left(m\right)/n$ per $d=6$,
$n=0,5\cdot10^{3}$: le linee di punti rappresentano $m=C_{6}$ (verticale),
$\Re_{n}^{2}\left(m\right)/n=1$ (orizzontale). La retta tratteggiata
è $D_{6}+\left(1-D_{6}/C_{6}\right)m$ ($D_{6}$ costante di diffusione
del SAW).}
\end{figure}

\begin{figure}
\begin{centering}
\caption{}
\includegraphics[scale=0.5]{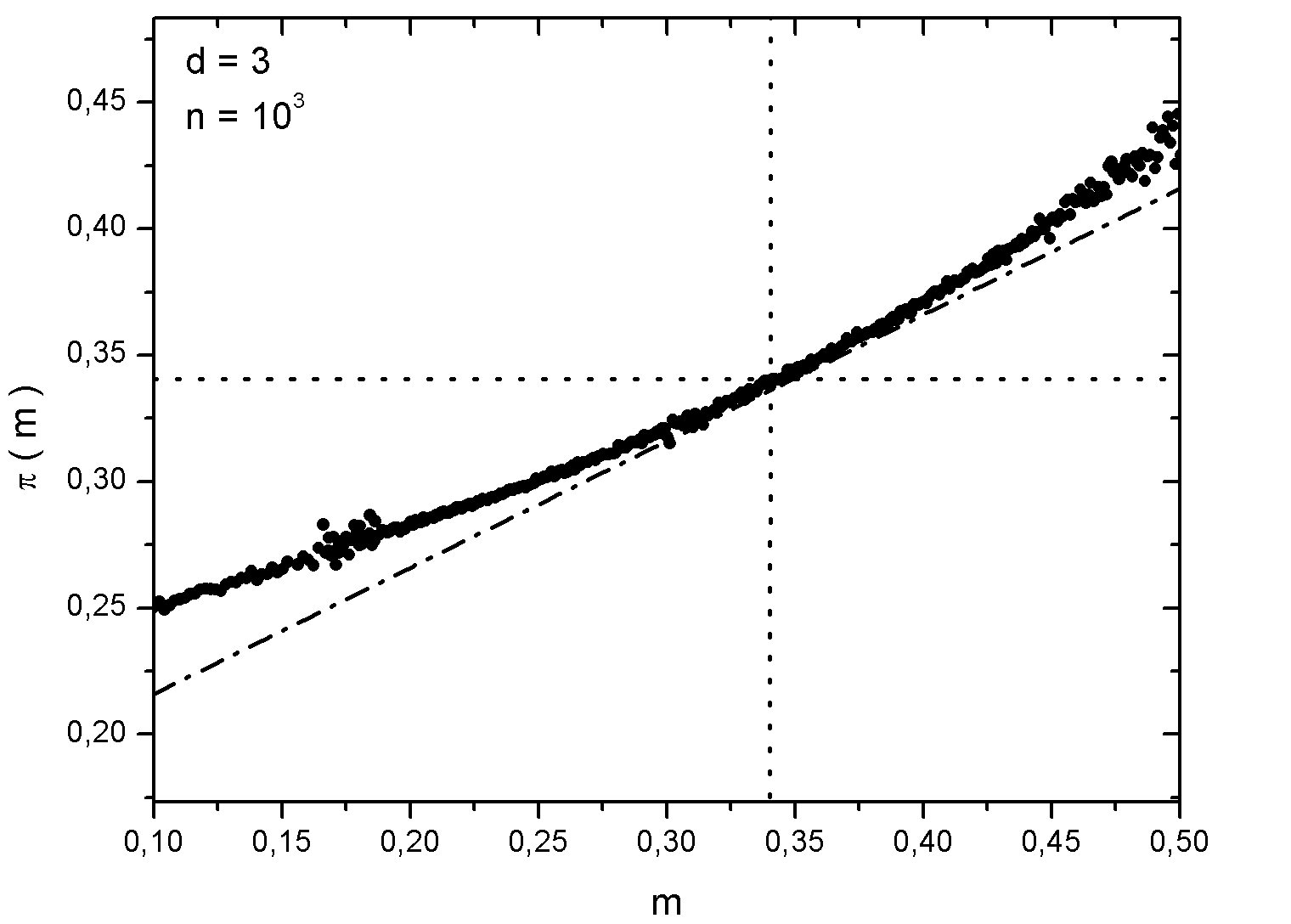}
\par\end{centering}
\begin{centering}
\label{Flo:g-3d-1000-p}
\par\end{centering}
\emph{Il grafico mostra $\pi_{n}\left(m\right)$, per $d=3$, $n=10^{3}$.}
\emph{Le linee di punti rappresentano $m=C_{3}$ (verticale), $\pi\left(m\right)=C_{3}$
(orizzontale), mentre la retta tratteggiata è l'espansione al primo
ordine di $\pi\left(m\right)$ in $m=C_{3}$: $\left(1-B_{3}\right)C_{3}+B_{3}m$}
\emph{(dove $B_{3}=1/2$). }
\end{figure}

\begin{figure}
\begin{centering}
\caption{}
\includegraphics[scale=0.5]{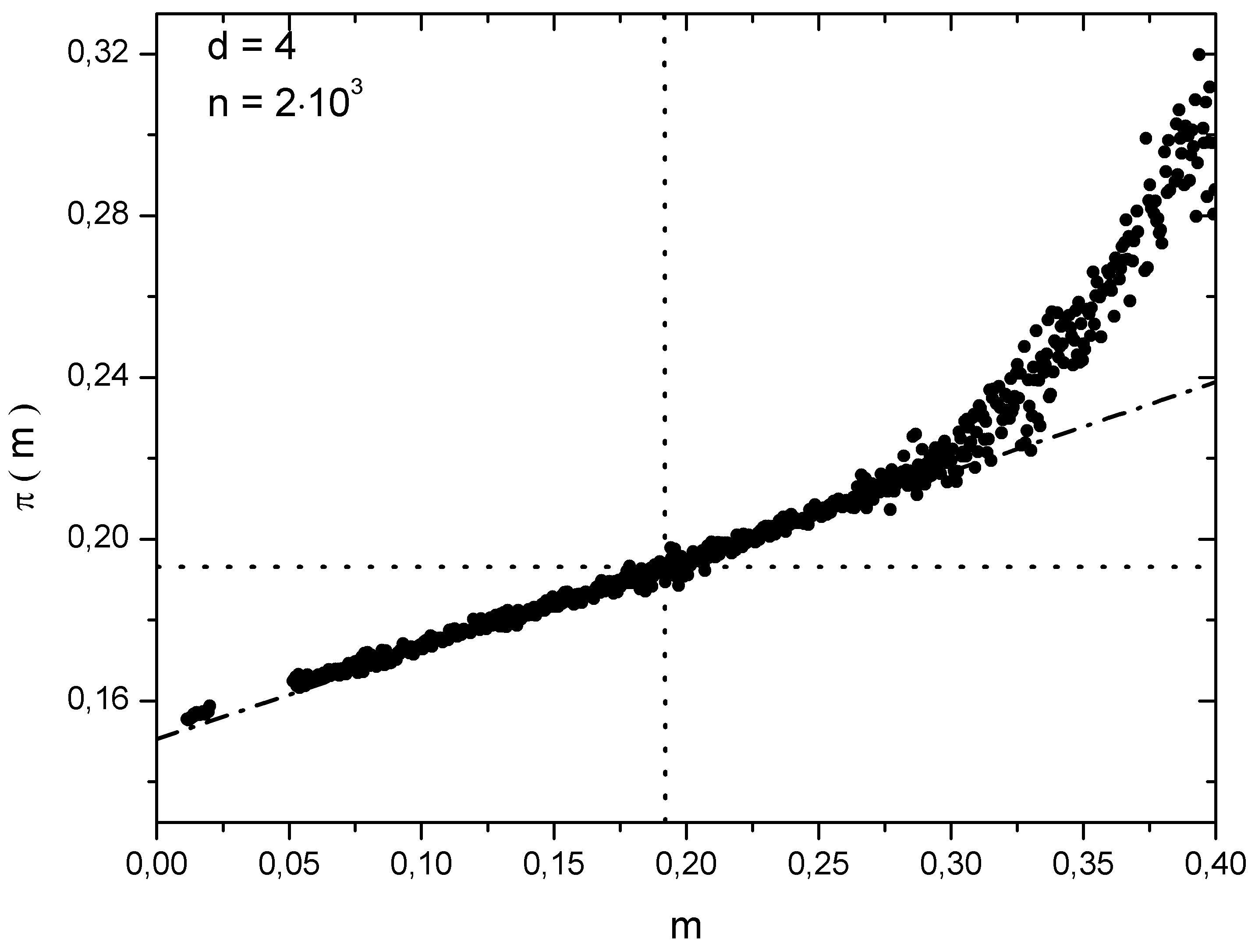}
\par\end{centering}
\label{Flo:p-M-4D}

\emph{Il grafico mostra $\pi_{n}\left(m\right)$, per $d=4$, $n=2\cdot10^{3}$.}
\emph{Le linee di punti rappresentano $m=C_{4}$ (verticale), $\pi\left(m\right)=C_{4}$
(orizzontale), mentre la retta tratteggiata è l'espansione al primo
ordine di $\pi\left(m\right)$ in $m=C_{4}$: $\left(1-B_{4}\right)C_{4}+B_{4}m$}.

\emph{La discontinuità prevista per $\pi\left(m\right)$ è netta,
tuttavia il punto di transizione appare consistentemente maggiore
di $C_{4}$. Questo è in contrasto con i risultati noti sulla WS,
ed anche con la Figura \prettyref{Flo:FIGURA2}, che mostra una transizione
molto più morbida. }

\emph{Dai confronti di questo e dei successivi grafici con la Figura
\prettyref{Flo:FIGURA1}, si nota che le discontinuità delle $\pi\left(m\right)$
corrispondono con buona approssimazione ai punti di flesso delle $\nu\left(m\right)$:
è sensato ritenere che, nel limite $n\rightarrow\infty$, questi andranno
ad assestarsi sui valori previsti di $m_{c}$.}
\end{figure}

\begin{figure}
\begin{centering}
\caption{}
\includegraphics[scale=0.5]{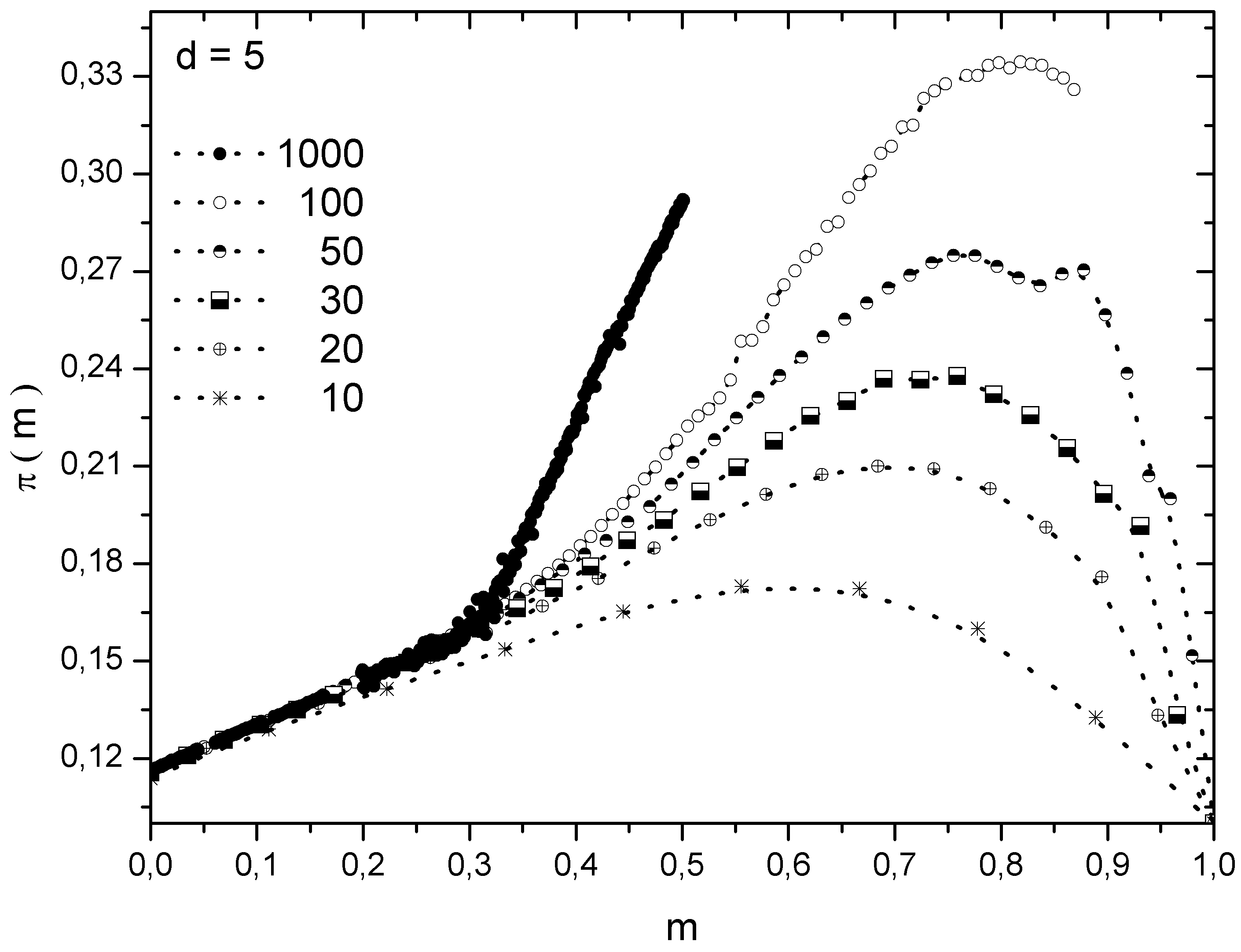}
\par\end{centering}
\begin{centering}
\label{Flo:5D-VARIE=000020n}
\par\end{centering}
\emph{Si riporta l'andamento di $\pi_{n}\left(m\right)$, $d=5$ per
alcuni valori di $n$. Il grafico mostra chiaramente l'esistenza della
transizione per un qualche $m=m_{c}$ (supposto maggiore di $C_{5}$).
I dati confermano l'esistenza del limite $\pi\left(m\right)$, almeno
per quanto riguarda $m<m_{c}$. Il grafico mostra inoltre il crollo
di $\pi_{n}\left(m\right)$ in prossimità di $m=1$, dovuto agli effetti
di taglia finita $\psi_{n}\left(m\right)$ del sistema: lo stesso
si osserva per tutte le $d$.}
\end{figure}

\begin{figure}
\begin{centering}
\caption{}
\includegraphics[scale=0.5]{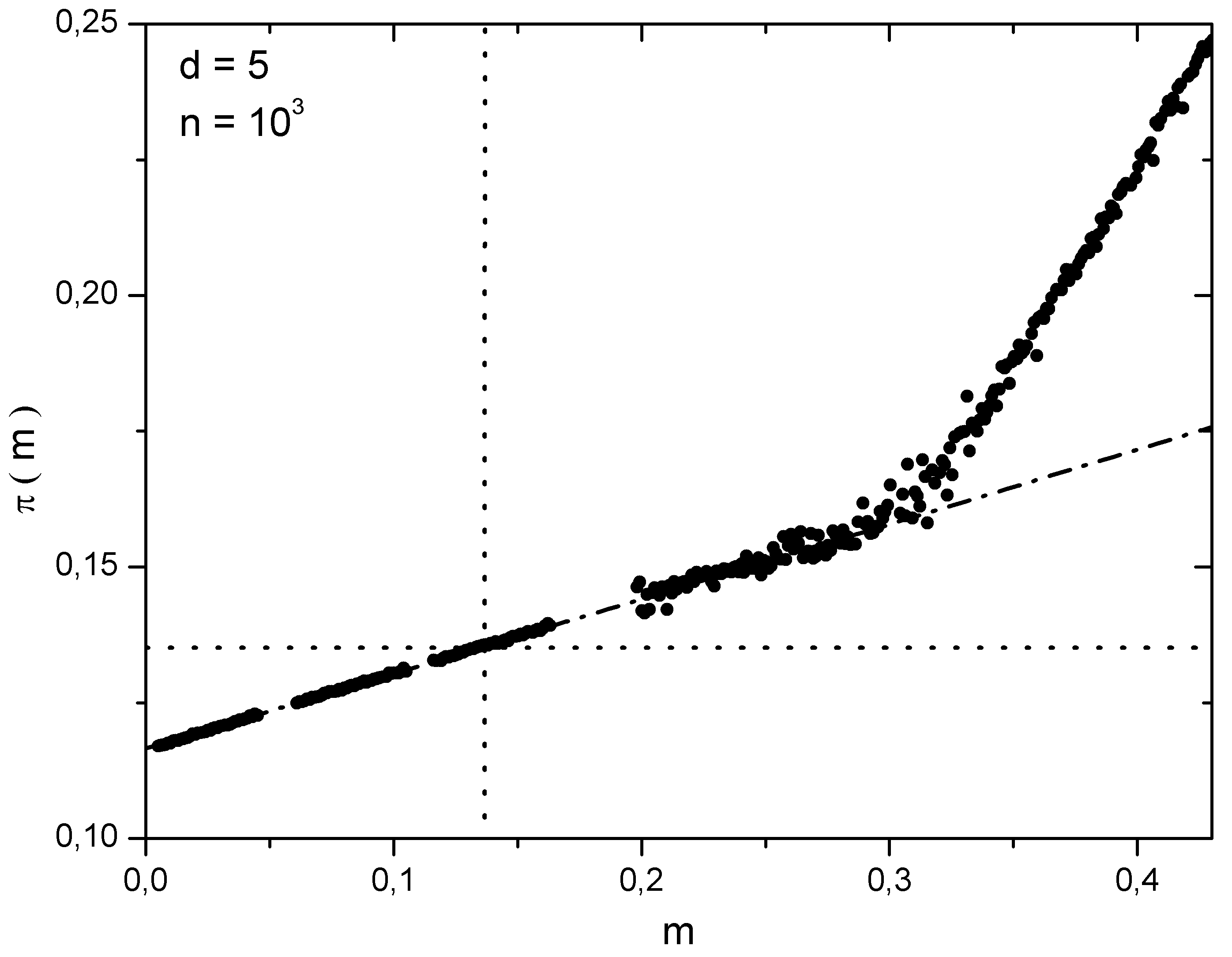}
\par\end{centering}
\label{Flo:pi-M-d5}

\emph{Il grafico mostra $\pi_{n}\left(m\right)$, per $d=5$, $n=10^{3}$.
Le linee di punti rappresentano $m=C_{5}$ (verticale), $\pi\left(m\right)=C_{5}$
(orizzontale), mentre la retta tratteggiata è l'espansione al primo
ordine di $\pi\left(m\right)$ in $m=C_{5}$: $\left(1-B_{5}\right)C_{5}+B_{5}m$}.
\end{figure}

\begin{figure}
\begin{centering}
\caption{}
\includegraphics[scale=0.5]{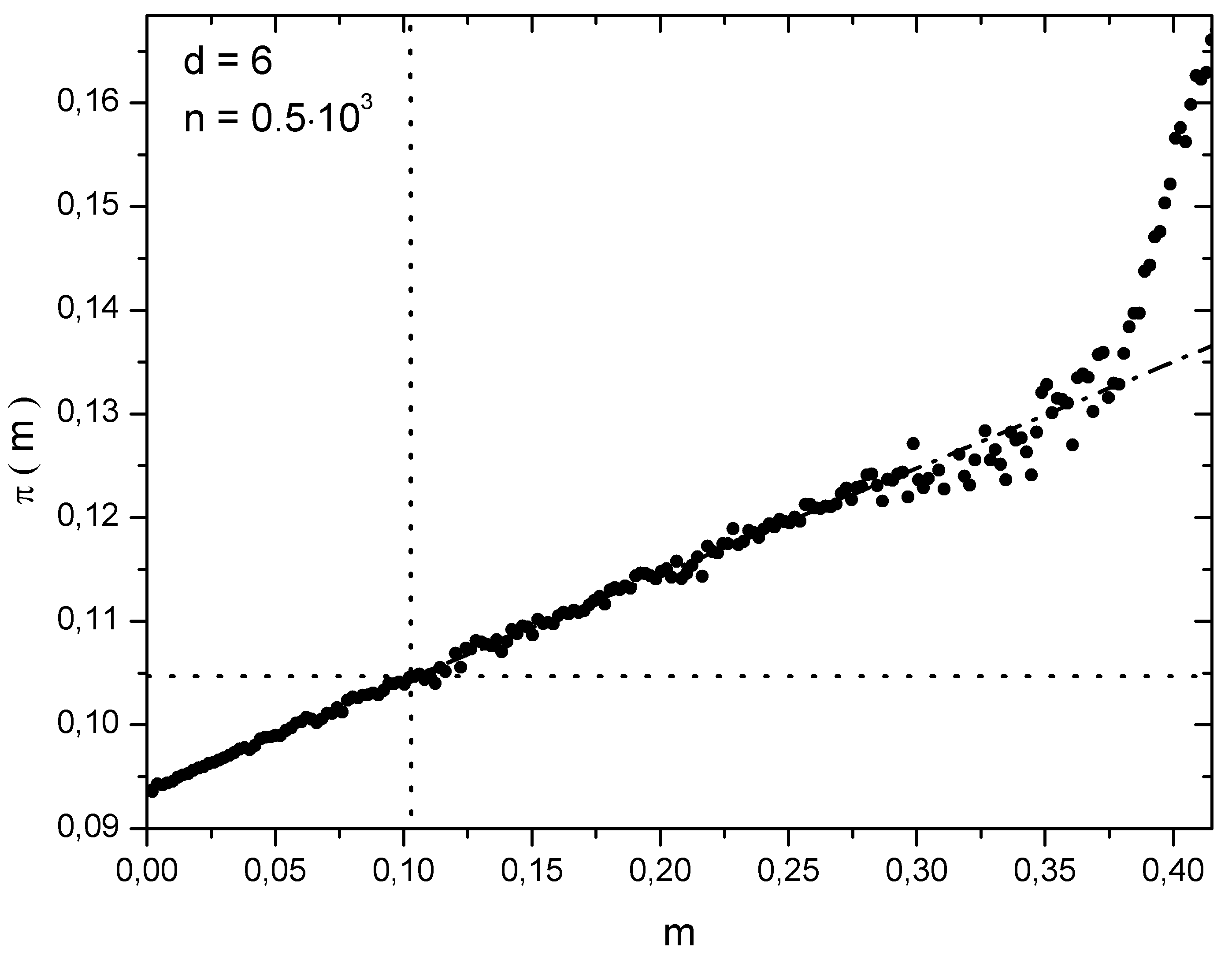}
\par\end{centering}
\label{Flo:pi-M-d6}

\emph{Il grafico mostra $\pi_{n}\left(m\right)$, per $d=6$, $n=0,5\cdot10^{3}$.
Le linee di punti rappresentano $m=C_{6}$ (verticale), $\pi\left(m\right)=C_{6}$
(orizzontale), mentre la retta tratteggiata è l'espansione al primo
ordine di $\pi\left(m\right)$ in $m=C_{6}$: $\left(1-B_{6}\right)C_{6}+B_{6}m$}.
\end{figure}

\chapter{Processo stocastico a due variabili}

Questo capitolo è dedicato al processo stocastico definito dalle \prettyref{eq:3.1.1-1},
\prettyref{eq:3.1.2-1}, allo scopo di dedurre alcune relazioni utili
per lo studio delle intersezioni. Per questo motivo prenderemo in
considerazione solo alcune tipologie di $\pi_{t}\left(M\right)$.
In particolare, tratteremo funzioni della forma $\pi_{t}\left(M\right)=\pi\left(M/t\right)$,
con $\pi\left(m\right)\in C^{\infty}$ invertibile nell'intervallo
$m\in\left[0,\,1\right]$. Per il momento faremo due ulteriori assunzioni:
$\pi\left(m\right)\in\left(0,\,1\right)$ in $m\in\left[0,\,1\right]$,
e $\exists!\,m_{0}\,|\,\pi\left(m_{0}\right)=m_{0},$ con $m_{0}\in\left(0,\,1\right)$. 

Sia dunque $P_{t}\left(M\right)$ la distribuzione al tempo $t$ della
variabile $M_{t}$, definita dal seguente processo stocastico:
\begin{equation}
M_{t+1}=M_{t}+\delta_{t}M_{t},\label{eq:3.1.1-4}
\end{equation}
\begin{equation}
\delta_{t}M_{t}=\theta\left(w_{t}-\pi\left(t^{-1}M_{t}\right)\right),\label{eq:3.1.1-5}
\end{equation}
\\
con $\theta\left(\cdot\right)$ funzione di Heaviside, e $w_{t}\in\left[0,\,1\right]$
distribuita uniformemente nell'intervallo secondo $\partial_{w}P\left(w\right)=0$.
Come anticipato, si richiede che, per $m\in\left[0,\,1\right],$ $\pi\left(m\right)$
goda delle seguenti proprietà: 
\begin{equation}
\pi\left(m\right)\in\left(0,\,1\right),
\end{equation}
\begin{equation}
\pi\left(m\right)\in C^{\infty},
\end{equation}
\begin{equation}
\exists!\,m_{0}\,|\,\pi\left(m_{0}\right)=m_{0}.
\end{equation}
$\,$

Benché non sia necessario nell'immediato, conviene introdurre subito
la variabile $m_{t}\equiv M/t$: per motivi che saranno evidenti più
avanti, da un certo punto in poi, al posto di $P_{t}\left(M\right)$
converrà lavorare con la distribuzione $P_{t}\left(m_{t}\right)$
(ottenibile da $P_{t}\left(M\right)$ attraverso il cambio di variabile
$M_{t}\rightarrow m_{t}\,t$ ).

\section{Momenti della distribuzione $P_{t}\left(M\right)$}

Iniziamo con il calcolare i primi due momenti della distribuzione.
Per semplicità risolveremo prima il caso 
\begin{equation}
\pi\left(m\right)=a+bm,\label{eq:3.1.P}
\end{equation}
con $a>0$, $\left(a+b\right)\in\left(0,1\right)$; vedremo poi come
sarà possibile generalizzare a \foreignlanguage{english}{$\pi\left(m\right)\in C^{\infty}$}.
Nei prossimi conti useremo la notazione
\begin{equation}
\delta_{t}z_{t}=z_{t+1}-z_{t},
\end{equation}
\begin{equation}
\Delta z_{t}=z_{t}-\langle z_{t}\rangle,
\end{equation}
con la media $\langle\cdot\rangle$ intesa sull'insieme $\Phi_{t}$
delle possibili realizzazioni della successione $w_{t}$. La media
sulla variabile $w_{t}$ sarà invece indicata con una barra; si noti
che la media su $w_{t}$ va eseguita prima di quella sull'ensemble
$\Phi_{t}$. 

Il calcolo di $\langle M_{t}\rangle$ è piuttosto semplice. Dalle
\prettyref{eq:3.1.1-4}, \prettyref{eq:3.1.1-5}, segue la relazione
\begin{equation}
\delta_{t}\langle M_{t}\rangle=\langle\overline{\delta_{t}M_{t}}\rangle=\langle\pi_{t}\left(M_{t}\right)\rangle.\label{eq:xasda}
\end{equation}
Sostituendo \prettyref{eq:3.1.P} in \prettyref{eq:xasda} otteniamo
la seguente equazione alle differenze finite: 
\begin{equation}
\delta_{t}\langle M_{t}\rangle=a+bt^{-1}\langle M_{t}\rangle.\label{eq:3.EQU1}
\end{equation}
Questa equazione ammette una soluzione $\langle M_{t}\rangle=m_{0}t$,
con $m_{0}=a/\left(1-b\right)$. Si noti che $m_{0}$ è il punto che
soddisfa l'equazione $\pi\left(m\right)=m$, e quindi la sua unicità
e garantita dalle ipotesi preliminari su $\pi\left(m\right)$. Dalla
relazione per $m_{0}$ segue anche la condizione $b<1$, garantita
delle ipotesi $a>0$, $\left(a+b\right)\in\left(0,1\right)$.

Attraverso un calcolo ulteriore (\cite{Balzan}) è possibile individuare
il momento secondo $\langle\Delta M_{t}^{2}\rangle$ (lavoriamo ancora
con $\pi\left(m\right)$ lineare). 

Scriviamo la variazione della $\langle\Delta M_{t}^{2}\rangle$: dalle
definizioni si ha 
\begin{equation}
\delta_{t}\langle\Delta M_{t}^{2}\rangle=\langle\overline{\delta_{t}\left(\Delta M_{t}^{2}\right)}\rangle=\langle\overline{\left[\Delta M_{t}+\delta_{t}\left(\Delta M_{t}\right)\right]^{2}}\rangle-\langle\Delta M_{t}^{2}\rangle.
\end{equation}
Svolgendo i quadrati troveremo 
\begin{equation}
\delta_{t}\langle\Delta M_{t}^{2}\rangle=2\langle\overline{\Delta M_{t}\delta_{t}\left(\Delta M_{t}\right)}\rangle+\langle\overline{\left[\delta_{t}\left(\Delta M_{t}\right)\right]^{2}}\rangle.\label{eq:3.1.4}
\end{equation}
Calcoliamo il termine $\langle\overline{\left[\delta_{t}\left(\Delta M_{t}\right)\right]^{2}}\rangle$:
come da notazione, svolgiamo prima la media su $w_{t}$ 
\begin{equation}
\overline{\left[\delta_{t}\left(\Delta M_{t}\right)\right]^{2}}=\pi\left(t^{-1}M_{t}\right)\left(1-\langle\delta_{t}M_{t}\rangle\right)^{2}+\left[1-\pi\left(t^{-1}M_{t}\right)\right]\langle\delta_{t}M_{t}\rangle^{2},
\end{equation}
poi applichiamo $\langle\cdot\rangle$. Notando che $\langle\delta_{t}M_{t}\rangle=\langle\pi\left(t^{-1}M_{t}\right)\rangle=m_{0}$,
troveremo 
\begin{equation}
\langle\overline{\left[\delta_{t}\left(\Delta M_{t}\right)\right]^{2}}\rangle=m_{0}\bar{m}_{0},
\end{equation}
dove abbiamo introdotto $\bar{m}_{0}=1-m$. Svolgiamo le media su
$w_{t}$ e (successivamente) sull'ensemble anche per il primo termine
della \prettyref{eq:3.1.4}: dopo qualche calcolo troveremo 
\begin{equation}
\langle\overline{\Delta M_{t}\delta_{t}\left(\Delta M_{t}\right)}\rangle=2\langle\Delta M_{t}\pi\left(t^{-1}M_{t}\right)\rangle,
\end{equation}
da cui, sostituendo la \prettyref{eq:3.1.P} ricaviamo:
\begin{equation}
\langle\Delta M_{t}\pi\left(t^{-1}M_{t}\right)\rangle=b\langle\Delta M_{t}\rangle.
\end{equation}
Riunendo le espressioni si trova una equazione alle differenze finite
per $\langle\Delta M_{t}^{2}\rangle$: 
\begin{equation}
\delta_{t}\langle\Delta M_{t}^{2}\rangle=2bt^{-1}\langle\Delta M_{t}^{2}\rangle+m_{0}\bar{m}_{0},\label{eq:3.EQU2}
\end{equation}
che ammette una soluzione lineare $\langle\Delta M_{t}^{2}\rangle=\sigma t$,
con $\sigma=m_{0}\bar{m}_{0}/\left(1-2b\right)$. 

\begin{figure}
\caption{}

\begin{centering}
\includegraphics[scale=0.5]{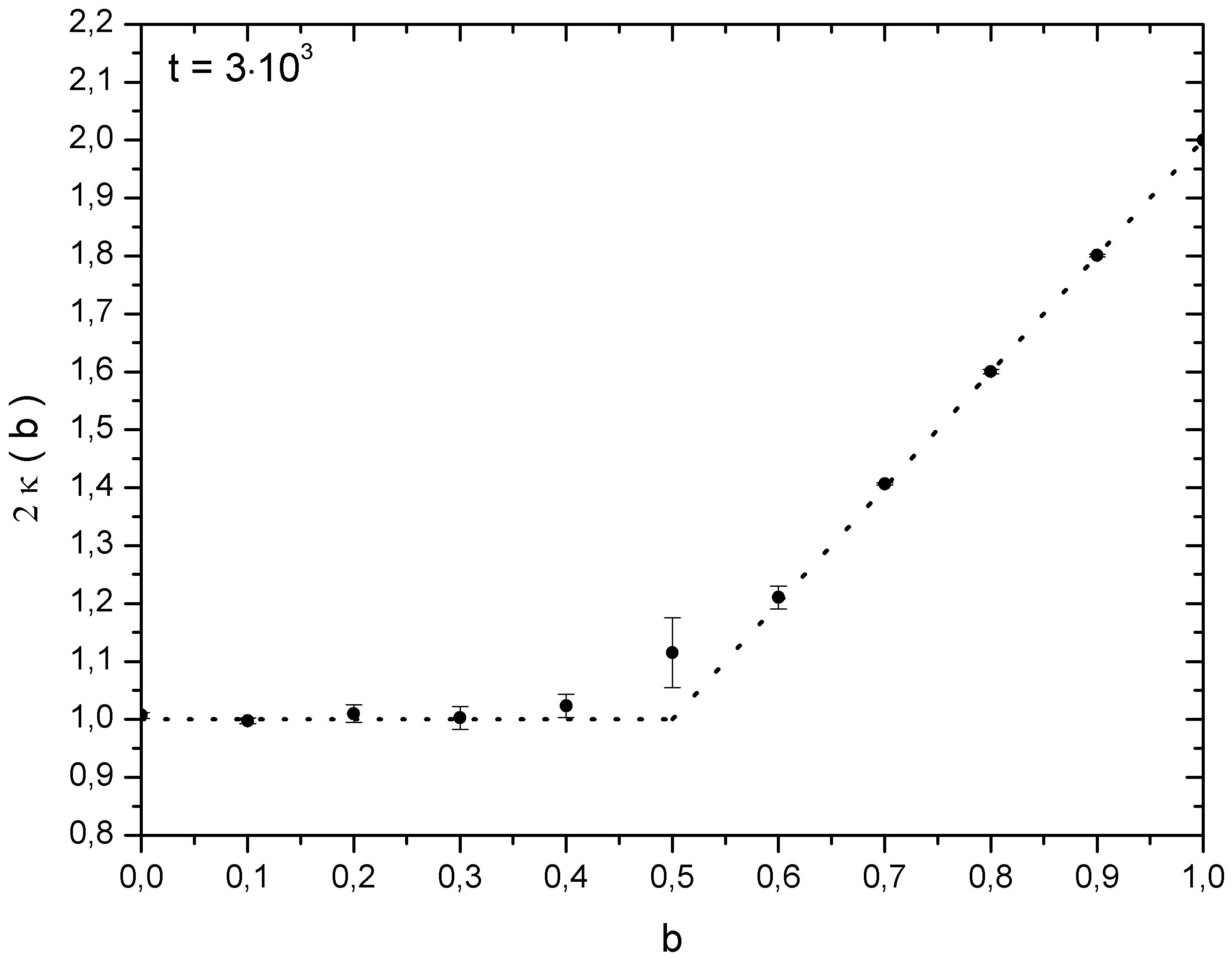}\label{Flo:figura3.1}
\par\end{centering}
\emph{Il processo \prettyref{eq:3.1.1-4}, \prettyref{eq:3.1.1-5},
con }$\pi\left(m\right)=1/2+b\left(m-1/2\right)$\emph{, è stato simulato
numericamente per alcuni valori di $b$. Nel grafico si riportano
i valori ottenuti al variare di $b$ per l'esponente di scaling della
varianza ($\langle\Delta M_{t}^{2}\rangle\propto t^{\kappa}$). I
valori di $\kappa$ provengono da fit di $\log\left[\langle\Delta M_{t}^{2}\rangle\right]$
vs $\log\left(t\right)$: il range per ogni fit è $t\in\left[0.8\,t_{M},\,t_{M}\right]$,
con $t_{M}=3\cdot10^{3}$. I dati ottenuti confermano la previsione
\prettyref{eq:2.1.9} (linea di punti).}
\end{figure}

Si nota subito come per $b>1/2$ la soluzione lineare non sia più
valida: per comprendere meglio cosa effettivamente avvenga, possiamo
operare un passaggio al continuo in $t$, sostituendo $\langle\Delta M_{t}^{2}\rangle$
con la funzione $V\left(t\right)$: 
\begin{equation}
\partial_{t}V\left(t\right)=2bt^{-1}V\left(t\right)+m_{0}\bar{m}_{0}.\label{eq:2.1.7-2}
\end{equation}
Si tratta di una semplice equazione differenziale del I ordine, le
cui soluzioni al variare di $b$ sono: 
\begin{equation}
V\left(t\right)\sim\left\{ \begin{array}{l}
\sigma t\\
\alpha t\log\left(t\right)\\
\beta t^{2b}
\end{array}\,\,\,\begin{array}{r}
b<1/2\\
b=1/2\\
b>1/2
\end{array}\right..\label{eq:2.1.9}
\end{equation}
\\
Dunque, per $b\geq1/2$ si troverà $\langle\Delta M_{t}^{2}\rangle\propto t^{2b}$
(Figura \prettyref{Flo:figura3.1}): come vedremo più avanti questo
indicherà un cambiamento qualitativo nel tipo di distribuzione. 

I risultati fin qui ottenuti possono essere applicati direttamente
ad una generica $\pi\left(m\right)\in C^{\infty}:$ attraverso una
espansione in serie di $\pi\left(m\right)$ nel punto $m_{0}=\langle\pi\left(t^{-1}M_{t}\right)\rangle$,
si ottiene 
\begin{equation}
\pi\left(t^{-1}M_{t}\right)=m_{0}+b\,t^{-1}\Delta M_{t}+\mathcal{O}\left(t^{-2}\Delta M_{t}^{2}\right).\label{eq:3.KKK}
\end{equation}
dove $b$ ora rappresenta la derivata di $\pi\left(m\right)$ in $m_{0}$:
\begin{equation}
b=\left[\partial_{m}\pi\left(m\right)\right]_{m_{0}}.
\end{equation}
Inserendo lo sviluppo \prettyref{eq:3.KKK} in \prettyref{eq:3.EQU1}
si avrà una correzione $\mathcal{O}\left(t^{-2}\langle\Delta M_{t}^{2}\rangle\right)$,
mentre per la \prettyref{eq:3.EQU2} la correzione sarà $\mathcal{O}\left(t^{-3}\langle\Delta M_{t}^{4}\rangle\right)$.
Poniamo $\langle\Delta M_{t}^{2}\rangle\propto t^{\alpha}$, $\langle\Delta M_{t}^{4}\rangle\propto t^{2\alpha}$:
perché le correzioni siano trascurabili si dovrà avere $\alpha<2$
(le condizioni sono $\alpha-2<0$ dal primo caso, $2\alpha-3<\alpha-1$
dal secondo). Questo è verificato dalle \prettyref{eq:2.1.9} per
$b<1$.

L'unicità della soluzione di $\pi\left(m\right)=m$, unita a $\pi\left(m\right)\in C^{\infty}$,
implica necessariamente $\left[\partial_{m}\pi\left(m\right)\right]_{m_{0}}\leq1$:
ne segue che i comportamenti asintotici di valor medio e varianza
di $M_{t}$, generati dalla \prettyref{eq:3.KKK}, sono gli stessi
del caso lineare, con $m_{0}\,|\,\pi\left(m_{0}\right)=m_{0}\in\left(0,\,1\right)$
e $b=\left[\partial_{m}\pi\left(m\right)\right]_{m_{0}}<1$.

Si sottolinea che nei momenti di $P_{n}\left(M\right)$ oltre il secondo,
gli ordini superiori dello sviluppo di $\pi\left(m\right)$ non saranno
trascurabili: il metodo usato per determinare i primi due momenti
di $P_{t}\left(M\right)$ può essere infatti applicato iterativamente,
per ottenere momenti di ordine superiore. A parte un notevole aumento
dell'algebra necessaria, non vi sono problemi di ordine concettuale
nella determinazione momenti di ordine arbitrario $s$, in cui, tuttavia,
la dipendenza funzionale da $\pi(m)$ sarà presente con derivate (calcolate
in $m_{0}$) fino all'ordine $\left(s/2\right)-$esimo.

\section{Master equation}

Un modo più diretto per derivare le proprietà della distribuzione
$P_{t}\left(M\right)$ viene dallo studio della \emph{master equation}.
Partiamo dalle \prettyref{eq:3.1.1-1}, \prettyref{eq:3.1.2-1} per
una generica $\pi_{t}\left(M\right)$: la \emph{master equation} per
$P_{t}\left(M\right)$ sarà, per definizione, 
\begin{equation}
P_{t+1}\left(M\right)=\pi_{t}\left(M-1\right)P_{t}\left(M-1\right)+\left[1-\pi_{t}\left(M\right)\right]P_{t}\left(M\right).\label{eq:mastereq1}
\end{equation}
Consideriamo ora le \prettyref{eq:3.1.1-4}, \prettyref{eq:3.1.1-5}:
all'inizio del capitolo avevamo introdotto la variabile $m_{t}=M/t$,
e la distribuzione associata $P_{t}\left(m_{t}\right)$ (del tutto
equivalente a $P_{t}\left(M\right)$). Come vedremo a breve, è conveniente
lavorare in termini di $P_{t}\left(m_{t}\right)$, la cui equazione
segue da \prettyref{eq:mastereq1}:

\begin{equation}
P_{t+1}\left(m_{t+1}\right)=\pi\left(m_{t}-t^{-1}\right)P_{t}\left(m_{t}-t^{-1}\right)+\left[1-\pi_{t}\left(m_{t}\right)\right]P_{t}\left(m_{t}\right).\label{eq:mastereq2}
\end{equation}
Siamo interessati a studiare il caso $M\propto t$: imponiamo dunque
che la successione $m_{t}$ converga ad un valore finito per $t\rightarrow\infty$:
\begin{equation}
\lim_{t\rightarrow\infty}m_{t}=m\in\left[0,\,1\right].\label{eq:succession}
\end{equation}
Le relazioni \prettyref{eq:1.2.2}, \prettyref{eq:1.2.3} per la distribuzione
del range, suggeriscono di cercare soluzioni di \prettyref{eq:mastereq2}
nella forma 

\begin{equation}
P_{t}\left(m_{t}\right)=\mathcal{N\,}e^{\varphi\left(m_{t}\right)t},\label{eq:n-form}
\end{equation}
con $\varphi\left(m\right)$ supposta di classe $C^{\infty}$. Sostituendo
nell'equazione troviamo 
\begin{equation}
e^{\varphi\left(m_{t+1}\right)\left(t+1\right)}=\pi\left(m_{t}-t^{-1}\right)e^{\varphi\left(m_{t}-t^{-1}\right)t}+\left[1-\pi\left(m_{t}\right)\right]e^{\varphi\left(m_{t}\right)t}.\label{eq:EQMAZ}
\end{equation}
Dalle ipotesi $\varphi\left(m\right)\in C^{\infty}$, $m_{t\rightarrow\infty}\in\left[0,\,1\right]$,
possiamo espandere $\varphi\left(m_{t}-t^{-1}\right)$ in serie di
$t^{-1}$ 
\begin{equation}
\varphi\left(m_{t}-t^{-1}\right)=\varphi\left(m_{t}\right)-\dot{\varphi}\left(m_{t+1}\right)\,t^{-1}+\mathcal{O}\left(t^{-2}\right),\label{eq:ESPA1}
\end{equation}
dove, per comodità, si è introdotta la notazione $\dot{\varphi}\left(m\right)=\partial_{m}\varphi\left(m\right)$. 

Dalla definizione di $m_{t}$, sempre condizionata $m_{t\rightarrow\infty}\in\left[0,\,1\right]$,
si può riscrivere $m_{t+1}$ espandendo $1/\left(1+t\right)$ in termini
$\mathcal{O}\left(t^{-1}\right)$: 
\begin{equation}
m_{t+1}=m_{t}-t^{-1}m_{t}+\mathcal{O}\left(t^{-2}\right).
\end{equation}
Inserendo in $\varphi\left(m_{t+1}\right)$ e proseguendo come sopra,
troveremo: 
\begin{equation}
\varphi\left(m_{t+1}\right)=\varphi\left(m_{t}\right)-m_{t}\,\dot{\varphi}\left(m_{t+1}\right)\,t^{-1}+\mathcal{O}\left(t^{-2}\right).\label{eq:ESPA2}
\end{equation}
Sostituiamo le relazioni appena trovate nella \prettyref{eq:EQMAZ}:

\begin{equation}
e^{\varphi\left(m_{t}\right)-m_{t}\,\dot{\varphi}\left(m_{t}\right)+\mathcal{O}\left(t^{-1}\right)}=\pi\left(m_{t}-t^{-1}\right)e^{-\dot{\varphi}\left(m_{t}\right)+\mathcal{O}\left(t^{-1}\right)}+1-\pi\left(m_{t}\right).
\end{equation}
In fine, prendendo il limite $t\rightarrow\infty$ si arriva alla
seguente equazione differenziale in $m$:
\begin{equation}
e^{\varphi\left(m\right)-m\,\dot{\varphi}\left(m\right)}=\pi\left(m\right)e^{-\dot{\varphi}\left(m\right)}+1-\pi\left(m\right),\label{eq:EQUAZIONEFONDAMENTALE}
\end{equation}
dove $\varphi\left(m\right)$ è identificabile con il limite 
\begin{equation}
\varphi\left(m\right)=\lim_{t\rightarrow\infty}\log\left(P_{t}\left(m_{t}\right)^{\frac{1}{t}}\right).\label{eq:VARFIFUNCT}
\end{equation}

Dalla equazione discreta \prettyref{eq:mastereq2} siamo dunque riusciti,
attraverso alcune ipotesi, a derivare una equazione differenziale
per $\varphi\left(m\right)$, valida nel limite asintotico. Purtroppo
la \prettyref{eq:EQUAZIONEFONDAMENTALE} è trascendente implicita
nella variabile $\dot{\varphi}\left(m\right)$: una sua soluzione
con strumenti analitici classici sembra dunque improbabile.

Come vedremo a breve, è comunque possibile estrarre dalla \prettyref{eq:EQUAZIONEFONDAMENTALE}
una grande quantità di informazioni: di seguito riporteremo alcuni
risultati ottenibili con metodi perturbativi.

\section{Soluzioni perturbative intorno al massimo}

Iniziamo col verificare che i primi due momenti della distribuzione
coincidano con quelli calcolati in precedenza. Si considerino dunque
le espansione al secondo ordine di $\pi\left(m\right)$, $\varphi\left(m\right)$
nel punto $m=m_{0}$: chiamando $\delta m=m-m_{0}$ si avrà 
\begin{equation}
\pi\left(m\right)=m_{0}+b\delta m+\mathcal{O}\left(\delta m^{2}\right),
\end{equation}
\begin{equation}
\varphi\left(m\right)=\varphi_{0}+\varphi_{1}\delta m-\varphi_{2}\delta m^{2}+\mathcal{O}\left(\delta m^{3}\right).
\end{equation}
Espandiamo al secondo ordine gli esponenziali in \prettyref{eq:EQUAZIONEFONDAMENTALE}
e sostituiamo le espressioni precedenti: trascurando i termini $\mathcal{O}\left(\delta m^{3}\right)$,
e annullando i coefficienti per uguali potenze di $\delta m$, troveremo:
$\varphi_{0}=\varphi_{1}=0$, $\varphi_{2}=\left(1-2b\right)/\left(2m_{0}\bar{m}_{0}\right)$,
ovvero 
\begin{equation}
\varphi\left(m\right)=-\frac{\delta m^{2}}{2\sigma}+\mathcal{O}\left(\delta m^{3}\right),
\end{equation}
dove $\sigma=m_{0}\bar{m}_{0}/\left(1-2b\right)$. Si nota immediatamente
che i primi due momenti della distribuzione limite $\exp\left[\varphi\left(t^{-1}M\right)\,t\right]$
(non normalizzata) sono

\begin{equation}
\langle M\rangle=m_{0}t+o\left(t\right),
\end{equation}
\begin{equation}
\langle\Delta M^{2}\rangle=\sigma t+o\left(t\right),
\end{equation}
e coincidono con quelli calcolati in precedenza per $P_{t}\left(M\right)$.
Si osservi come, aggiungendo termini fino all'ordine $s-$esimo, i
momenti di ordine superiore possano essere facilmente calcolati al
computer anche quando necessitano di $s$ molto grandi, impostando
un sistema di $s$ equazioni in $s$ variabili con un qualunque sistema
di algebra computazionale (CAS).

Lo studio perturbativo della \prettyref{eq:EQUAZIONEFONDAMENTALE}
permette inoltre di spiegare il caso $b>1/2$: per fare questo dobbiamo
considerare una $\varphi\left(m\right)$ del tipo 
\begin{equation}
\varphi\left(m\right)=-\mathcal{A}\left|\delta m\right|^{\beta}+o\left(\left|\delta m\right|^{\beta}\right),\label{eq:betabtera}
\end{equation}
con $\beta\in\left(2^{+},\,\infty\right)$. 

Contrariamente a quanto richiesto fino ad ora, questa $\varphi\left(m\right)$
non è sempre $C^{\infty}$: per $\beta\neq2k$, $k\in\mathbb{N}^{+}$,
le derivate in $m_{0}$ esistono fino all'ordine $\left\lfloor \beta\right\rfloor $,
ma quelle di oridini superiori divergono. In questa situazione le
espansioni \prettyref{eq:ESPA1}, \prettyref{eq:ESPA2} non sono più
valide in $m_{0}$, ma sussistono ancora nel suo intorno $\left\{ m_{0}-\epsilon,\,m_{0}+\epsilon\right\} \setminus m_{0}$,
in cui la \prettyref{eq:EQUAZIONEFONDAMENTALE} resta accurata. 

Linearizzando gli esponenziali, e annullando i coefficienti con esponenti
uguali, troveremo l'equazione 
\begin{equation}
\mathcal{A}\left(1-\beta+b\beta\right)=0,\label{eq:singolar}
\end{equation}
che lega $b$ e $\beta$ secondo la relazione: 
\begin{equation}
\beta=\left(1-b\right)^{-1}.
\end{equation}
Si noti che la \prettyref{eq:singolar} lascia indeterminata l'ampiezza
$\mathcal{A}$: se $\beta\neq2k$, $k\in\mathbb{N}^{+}$, questo avviene
per qualunque ordine di sviluppo degli esponenziali, ed è imputabile
alla singolarità in $m_{0}$. A titolo speculativo, si osserva che
una estensione delle \prettyref{eq:ESPA1}, \prettyref{eq:ESPA2}
a forme più generali di $\varphi\left(m\right)$ (o di $P_{t}\left(m_{t}\right)$
in generale) potrebbe essere possibile attraverso uno sviluppo in
serie con esponenti razionali (\emph{Fractional Taylor Series} FTS),
che tuttavia necessita di un massiccio uso del calcolo frazionario. 

\begin{figure}
\caption{}

\begin{centering}
\includegraphics[scale=0.5]{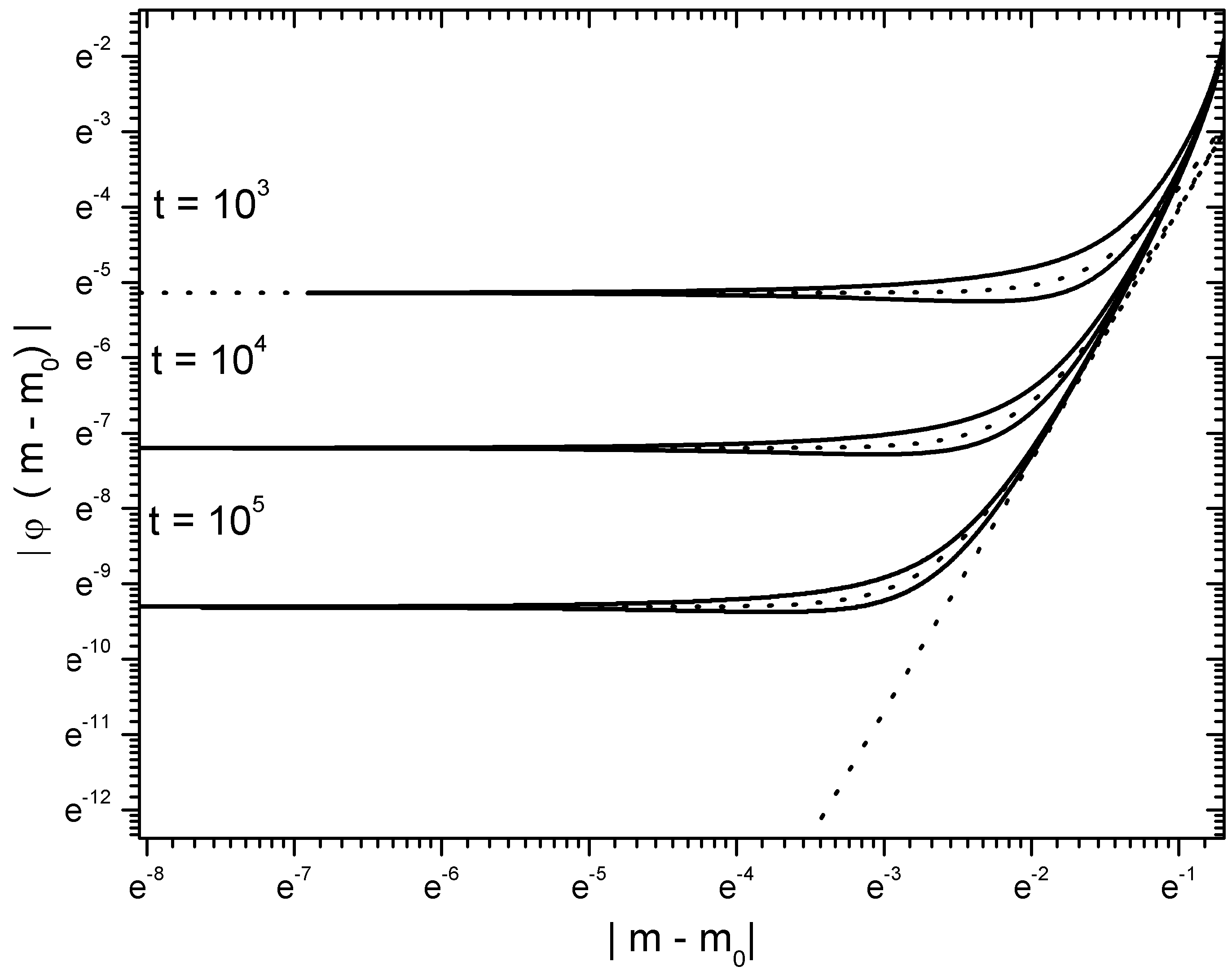}\label{Flo:figura3.2}
\par\end{centering}
\emph{Il grafico mostra $\left|\varphi_{t}\left(\delta m\right)\right|$
vs $\left|\delta m\right|$, ottenuto simulando il processo \prettyref{eq:3.1.1-4},
\prettyref{eq:3.1.1-5}, con $\pi\left(m\right)=1/2+b\left(m-1/2\right)$,
$b=7/10$. I dati sono (dall'alto verso il basso) per $t=10^{3}$,
$t=10^{4}$, $t=10^{5}$. Le linee di punti rappresentano invece le
previsioni $\varphi_{t}\left(m\right)=-c_{t}-\mathcal{A}\left|\delta m\right|^{\beta}$,
con $\beta=1/\left(1-b\right)=10/3$, }\foreignlanguage{english}{\emph{$\mathcal{A}=0.5$}}\emph{,
$c_{10^{3}}=5,857\cdot10^{-2}$, $c_{10^{4}}=7,460\cdot10^{-4}$,
$c_{10^{5}}=9,091\cdot10^{-6}$, $c_{\infty}=0$. La \prettyref{eq:betabtera}
con $\beta=1/\left(1-b\right)$ sembra confermata.}
\end{figure}

In ogni caso, è facile dimostrare come la distribuzione associata
alla \prettyref{eq:betabtera} dia un momento secondo $\langle\Delta M^{2}\rangle\propto t^{\delta}$,
con 
\begin{equation}
\delta=2\left(1-\beta^{-1}\right)=2b
\end{equation}
uguale a quello previsto dalla \prettyref{eq:2.1.9} per $b>1/2$.
La \prettyref{eq:singolar} è inoltre avallata da un confronto con
la soluzione numerica della \prettyref{eq:mastereq2} (Figura \prettyref{Flo:figura3.2}).

Il caso $b=1/2$ rappresenta una situazione particolare. A causa della
correzione logaritmica per l'andamento di $\langle\Delta M^{2}\rangle$,
non è possibile esprimere perturbativamente il limite di $\log\left(P_{t}\left(m_{t}\right)\right)$
in $m_{0}$ con termini dipendenti solo dal rapporto $n^{-1}M$, dunque
l'equazione \prettyref{eq:EQUAZIONEFONDAMENTALE} non è valida: è
comunque possibile affermare che il limite di $\log\left(P_{t}\left(m_{t}\right)\right)$
per $t\rightarrow\infty$, $m_{t}\rightarrow m_{0}+\delta m$ è sempre
compreso fra due funzioni $c_{1}\left|\delta m\right|^{2}$ e $c_{2}\left|\delta m\right|^{2+\epsilon}$,
con $\epsilon\rightarrow0^{+}$.

\section{Due casi specifici }

Come ultimo argomento, affrontiamo un caso particolare che ci sarà
utile più avanti. Consideriamo due funzioni, $\tilde{\pi}\left(m\right)$
e $\mu\left(m\right)$: la prima gode delle proprietà richieste fino
ad ora a $\pi\left(m\right)$, ovvero che sia invertibile, che $\tilde{\pi}\left(m\right)\in\left(0,\,1\right)$,
$\tilde{\pi}\left(m\right)\in C^{\infty}$ per $m\in\left[0,\,1\right]$,
e che 
\begin{equation}
\exists!\,m_{0}\,|\,\tilde{\pi}\left(m_{0}\right)=m_{0}.
\end{equation}
La seconda invece, sia una funzione invertibile di classe $C^{\infty}$,
con $\mu\left(1\right)=1$, $\mu\left(m\right)<m$ nell'intervallo
$m\in\left(m_{0},\,1\right)$. Date le proprietà richieste alle due
funzioni, segue che 
\begin{equation}
\exists!\,m_{c}\in\left[m_{0},\,1\right)\,|\,\tilde{\pi}\left(m_{c}\right)=\mu\left(m_{c}\right).
\end{equation}
 Detto questo, definiamo la funzione $\pi\left(m\right)$ come segue:
\begin{equation}
\pi\left(m\right)=\left\{ \begin{array}{l}
\tilde{\pi}\left(m\right)\\
\mu\left(m\right)
\end{array}\,\,\,\begin{array}{r}
m\geq m_{c}\\
m<m_{c}
\end{array}\right..\label{eq:BARBABLU}
\end{equation}
Questa funzione è globalmente $C^{0}$, con un punto di discontinuità
della derivata in $m_{c}$, e due soluzioni di $\pi\left(m\right)=m$
(in $m_{0}$ e $1$). Si tratta essenzialmente della $\pi\left(m\right)$
descritta nel capitolo precedente , con l'ulteriore assunzione che
$\pi\left(m\right)<m$ nell'intervallo $m\in\left(m_{0},\,1\right)$
(il motivo sarà chiaro più avanti).

Vogliamo trovare la distribuzione generata dalla \prettyref{eq:BARBABLU}
nella forma $\exp\left[\varphi\left(m\right)t\right]$, con $\varphi\left(m\right)\in C^{1}$
soluzione della \prettyref{eq:EQUAZIONEFONDAMENTALE}.

La \prettyref{eq:BARBABLU} ha per definizione due intersezioni con
la retta $m$, in $m=m_{0}$ e $m=1$; ma, mentre $b_{0}=\dot{\pi}\left(m_{0}\right)<1$,
per $m=1$ le proprietà richieste a $\mu\left(m\right)$ impongono
necessariamente $b_{1}=\dot{\pi}\left(1\right)>1$. 

Dalla \prettyref{eq:2.1.9} sappiamo che la varianza $\langle\Delta M_{t}^{2}\rangle$
scala come una potenza di $t$, con un esponente $2b$. Se $b>1$
le fluttuazioni sono dunque su una scala $\mathcal{O}\left(t^{b}\right)$,
e divergono rispetto alla media ($\langle\Delta M_{t}^{2}\rangle/\langle M_{t}\rangle^{2}\rightarrow\infty$),
coprendo un intorno $m\in\left[1-\epsilon,1\right]$ con $\epsilon>0$.

L'unica $\varphi\left(m\right)\in C^{\infty}$ compatibile con questa
situazione, che sia anche soluzione di \prettyref{eq:EQUAZIONEFONDAMENTALE},
è $\varphi\left(m\right)=0$ (che risolve la \prettyref{eq:EQUAZIONEFONDAMENTALE}
indipendentemente da $\pi\left(m\right)$).

Tuttavia se $\pi\left(0\right)>0$, segue necessariamente $\varphi\left(0\right)<0$;
abbiamo dunque bisogno di una funzione $C^{\infty}$, soluzione di
\prettyref{eq:EQUAZIONEFONDAMENTALE}, che connetta $\varphi\left(0\right)<0$
con $\varphi\left(m\in\left[1-\epsilon,1\right]\right)=0$, dando
globalmente una $\varphi\left(m\right)\in C^{1}$ nel range $m\in\left[0,\,1\right]$. 

\begin{figure}
\caption{}

\begin{centering}
\includegraphics[scale=0.5]{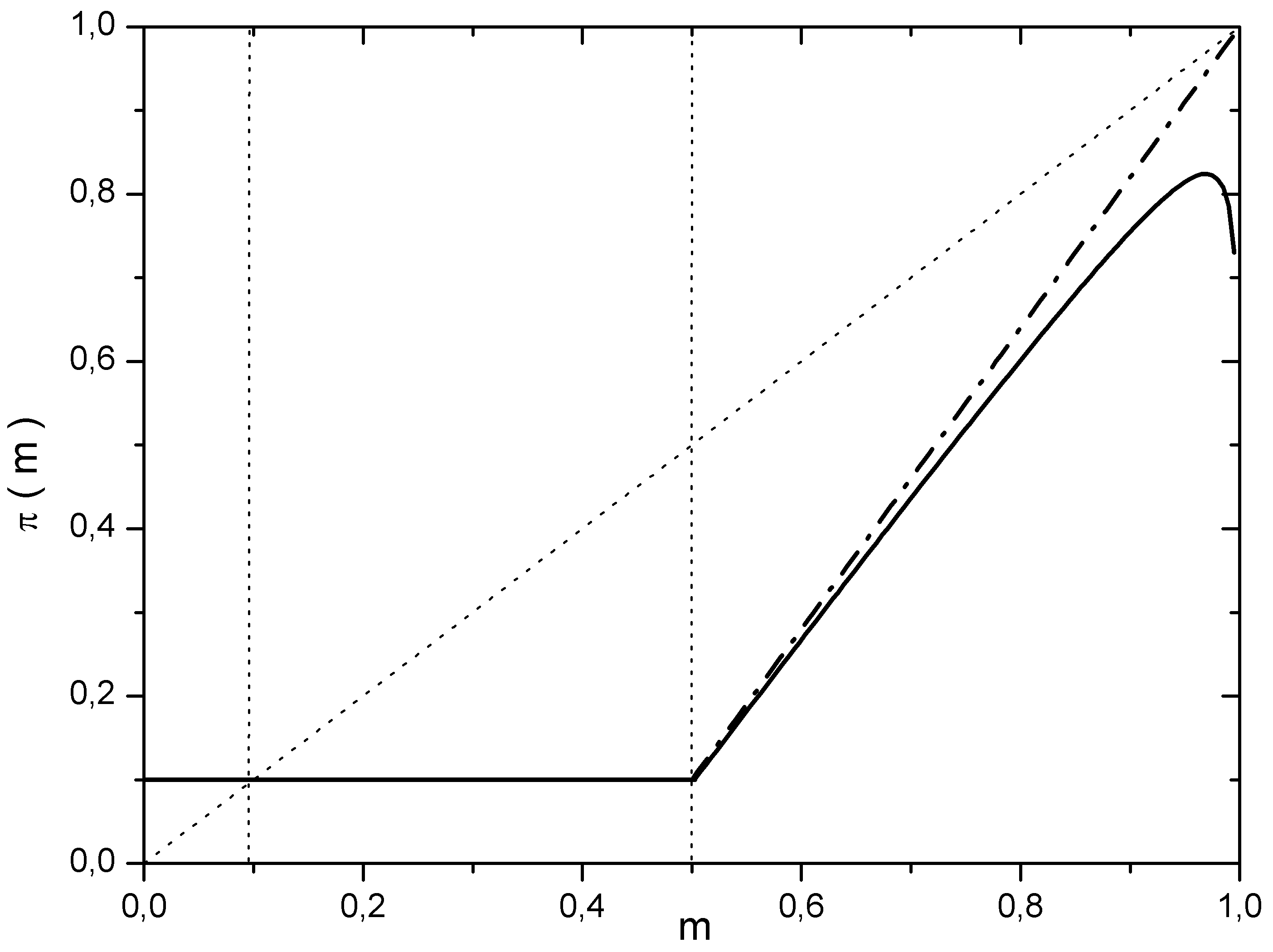}\label{Flo:figura3.3}
\par\end{centering}
\emph{Generatori $\pi\left(m\right)$ e $\pi_{t}\left(m\right)$ per
le distribuzioni nelle successive Figure \prettyref{Flo:figura3.4},
\prettyref{Flo:figura3.5}. La linea interrotta da punti rappresenta
$\pi\left(m\right)$ data dalla \prettyref{eq:BARBABLU}, con $\tilde{\pi}\left(m\right)=0.1$
costante, $m_{c}=0.5$ e $\mu\left(m\right)=1-1.8\left(1-m\right)$.
La linea continua è invece è la \prettyref{eq:baOBAB}, con $\psi_{t}\left(m\right)=1-0.2\left[\left(1-m\right)t\right]^{-s}$,
$s=0.4$, $t=10^{2}$. }
\end{figure}

\begin{figure}
\label{Flo:figura3.4}\caption{}

\begin{centering}
\includegraphics[scale=0.5]{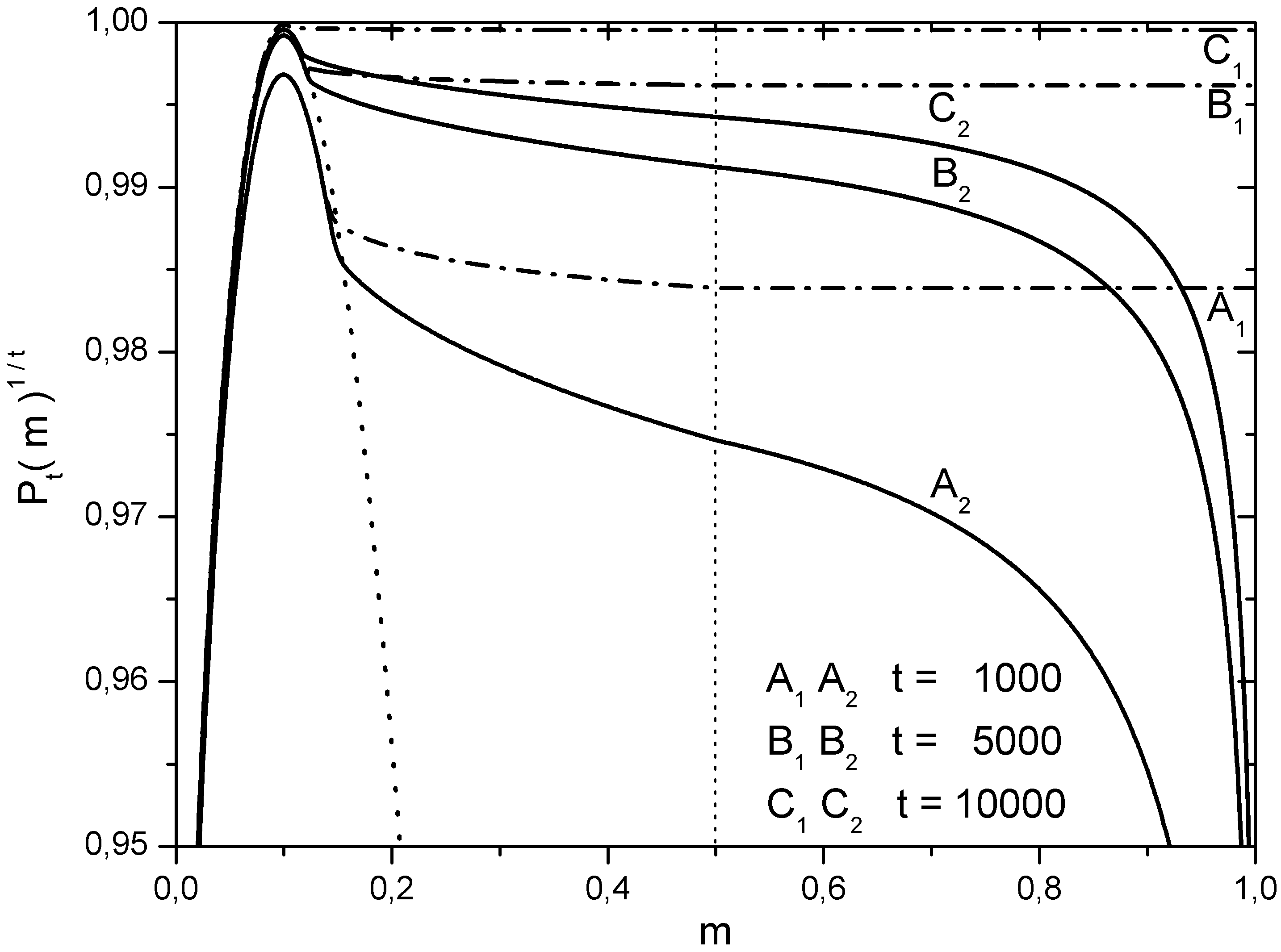}
\par\end{centering}
\emph{Soluzioni numeriche della \prettyref{eq:mastereq1}, generate
dalle $\pi\left(m\right)$, $\pi_{t}\left(m\right)$ descritte nella
precedente Figura \prettyref{Flo:figura3.3}. Le linee interrotte
da punti sono le distribuzioni dalla $\pi\left(m\right)$, per (partendo
dal basso) $t=10^{3}$, $t=5\cdot10^{3}$, $t=10^{4}$. Le linee piene
sono relative alla $\pi_{t}\left(m\right)$, sempre per (dal basso)
$t=10^{3}$, $t=5\cdot10^{3}$, $t=10^{4}$. In fine, la linea di
punti rappresenta $\tilde{\varphi}\left(m\right)$, soluzione della
\prettyref{eq:EQUAZIONEFONDAMENTALE} per $\tilde{\pi}\left(m\right)$.}

\emph{Il grafico mostra il comportamento predetto nella \prettyref{eq:forma1}.
Si noti che per tutte le curve è presente una discontinuità della
derivata in $m=m_{c}=0.5$. }
\end{figure}

Sia $\tilde{\varphi}\left(m\right)$ la soluzione della \prettyref{eq:EQUAZIONEFONDAMENTALE}
con $\pi\left(m\right)=\tilde{\pi}\left(m\right)$: è facile verificare
che 

\begin{equation}
\varphi\left(m\right)=\left\{ \begin{array}{l}
\tilde{\varphi}\left(m\right)\\
0
\end{array}\,\,\,\begin{array}{r}
m\in\left[0,\,m_{0}\right]\\
m\in\left(m_{0},\,1\right]
\end{array}\right.\label{eq:forma1}
\end{equation}
è l'unica funzione con le proprietà richieste. Una verifica di \prettyref{eq:forma1}
viene dal confronto con la soluzione numerica della \prettyref{eq:mastereq1}
(Figure \prettyref{Flo:figura3.3}, \prettyref{Flo:figura3.4}): da
questi dati si nota anche una discontinuità di $t^{-1}\log\left(P_{t}\left(m\right)\right)$
nel punto $m_{c}$.

Si conclude il capitolo enunciando due congetture circa la forma di
$\log\left(P_{t}\left(m\right)\right)$, basate essenzialmente sullo
studio numerico della \prettyref{eq:mastereq1}. 

La prima è una considerazione sulla generica $\pi_{t}\left(m\right)\in\left[0,\,1\right]$:
si ipotizza che, ogni qualvolta esista un $m_{i}$ con $\pi_{t}\left(m_{i}\right)=m_{i}$,
$\dot{\pi}_{t}\left(m_{i}\right)<1$, il punto $m_{i}$ sarà un massimo
locale di $\log\left(P_{t}\left(m\right)\right)$.

\begin{figure}
\caption{}

\begin{centering}
\includegraphics[scale=0.5]{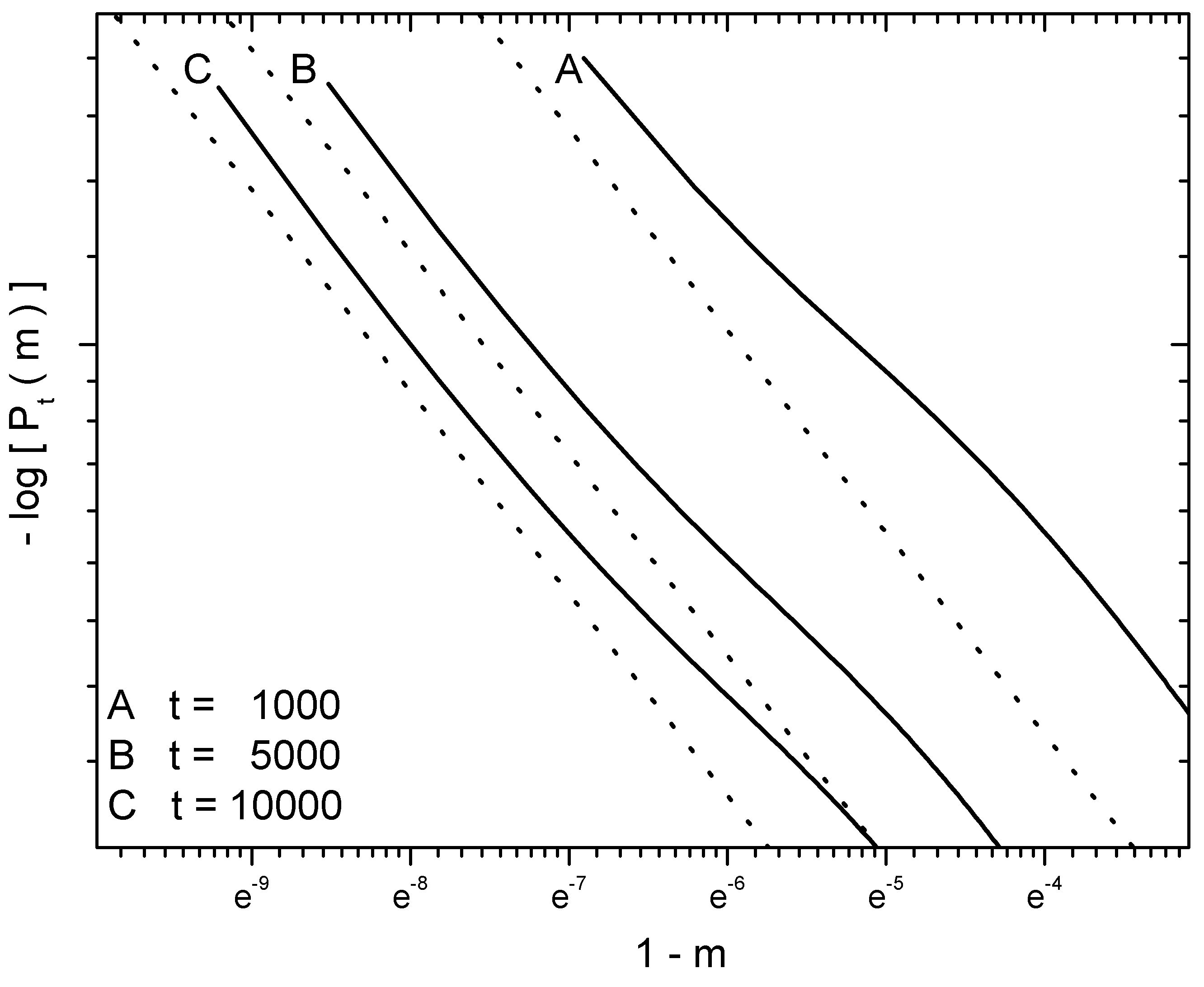}\label{Flo:figura3.5}
\par\end{centering}
\emph{La figura mostra, in scala doppio logaritmica, un particolare
di $-\log\left(P_{t}\left(m\right)\right)$ vs $\left(1-m\right)$
per le distribuzioni nella precedente Figura \prettyref{Flo:figura3.4}.
Le linee piene sono le distribuzioni calcolate con la $\pi_{n}\left(m\right)$
descritta in Figura \prettyref{Flo:figura3.3}, con (da destra) }$t=10^{3}$,
$t=5\cdot10^{3}$, $t=10^{4}$.\emph{ Le linee di punti rappresentano
invece }$\bar{\psi}_{t}\left(m\right)=0.2\left[\left(1-m\right)t\right]^{-s}$,
$s=0.4$\emph{, per i corrispondenti $t$ (partendo dall'alto }$t=10^{3}$,
$t=5\cdot10^{3}$, $t=10^{4}$\emph{).}
\end{figure}

La seconda riguarda la $\pi_{t}\left(m\right)$ descritta nel capitolo
precedente: si considerino le $\psi_{t}\left(m\right)=1-\bar{\psi}_{t}\left(m\right)$,
$\mu_{t}\left(m\right)$ definite attraverso le relazioni 
\begin{equation}
\bar{\psi}_{t}\left(m\right)=\psi_{0}\left(\bar{m}t\right)^{-s},
\end{equation}
\begin{equation}
\mu_{t}\left(m\right)=\mu\left(m\right)\psi_{t}\left(m\right).
\end{equation}
Definiamo la funzione $\pi_{t}\left(m\right)$ come 
\begin{equation}
\pi_{t}\left(m\right)=\left\{ \begin{array}{l}
\tilde{\pi}\left(m\right)\\
\mu_{t}\left(m\right)
\end{array}\,\,\,\begin{array}{r}
\tilde{\pi}\left(m\right)\geq\mu_{t}\left(m\right)\\
\tilde{\pi}\left(m\right)<\mu_{t}\left(m\right)
\end{array}\right..\label{eq:baOBAB}
\end{equation}
Nel limite $t\rightarrow\infty$, la \prettyref{eq:baOBAB} tende
alla \prettyref{eq:BARBABLU}, tuttavia la presenza di $\psi_{t}\left(m\right)$
fa in modo che $\pi_{t}\left(1\right)<1$ per $t<\infty$: basandoci
su una analisi numerica (di cui si riporta un esempio nelle Figure
\prettyref{Flo:figura3.3}, \prettyref{Flo:figura3.4}, \prettyref{Flo:figura3.5}),
si congettura che $\log\left(P_{t}\left(m\right)\right)$ per $m_{t}\in\left[m_{0},\,1\right]$
sia una funzione con $\partial_{m}\log\left(P_{t}\left(m\right)\right)\leq0$
(non crescente), il cui limite $t\rightarrow\infty$ per $m\sim1$
è 
\begin{equation}
\lim_{t\rightarrow\infty}\log\left(P_{t}\left(m\right)\right)\propto-\lambda\bar{\psi}_{t}\left(m\right)t\label{eq:prevision}
\end{equation}
con $\lambda>0$ (si veda la Figura \prettyref{Flo:figura3.5}). Si
osservi che $t^{-1}\log\left(P_{t}\left(m\right)\right)$ si annulla
nel limite $t\rightarrow\infty$, in conformità con la \prettyref{eq:forma1}.

Data la prima congettura, appare chiaro che, senza l'assunzione $\mu\left(m\right)<m$
per $m_{i}\in\left(m_{0},\,1\right)$, $\log\left(P_{t}\left(m\right)\right)$
può avere anche $\partial_{m}\log\left(P_{t}\left(m\right)\right)>0$
per alcuni $m$: se esistono punti $m_{i}\in\left(m_{0},\,1\right)$
per cui $\mu\left(m_{i}\right)=m_{i}$, per continuità di $\mu\left(m\right)$
ne deve esistere almeno uno con $\dot{\mu}\left(m_{i}\right)<1$.
Se invece $\mu\left(m\right)=1$ solo per $m=1$, l'applicazione di
$\psi_{t}\left(m\right)$ eliminerà l'intersezione di $\mu_{t}\left(m\right)$
con la retta $m$, e il conseguente massimo locale previsto per $m=1$.\pagebreak{}

\chapter{Forma dell'\emph{ambiente} $\pi\left(m\right)$ }

Nel secondo capitolo abbiamo discusso $\pi_{n}\left(m\right)$, separando
il contributo di taglia finita $\psi_{n}\left(m\right)$ dalla parte
$\pi\left(m\right)$, dipendente solo dalla concentrazione delle intersezioni
$m$. Vogliamo ora discutere in dettaglio la forma di $\pi\left(m\right).$

Abbiamo già individuato due valori particolari, corrispondenti a $\pi\left(C_{d}\right)=C_{d}$,
$\pi\left(0\right)=\eta_{d}$; come anticipato, è possibile imporre
una terza condizione legata alla varianza della distribuzione $P_{n}\left(m\right)$.

Nel capitolo precedente abbiamo visto come per $\pi\left(m\right)\in C^{\infty}$,
con un'unica soluzione $m_{0}$ per l'equazione $\pi\left(m\right)=m$,
la varianza della distribuzione definita dal processo \prettyref{eq:3.1.1-4},
\prettyref{eq:3.1.1-5} al tempo $t$ dipendesse solo da $m_{0}$
e $b=\dot{\pi}\left(m_{0}\right)$, secondo la relazione $\langle\Delta M_{t}^{2}\rangle=\sigma t$,
$\sigma=m_{0}\bar{m}_{0}/\left(1-2b\right)$. 

Invertendo la relazione per $\sigma$, è possibile risalire al valore
$B_{d}$ della derivata di $\pi\left(m\right)$ in $C_{d}$: ponendo
$m_{0}=C_{d}$, $\sigma=\sigma_{d}$ (tabella \prettyref{Flo:tab1}),
e $b=B_{d}$, si ottiene la relazione. 
\begin{equation}
2B_{d}=1-\sigma_{d}^{-1}C_{d}\bar{C}_{d}.
\end{equation}
E' opportuno prestare particolare attenzione al caso $d=3$: confrontando
la \prettyref{eq:2.1.9} con l'espressione di $\langle\Delta R_{n}^{2}\rangle$
per $d=3$ , si nota immediatamente che il corretto andamento è ottenuto
per $B_{3}=1/2$. Con questo valore si ottiene infatti $\langle\Delta M_{n}^{2}\rangle=\alpha n\log n$,
conformemente con la (1.19). In realtà l'ampiezza $\alpha=C_{d}\bar{C_{d}}$,
che troviamo con la \prettyref{eq:2.1.7-2} per $B_{d}=1/2$, è diversa
dal valore previsto $\sigma_{3}$ (tabella \prettyref{Flo:tab1}). 

Il risultato corretto si ottiene includendo una fluttuazione $\mathcal{O}\left(1/\log\left(n\right)\right)$,
determinata dalla successione $a_{n}=\sigma_{3}\log\left(n\right)$
(per una definizione di $a_{n}$ si veda l'appendice relativa al teorema
di Jain-Pruitt, o la referenza \cite{Jain-Pruitt} per il caso $d=3$): 

\begin{equation}
2B_{3}\left(n\right)\sim1+\left(1-\sigma_{3}^{-1}C_{3}\bar{C}_{3}\right)\log\left(n\right)^{-1}+\mathcal{O}\left(n^{-1}\right).
\end{equation}
Nel limite $n\rightarrow\infty$ questa espressione converge ad $1/2$,
ma la fluttuazione $\mathcal{O}\left(1/\log\left(n\right)\right)$
si annulla così lentamente da non poter essere trascurata. Si può
facilmente verificare come, sostituendo $B_{3}\left(n\right)$ in
\prettyref{eq:2.1.7-2}, si ottenga il risultato corretto per la varianza
in $d=3$.

Fino ad ora abbiamo studiato $\pi\left(m\right)$ da un punto di vista
più o meno fenomenologico: nella prossima sezione proveremo a ragionare
partendo dalle proprietà geometriche dei cammini. 

Abbiamo introdotto l'esistenza di un valore critico $m_{c}$, oltre
il quale le catene collassano in cluster compatti: per ora, il nostro
interesse sarà rivolto al caso opposto, $m<m_{c}$, in cui le catene
sono estese e le correlazioni tra monomeri più deboli: come vedremo
a breve, quest'ultimo fatto sarà cruciale per ottenere una stima di
$\pi\left(m\right)$. 

\section{Decomposizione di una catena $m<m_{c}$}

Prima di discutere la forma della funzione \emph{ambiente} per $m<m_{c}$
è opportuno fare alcune considerazioni preliminari. Immaginiamo di
decomporre la catena $\omega=\left\{ S_{0},\,S_{1},\,...\,,\,S_{n}\right\} $
in $n_{\tau}+1$ porzioni $\omega_{i}$ , composte da $\tau=n/\left(n_{\tau}+1\right)$
monomeri consecutivi. 
\begin{equation}
\omega_{i}\equiv\left\{ S_{i\tau},\,...\,,\,S_{\left(i+1\right)\tau-1}\right\} ,
\end{equation}
\begin{equation}
\left\{ S_{0},\,...\,,\,S_{n}\right\} \rightarrow\left\{ \omega_{0}\,,\,...\,,\,\omega_{n_{\tau}}\right\} .
\end{equation}
Dalle definizioni contenute nel primo capitolo possiamo riscrivere
le quantità $\pi\left[\omega\right]$, $M\left[\omega\right]$ come
\begin{equation}
\pi\left[\omega\right]=\pi\left[\omega_{0}\right]+\sum_{p=1}^{n_{\tau}}\sum_{k=0}^{p}\frac{\left(-1\right)^{k}}{2d}\sum_{\left|x\right|=1}\prod_{\alpha=p-k}^{p}R_{x}\left[\omega_{\alpha}\right],\label{eq:3.4.4}
\end{equation}
\begin{equation}
M\left[\omega\right]=\sum_{h_{0}=0}^{n_{\tau}}M\left[\omega_{h_{0}}\right]+\sum_{p=1}^{n_{\tau}}\sum_{\left\{ h,p\right\} }\left(-1\right)^{p+1}\sum_{x}\prod_{\alpha=0}^{p}R_{x}\left[\omega_{h_{\alpha}}\right],\label{eq:3.4.5}
\end{equation}
dove $R_{x}\left[\omega_{i}\right]$ è la funzione caratteristica
della $i-$esima porzione $\omega_{i}$, ed il vincolo $\left\{ h,p\right\} $
sulla sommatoria della \prettyref{eq:3.4.5} è 
\begin{equation}
\sum_{\left\{ h,p\right\} }=\sum_{h_{0}>0}\sum_{h_{1}>h_{0}}\sum_{h_{2}>h_{1}}\,...\,\sum_{h_{p}>h_{p-1}}.
\end{equation}

La \prettyref{eq:3.4.4} separa il contributo a $\pi\left[\omega\right]$
della porzione contenente l'origine da quello delle componenti più
lontane, mentre la scrittura \prettyref{eq:3.4.5} per $M\left[\omega\right]$
separa il contributo di auto-intersezione dei singoli componenti $\omega_{i}$
da quello di interazione tra componenti diverse.

Consideriamo per ora la sola espressione di $\pi\left[\omega\right]$.
Prendiamo la media della quantità \prettyref{eq:3.4.4}, sull'insieme
$\Omega_{m}$, con $m<m_{c}$ e $d\geq3$: 
\begin{equation}
\pi\left(m\right)=\langle\pi\left[\omega_{0}\right]\rangle_{\Omega_{m}}+\sum_{p=1}^{n_{\tau}}\sum_{k=0}^{p}\frac{\left(-1\right)^{k}}{2d}\sum_{\left|x\right|=1}\langle\prod_{\alpha=p-k}^{p}R_{x}\left[\omega_{\alpha}\right]\rangle_{\Omega_{m}}\label{eq:3.4.5-1}
\end{equation}

La sommatoria nella \prettyref{eq:3.4.5-1} rappresenta l'interazione
delle porzioni che si trovano ad una distanza maggiore di $\tau$
percorrendo la catena: si osserva che questi contributi sono tutti
del tipo 
\begin{equation}
\sum_{\left|x\right|=1}\langle R_{x}\left[\omega_{i\neq0}\right]R_{x}\left[\omega_{\alpha}\right]R_{x}\left[\omega_{\beta}\right]R_{x}\left[\omega_{\delta}\right]...\rangle_{\Omega_{m}}\leq\sum_{\left|x\right|=1}\langle R_{x}\left[\omega_{i\neq0}\right]\rangle_{\Omega_{m}}.
\end{equation}
Consideriamo l'interazione $\sum_{\left|x\right|=1}\langle R_{x}\left[\omega_{i}\right]\rangle_{\Omega_{m,n}}$
di un elemento $\omega_{i\neq0}$ con l'intorno dell'origine: come
prima approssimazione possiamo considerare che il contributo sia non
nullo solo se il cammino che separa $\omega_{i}$ dall'origine è chiuso
(sia un \emph{loop}), ovvero solo quando anche $\omega_{i}$ contiene
l'origine. Detta $f_{t}\left(m\right)$ la probabilità di avere un
loop chiuso sull'origine di lunghezza $t$, possiamo introdurre la
relazione

\begin{equation}
\sum_{\left|x\right|=1}\langle R_{x}\left[\omega_{i}\right]\rangle_{\Omega_{m}}\propto\tau\,f_{i\tau}\left(m\right),\label{eq:3.4.5.2}
\end{equation}
dove la moltiplicazione per $\tau$ tiene conto del fatto che il cammino
può essere chiuso su uno qualunque dei monomeri di $\omega_{i}$. 

La probabilità di avere un loop è proporzionale alla densità di monomeri
per unità di volume (nel caso del SRW questo è confermato dalla derivata
temporale della \prettyref{eq:1.3.8}): poniamo dunque $f_{t}\left(m\right)\propto t^{-\nu d}$.
Da questo, dalla \prettyref{eq:3.4.5.2}, e dal fatto che $\nu d>1$
per $d\geq3$, si ottiene la relazione:\\
\begin{equation}
\sum_{i=1}^{n_{\tau}}\sum_{\left|x\right|=1}\langle R_{x}\left[\omega_{i}\right]\rangle_{\Omega_{m}}=\mathcal{O}\left(\tau^{1-\nu d}\right).\label{eq:3.4.5.3}
\end{equation}
La \prettyref{eq:3.4.5.3} è una sovrastima della sommatoria in \prettyref{eq:3.4.5-1}:
ne deriva che la \prettyref{eq:3.4.5-1} può essere riscritta come
\begin{equation}
\pi\left(m\right)=\langle\pi\left[\omega_{0}\right]\rangle_{\Omega_{m}}+\mathcal{O}\left(\tau^{1-\nu d}\right).\label{eq:3.4.5.4}
\end{equation}
Applicando lo stesso ragionamento sulla relazione
\begin{equation}
n\,m=\sum_{h_{0}=0}^{n_{\tau}}\langle M\left[\omega_{h_{0}}\right]\rangle_{\Omega_{m}}+\sum_{p=1}^{n_{\tau}}\sum_{\left\{ h,p\right\} }\left(-1\right)^{p+1}\sum_{x}\langle\prod_{\alpha=0}^{p}R_{x}\left[\omega_{h_{\alpha}}\right]\rangle_{\Omega_{m}}
\end{equation}
si ottiene una espressione analoga:
\begin{equation}
m=n^{-1}\sum_{i}\langle M\left[\omega_{i}\right]\rangle_{\Omega_{m}}+\mathcal{O}\left(\tau^{1-\nu d}\right).\label{eq:3.4.5.5}
\end{equation}
Prima di proseguire è necessaria una precisazione. In tutte le relazioni
precedenti abbiamo mediato le quantità relative alle $\omega_{i}$
sull'insieme $\Omega_{m}$, anche se $\left|\omega_{i}\right|=\tau$:
in questi casi la media è da intendersi come su tutte le possibili
realizzazioni della $i-$esima porzione di $\omega$. E' chiaro come
questa media non consideri solo cammini $\left|\omega_{i}\right|=\tau$
con esattamente $\tau m$ intersezioni, ma un insieme più grande,
in cui i cammini hanno solo in media $\tau m$ intersezioni.

\section{Approssimazione lineare per $\pi\left(m\right)$}

Dalle informazioni raccolte è noto che per i cammini in fase estesa
$\nu\geq1/2$, da cui per $d\geq3$ l'esponente $1-\nu d$ è negativo.
Supponiamo dunque di prendere $\tau=\mathcal{O}\left(n^{\epsilon}\right)$:
scegliendo $\epsilon>0$, gli ordini nelle \prettyref{eq:3.4.5.4},
\prettyref{eq:3.4.5.5} diventano trascurabili nel limite di grandi
$n$. Da un punto di vista geometrico questo equivale a dire che,
per cammini con $d\geq3$ in fase estesa, la maggior parte delle intersezioni
è data da autointersezioni dei singoli elementi $\omega_{i}$, mentre
il contributo alla densità nell'intorno dell'origine viene principalmente
dalla porzione più prossima ad essa, che é appunto $\omega_{0}$.
Detto questo, proviamo ad omettere in prima approssimazione i contributi
di ordine $\mathcal{O}\left(\tau^{1-\nu d}\right)$, scrivendo:
\begin{equation}
\pi\left(m\right)\simeq\langle\pi\left[\omega_{0}\right]\rangle_{\Omega_{m}}\label{eq:REL1}
\end{equation}
\begin{equation}
n\,m\simeq\sum_{i}\langle M\left[\omega_{i}\right]\rangle_{\Omega_{m}}.\label{eq:REL2}
\end{equation}

Ora, consideriamo due cammini $\omega$, $\omega^{+},$appartenenti
rispettivamente agli insiemi $\Omega_{m}$, con $m\leq C_{d}$, e
$\Omega_{m}^{+}$, che rappresenta l'insieme dei cammini con almeno
$m\,n$ intersezioni: 
\begin{equation}
\Omega_{m}^{+}\equiv\left\{ \left|\omega\right|=n,\,M\left[\omega\right]\geq m\,n\right\} .
\end{equation}
Applichiamo a entrambi la decomposizione in porzioni di lunghezza
$\tau$ e immaginiamo di sostituire una delle $n_{\tau}$ porzioni
di $\omega$ con una di $\omega^{+}$. 

I cambiamenti globali dovuti alle differenze di estensione media sono
dello stesso ordine della probabilità di interazione, e in questa
approssimazione vengono trascurati. 

Indichiamo con $\tilde{\omega}_{i\rightarrow j}$ il cammino risultante
dalla sostituzione della porzione $\omega_{i}\in\omega$ con una la
porzione $\omega_{j}^{+}\in\omega^{+}$, e consideriamo la quantità
$\langle\pi\left[\tilde{\omega}_{i\rightarrow j}\right]\rangle_{\Omega_{m},\Omega_{m}^{+}}$,
che rappresenta l'ambiente di $\tilde{\omega}_{i\rightarrow j}$ ,
mediato sulle realizzazioni di $\omega$, $\omega^{+}$.

Dall'approssimazione \prettyref{eq:REL1} segue che la porzione contenente
l'origine è l'unica significativa per calcolo di $\pi\left(m\right)$,
dunque $\langle\pi\left[\tilde{\omega}_{i\rightarrow j}\right]\rangle_{\Omega_{m},\Omega_{m}^{+}}$
sarà diverso da $\langle\pi\left[\omega\right]\rangle_{\Omega_{m}}$
solo se si sceglie $i=0$. La probabilità di sostituire la porzione
$\omega_{0}$ è $1/n_{\tau}$ : dunque, mediando sulle possibili scelte
di $i$, $j$ otterremo la relazione 
\begin{equation}
\langle\pi\left[\tilde{\omega}\right]\rangle_{\Omega_{m},\Omega_{m}^{+}}\simeq\langle\pi\left[\omega\right]\rangle_{\Omega_{m}}+n_{\tau}^{-1}\left(\langle\pi\left[\tilde{\omega}\right]\rangle_{\Omega_{m}^{+}}-\langle\pi\left[\omega\right]\rangle_{\Omega_{m}}\right),
\end{equation}
dove con $\tilde{\omega}$ si intende il cammino ottenuto cambiando
una qualunque porzione di $\omega$ con una di $\omega^{+}$. Notiamo
che, per $m\leq C_{d}$ il valore di $\langle\pi\left[\tilde{\omega}\right]\rangle_{\Omega_{m}^{+}}$
tende a quello di $\langle\pi\left[\omega\right]\rangle_{\Omega_{n}}=C_{d}$
($\Omega_{n}$ insieme dei SRW) nel limite $n\rightarrow\infty$:
ne segue che 
\begin{equation}
\langle\pi\left[\tilde{\omega}\right]\rangle_{\Omega_{m},\Omega_{m}^{+}}\simeq\pi\left(m\right)+n_{\tau}^{-1}\left(C_{d}-\pi\left(m\right)\right).\label{eq:AAAAAAAA}
\end{equation}

Applichiamo un ragionamento simile per $M\left[\tilde{\omega}\right]$:
in questo caso la una sostituzione modificherà solo una frazione $1/n_{\tau}$
dell'intero cammino, qualunque porzione si cambi. Scegliamo di cambiare
la $i-$esima porzione: dalla \prettyref{eq:REL2} 
\begin{equation}
\langle M\left[\tilde{\omega}\right]\rangle_{\Omega_{m},\Omega_{m}^{+}}\simeq\langle M\left[\omega\right]\rangle_{\Omega_{m}}+\left(\langle M\left[\tilde{\omega}_{i}\right]\rangle_{\Omega_{m}^{+}}-\langle M\left[\omega_{i}\right]\rangle_{\Omega_{m}}\right).
\end{equation}
Considerando che, in approssimazione $\tau\rightarrow\infty$, si
ha (sempre da \prettyref{eq:REL2})
\begin{equation}
\tau^{-1}\langle M\left[\omega_{i}\right]\rangle_{\Omega_{m}}\simeq n^{-1}\langle M\left[\omega\right]\rangle_{\Omega_{m}}=m,
\end{equation}
 e che, anche in questo caso, per $m<C_{d}$ il limite di $n^{-1}\langle M\left[\tilde{\omega}\right]\rangle_{\Omega_{m}^{+}}$
sarà $C_{d}$:
\begin{equation}
n^{-1}\langle M\left[\tilde{\omega}\right]\rangle_{\Omega_{m},\Omega_{m}^{+}}\simeq m+n_{\tau}^{-1}\left(C_{d}-m\right).\label{eq:BBBBBBBBBB}
\end{equation}
Ora, indichiamo le variazioni ottenute attraverso la sostituzione
con
\begin{equation}
\delta m=n^{-1}\langle M\left[\tilde{\omega}\right]\rangle_{\Omega_{m},\Omega_{m}^{+}}-m,
\end{equation}
\begin{equation}
\delta\pi\left(m\right)=\langle\pi\left[\tilde{\omega}\right]\rangle_{\Omega_{m},\Omega_{m}^{+}}-\pi\left(m\right):
\end{equation}
combinando con \prettyref{eq:AAAAAAAA} e \prettyref{eq:BBBBBBBBBB},
otterremo l'equazione differenziale di primo grado 
\begin{equation}
\frac{\delta\pi\left(m\right)}{\delta m}\simeq\left(\frac{C_{d}-\pi\left(m\right)}{C_{d}-m}\right),\label{eq:3.4.6}
\end{equation}
la cui soluzione (imponendo la condizione $\left[\partial_{m}\pi\left(m\right)\right]_{C_{d}}=B_{d}$)
è 
\begin{equation}
\pi\left(m\right)\simeq C_{d}\left(1-B_{d}\right)+B_{d}\,m.\label{eq:3.4.6.1}
\end{equation}
La relazione trovata corrisponde all'espansione al primo ordine di
$\pi\left(m\right)$ in $m=C_{d}$. Si noti che i limiti $\langle\pi\left[\tilde{\omega}\right]\rangle_{\Omega_{m}^{+}}\rightarrow C_{d}$
e $n^{-1}\langle M\left[\tilde{\omega}\right]\rangle_{\Omega_{m}^{+}}\rightarrow C_{d}$
sono validi solo per $m\leq C_{d}$, dunque la forma di $\pi\left(m\right)$
trovata è giustificata solo per $m\leq C_{d}$. Per calcolare il ramo
con $m\in\left(C_{d},\,m_{c}\right)$ è sufficiente applicare lo stesso
procedimento, con $\omega^{-}\in\mbox{\ensuremath{\Omega}}_{m}^{-}$
al posto di $\omega_{+}$, dove 
\begin{equation}
\Omega_{m}^{-}\equiv\left\{ \left|\omega\right|=n,\,M\left[\omega\right]\leq m\,n\right\} .
\end{equation}
Con questo insieme i due limiti tornano ad essere validi, e si arriva
alla stessa $\pi\left(m\right)$ in \prettyref{eq:3.4.6.1}. Si ricorda
che questo discorso, basandosi su proprietà esclusive delle catene
estese, è valido solo per $m<m_{c}$, 

E' facile verificare come la \prettyref{eq:3.4.6.1} sia inconsistente
con le informazioni raccolte fino ad ora su $\pi\left(m\right)$:
abbiamo infatti individuato tre condizioni, che corrispondono ai valori
esatti $\pi\left(0\right)=\eta_{d}$, $\pi\left(C_{d}\right)=C_{d}$
e $\left[\partial_{m}\pi\left(m\right)\right]_{C_{d}}=B_{d}$, e il
valore $\pi\left(0\right)=C_{d}\left(1-B_{d}\right)$ predetto dalla
\prettyref{eq:3.4.6.1} è certamente diverso da $\eta_{d}$. 

Una spiegazione può essere trovata nel fatto di aver costruito i cammini
$\omega$ combinando pezzi di lunghezza divergente nel limite $n\rightarrow\infty$:
per avere cammini con misura di probabilità non nulla è infatti necessario
usare delle $\tau$ dell'ordine del tempo di intersezione, che però
è una costante di ordine $\tau_{c}\sim\mathcal{O}\left(C_{d}^{-1}\right)<\infty$
. Dato che la suddivisione in porzioni finite comporta immancabilmente
un errore di ordine costante nel trascurare i termini $\mathcal{O}\left(\tau^{1-\nu d}\right)$,
è chiaro come la \prettyref{eq:3.4.6.1} non possa essere esatta. 

Tuttavia, per $d\geq4$ le simulazioni presentate nelle Figure 2.10,
211, 2.12, 2.13 indicano che la \prettyref{eq:3.4.6.1} rappresenta
una buona (se non ottima) approssimazione per $\pi\left(m\right)$.

In tabella \prettyref{Flo:tab3} sono riportate le discrepanze tra
la $\eta_{p}^{\left(l\right)}=C_{d}\left(1-B_{d}\right)$, calcolato
attraverso la \prettyref{eq:3.4.6.1}, e i valori reali $\eta_{d}$
ottenuti da simulazioni (si veda la sezione sui SAW presente in appendice):
mentre il caso $d=3$ presenta una notevole sottostima (oltre il $20\%$),
già per $d=4$ si ha un errore assoluto sotto il $2\%$, mentre i
casi $d\geq5$ danno una sovrastima al di sotto del punto percentuale.

Ad ogni modo, mentre per $d=3$ si osserva una discrepanza notevole
(evidente anche in Figura 2.9), dovuta al fatto che $d=3$ è solo
debolmente transiente (il tempo di intersezione è molto piccolo) già
per $d\geq4$ si ha una approssimazione piuttosto fedele dei risultati
esatti: in effetti $C_{d}$ decresce in $d$, e dunque anche l'errore
commesso nel trascurare le interazioni a lungo raggio sarà $\mathcal{O}\left(C_{d}^{\nu d-1}\right)$
decrescente in $d$. 

\begin{table}
\begin{centering}
\caption{}
\par\end{centering}
~~~
\begin{centering}
\begin{tabular}{|c|c||c|c|c|}
\hline 
\noalign{\vskip\doublerulesep}
$d$ & $B_{d}$ & $\eta_{p}^{\left(l\right)}$ & $\eta_{d}$ & $\eta_{p}^{\left(l\right)}/\eta_{d}-1$\tabularnewline[\doublerulesep]
\hline 
\hline 
$3$ & $1/2$ & $0.17026\left(9\right)$ & $0.2193\left(5\right)$ & $-22.38\%$\tabularnewline
\hline 
$4$ & $0.22080\left(9\right)$ & $0.15054\left(1\right)$ & $0.1532445\left(6\right)$ & $-1.76\%$\tabularnewline
\hline 
$5$ & $0.13767\left(2\right)$ & $0.11656\left(8\right)$ & $0.1161456\left(3\right)$ & $0.36\%$\tabularnewline
\hline 
$6$ & $0.10266\left(0\right)$ & $0.09396\left(5\right)$ & $0.0934921\left(3\right)$ & $0.51\%$\tabularnewline
\hline 
$7$ & $0.08291\left(2\right)$ & $0.078727\left(3\right)$ & $0.07837021\left(4\right)$ & $0.46\%$\tabularnewline
\hline 
$8$ & $0.07030\left(6\right)$ & $0.067786\left(5\right)$ & $0.0675464\left(2\right)$ & $0.36\%$\tabularnewline
\hline 
\end{tabular}\label{Flo:tab3}
\par\end{centering}
\end{table}

\section{Caso collassato $m>m_{c}$ e distribuzione $P_{n}\left(m\right)$}

Per il caso collassato non è possibile aggiungere molto rispetto a
quanto detto nel secondo capitolo. Per $m>m_{c}$ non è infatti possibile
ripetere il discorso precedente, visto che $1-\nu d=0$ e le fluttuazioni
non sono trascurabili in nessun caso. 

Ad ogni modo, i dati in nostro possesso ci permettono comunque di
arrivare ad alcune conclusioni circa il comportamento limite di $P_{n}\left(m\right)$. 

Con una argomentazione euristica eravamo arrivati a definire che $\pi\left(1\right)=1$,
mentre dai dati si era ipotizzato che $\pi\left(m\right)<1$ per $m\in\left(m_{c},\,1\right)$. 

Queste due informazioni ci permettono di applicare direttamente i
risultati ottenuti alla fine del terzo capitolo: identificando $\pi\left(m\leq m_{c}\right)$
con $\tilde{\pi}\left(m\right)$, e $\pi\left(m>m_{c}\right)$ con
$\mu\left(m\right)$, si ottiene la $\pi\left(m\right)$ descritta
in \prettyref{eq:BARBABLU}, per cui il limite 
\begin{equation}
\lim_{n\rightarrow\infty}n^{-1}\log\left(P_{n}\left(m\right)\right)=\varphi\left(m\right),
\end{equation}
esiste, ed è dato dalla funzione \prettyref{eq:forma1}: 
\begin{equation}
\varphi\left(m\right)=\left\{ \begin{array}{l}
\tilde{\varphi}\left(m\right)\\
0
\end{array}\,\,\,\begin{array}{r}
m\in\left[0,\,C_{d}\right]\\
m\in\left(C_{d},\,1\right]
\end{array}\right.,\label{eq:MIG-21}
\end{equation}
dove $\tilde{\varphi}\left(m\right)$ è soluzione della \prettyref{eq:EQUAZIONEFONDAMENTALE}
per $\tilde{\pi}\left(m\right)=\pi\left(m<m_{c}\right)$. 

Un ultima congettura circa le proprietà di $\pi\left(m\right)$ nel
caso collassato può comunque essere introdotta ipotizzando l'analogia
con la WS. 

Sotto questa condizione, la distribuzione $P_{n}\left(m\right)$ per
$m>C_{d}$ dovrebbe godere delle seguenti proprietà (si veda la sezione
in appendice sulla WS, dove si riassumono i risultati di \cite{Den=000020Hollander}):

\begin{equation}
\lim_{n\rightarrow\infty}n^{-1+\frac{2}{d}}\log\left(P_{n}\left(m\right)\right)=\phi\left(m\right),\label{eq:BB1}
\end{equation}

\begin{equation}
\lim_{m\rightarrow1}\left(1-m\right)^{\frac{2}{d}}\phi\left(m\right)=\lambda\in\left(0,\,\infty\right),\label{eq:BB2}
\end{equation}

\begin{equation}
\partial_{m}\phi\left(m\right)<0,\,m\in\left(C_{d},\,1\right),\label{eq:BB3}
\end{equation}

\begin{equation}
\lim_{m\rightarrow m_{c}^{+}}\partial_{m}\phi\left(m\right)\neq\lim_{m\rightarrow m_{c}^{-}}\partial_{m}\phi\left(m\right).\label{eq:BB4}
\end{equation}

Stanti le congetture fatte nell'ultima sezione del secondo capitolo,
queste caratteristiche sono ottenibili imponendo che per $m\in\left(m_{c},\,1\right)$
valga la relazione $\pi\left(m\right)<m$. Combinando la \prettyref{eq:baOBAB}
con gli effetti di bordo $\bar{\psi}_{n}\left(m\right)$ dati dall'equazione
\prettyref{eq:FROV}, possiamo scrivere direttamente la \prettyref{eq:prevision}
come 
\begin{equation}
\lim_{n\rightarrow\infty}\log\left(P_{n}\left(m\right)\right)\propto-\left(1-m\right)^{-\frac{2}{d}}n^{1-\frac{2}{d}}.\label{eq:CONGETTURONA}
\end{equation}
Questa funzione verifica le caratteristiche \prettyref{eq:BB1}, \prettyref{eq:BB2}.
La \prettyref{eq:BB3} segue invece dalla congettura per cui se $\pi_{n}\left(m_{i}\right)=m_{i}$,
$\dot{\pi}_{n}\left(m_{i}\right)<1$, allora $m_{i}$ è un massimo
locale di $\log\left(P_{n}\left(m\right)\right)$: imponendo $\pi\left(m\in\left(m_{c},\,1\right)\right)<m$
non possono infatti esserci intersezioni di $\pi_{n}\left(m\right)$
con la retta $m$ nel range considerato. 

In fine, studi numerici della \prettyref{eq:mastereq1} (di cui è
riportato un esempio in Figura \prettyref{Flo:figura3.4}) confermano
la discontinuità della derivata di $\phi\left(m\right)$ in $m_{c}$,
se questa è presente anche in $\pi\left(m\right)$.

\pagebreak{}

\chapter{Modelli termodinamici}

Mostreremo ora come i risultati fin qui ottenuti possano essere impiegati
per giustificare alcune proprietà matematiche dei modelli di polimeri
interagenti: in particolare, ci occuperemo di un modello introdotto
da Stanley et al. alcuni decenni fa (\cite{Stanley}). Benché il modello
di Stanley (SM) risulti piuttosto astratto, molte sue proprietà sono
(almeno indirettamente) di notevole interesse nell'attuale panorama
matematico. Inoltre, ci consentirà di formulare alcune congetture
sul parametro d'ordine \prettyref{eq:VARRO} in prossimità del punto
di transizione. 

La nostra trattazione dello SM consiste essenzialmente nella riscrittura
del problema \prettyref{eq:EQUAZIONEFONDAMENTALE} in termini di funzioni
generatrici: come vedremo, questo ci consentirà di costruire una equazione
risolvente più abbordabile per il processo stocastico \prettyref{eq:3.1.1-4},
\prettyref{eq:3.1.1-5}, e di risolverne esattamente il caso lineare.

\section{Catene interagenti}

I modelli di polimeri lineari che, come il SAW, includono interazioni
fra monomeri mostrano una forte autocorrelazione, e sono dunque estremamente
difficili da analizzare. Queste situazioni vengono ottenute, in termini
termodinamici, associando un energia alle possibili configurazioni
di un SRW. Indicando con $\omega$ una generica configurazione appartenente
all'insieme $\Omega_{n}$ dei SRW di $n-$passi, possiamo associargli
una energia $E\left[\omega\right]$. Usando l'usuale ensemble canonico,
possiamo definire una probabilità di ensemble dipendente dalla temperatura
come 
\begin{equation}
\mbox{P}_{\beta}\left[\omega\in\Omega_{n}\right]\equiv Z_{n}\left(\beta\right)^{-1}\exp\left(-\beta E\left[\omega\right]\right),\label{eq:3.1.1}
\end{equation}
dove $\beta=1/kT$, $k$ è la costante di Boltzmann e $T$ la temperatura
assoluta. La costante di normalizzazione è ovviamente la \emph{funzione
di partizione canonica} del sistema, 
\begin{equation}
Z_{n}\left(\beta\right)\equiv\sum_{\omega\in\Omega_{n}}\exp\left(-\beta E\left[\omega\right]\right),\label{eq:3.1.2}
\end{equation}
dalla quale possiamo calcolare le funzioni termodinamiche. Si noti
che nel limite di alte temperature $\beta\rightarrow0$ otteniamo
$Z_{n}\left(\beta\right)=\left(2d\right)^{n}$, $\mbox{P}_{0}\left[\omega\right]=\left(2d\right)^{-n}$
corrispondente al SRW, in cui tutte le configurazioni hanno eguale
peso statistico. 

Tuttavia, per temperature finite le probabilità di ensemble legate
alle configurazioni differiscono dalla distribuzione piatta. Se introduciamo
l'interazione fra monomeri attraverso le auto-intersezioni, scegliendo
ad esempio \foreignlanguage{english}{$E\left[\omega\right]$} crescente
con $M$, otterremo un peso statistico maggiore per i cammini con
molte intersezioni, mentre per $E\left[\omega\right]$ decrescente
in $M$ sarà maggiore la probabilità di avere meno intersezioni della
media di un SRW. Fisicamente questo corrisponde a scoraggiare (o incoraggiare)
termodinamicamente le configurazioni a seconda del loro numero di
auto-intersezioni. 

Si noti che il sistema considerato è, per postulato, in \emph{equilibrio
termodinamico}, dunque l'energia $E\left[\omega\right]$ deve essere
sempre calcolata sulle configurazioni di lunghezza $n$: in nessun
modo le probabilità di transizione usate per generare il cammino devono
venire modificate durante la crescita dello stesso. 

\section{Modello di Stanley}

Un modello canonico particolarmente semplice è il modello di Stanley
(\cite{Stanley}), in cui le intersezioni sono scoraggiate o incoraggiate
attraverso una hamiltoniana proporzionale ad $M$, con un prefattore
$\epsilon$ che può essere sia positivo (intersezioni scoraggiate)
che negativo (intersezioni incoraggiate).

Detta $E\left[\omega\right]$ l'hamiltoniana del sistema per una determinata
configurazione $\omega$, questa è definita come 
\begin{equation}
E\left[\omega\right]=\epsilon M\left[\omega\right]=\epsilon\lim_{\alpha\rightarrow0^{+}}\sum_{i=0}^{n-1}\left(g\left(S_{i}\right)-1\right)^{\alpha},\label{eq:3.2.1}
\end{equation}
dove $g\left(S_{i}\right)$ corrisponde al numero di occupazione del
sito $S_{i}$: funzione di partizione e probabilità dell'ensemble
seguono dalle definizioni \prettyref{eq:3.1.1}, \prettyref{eq:3.1.2}. 

Poniamo per semplicità $\left|\epsilon\right|=1$: possiamo trattare
entrambi i casi con un unico parametro $q=e^{\pm\beta}$, ottenendo
il modello di Stanley repulsivo per $q<1$, il SRW per $q=1$, e il
caso attrattivo per $q>1$. In particolare $q=0$ corrisponde al SAW.

Vogliamo esprimere lo spettro energetico del modello in funzione di
$q$. Da semplici manipolazioni è possibile riscrivere la funzione
di partizione in termini della funzione $\mathrm{N}_{n}\left(M\right)=\left(2d\right)^{n}P_{n}\left(M\right)$,
rappresentante il numero di cammini di lunghezza $n$ con esattamente
$M$ intersezioni:

\begin{equation}
Z_{n}\left(q\right)=\sum_{M=0}^{n-1}\mathrm{N}_{n}\left(M\right)\,q^{M}=\left(2d\right)^{n}\sum_{M=0}^{n-1}P_{n}\left(M\right)\,q^{M}.\label{eq:3.2.3}
\end{equation}
Dalla definizione di distribuzione dell'ensemble ricaviamo direttamente
lo spettro energetico (normalizzato all'unità) 
\begin{equation}
P_{n}^{\left(q\right)}\left(M\right)=\left(2d\right)^{n}Z_{n}\left(q\right)^{-1}\,P_{n}\left(M\right)\,q^{M}.\label{eq:3.2.4}
\end{equation}
Come possiamo notare, la distribuzione libera $P_{n}\left(M\right)$
contiene tutte le informazioni necessarie per descrivere il modello
dal punto di vista energetico.

Nel caso $\epsilon<0$, si ritiene (\cite{Huges}) che il modello
di Stanley abbia comportamenti del tutto simili al modello di Domb-Joyce
(\cite{Domb-Joyce}), corrispondente al caso $\alpha=2$ della \prettyref{eq:3.2.1}.
La funzione di partizione per il modello di Domb-Joyce è: 
\begin{equation}
Z_{n}\left(\beta\right)=\sum_{\omega\in\Omega_{n}}\prod_{i=0}^{n-1}\prod_{j=i+1}^{n}\left[1+\left(e^{-\beta}-1\right)\delta_{S_{i},S_{j}}\right],\label{eq:3.2.2}
\end{equation}
ed è facile dimostrare l'equivalenza con quella derivante dalle \prettyref{eq:3.1.2},
\prettyref{eq:3.2.1}, con $\alpha=2$ (a meno di un fattore dipendente
solo da $n$). 

Vediamo ora i risultati noti nel caso $q\in\left[0,\,1\right)$, $d\geq3$:
come per il SAW, la maggior parte dei risultati rigorosi si ha per
$d\geq5$. Attraverso la \emph{Lace-Expansion} (\cite{Bridges-Spencer})
è infatti stato possibile dimostrare che, per $d\geq5$, la distribuzione
$P_{n}^{\left(q\right)}\left(\left|S_{0}-S_{n}\right|\right)$ converge
ad una gaussiana per $n\rightarrow\infty$, e il raggio quadratico
medio 
\begin{equation}
\Re_{n}^{2}\left(q\right)=\langle\,\,\left|S_{n}-S_{0}\right|^{2}\rangle_{q}
\end{equation}
ha $2\nu_{d}\left(q\right)=1$ come esponente caratteristico di scaling
(con $\langle\cdot\rangle_{q}$ si indica la media di ensemble).

Oltre la dimensione critica $d_{c}=4$ il modello scala dunque come
il SRW, con un diverso valore della costante di diffusione. Quest'ultimo
risultato è valido anche nel caso $q=0$, $d\geq5$, dove è in effetti
dimostrato (sempre con la \emph{Lace-Expansion}) che $2\nu_{d}=1$.
Si noti che i risultati rigorosi riportati sono in accordo con la
formula di Fisher-Flory, che prevede appunto $2\nu_{d}\left(q\right)=1$
oltre la dimensione critica.

Per $d=3$, calcoli non rigorosi di teoria dei campi (oltre che da
argomentazioni tipo Fisher-Flory) suggeriscono un comportamento analogo
al caso del SAW per $q\in\left(0,\,1\right)$, con esponente $\nu_{3}\left(q\right)=\nu_{3}$.
Queste congetture sono supportate da precise simulazioni numeriche
e sono generalmente accettate come esatte. Anche il caso $d=4$ mostra
uno scaling identico a quello del SAW, per cui è previsto $\nu_{4}=1/2$
con una debole correzione logaritmica (\cite{Bridges-Slade,Bauerschmidt}). 

Riassumendo, per $q<1$ la distanza quadratica media $\Re_{n}^{2}\left(q\right)$
scala come :
\begin{equation}
\Re_{n}^{2}\left(q\right)\sim\left\{ \begin{array}{l}
n^{2\nu_{d}}\\
n\,\log\left(n\right)^{\frac{1}{4}}
\end{array}\,\,\,\begin{array}{r}
d\neq4\\
d=4
\end{array}\right.
\end{equation}
\\
dove $\nu_{d}$ sono gli esponenti critici di correlazione del Self-Avoiding
Walk (si veda in appendice la sezione relativa).

Non sono noti risultati rigorosi sull'andamento della distanza quadratica
media nel caso $q>1$, a parte una argomentazione euristica in \cite{Huges}
basata sulla \prettyref{eq:EQUVALOR}: la stima fornita per $d\geq3$
è
\begin{equation}
\Re_{n}^{2}\left(q\right)\sim n^{\frac{2}{d+2}}.
\end{equation}

L'andamento del range nel modello di Stanley per $q>1$, $d\neq2$,
è stato invcece dimostrato in maniera rigorosa da Donsker e Varadhan
(\cite{Donsker-Varadhan}), nell'ambito della \emph{teoria delle grandi
deviazioni}. La complessità della dimostrazione non consente una trattazione
completa dei risultati (che in gran parte esulano da questo lavoro),
tuttavia possiamo riassumere la parte relativa al modello di Stanley
nella relazione seguente:

\begin{equation}
\lim_{n\rightarrow\infty}\left[n^{-\delta}\langle R\,\rangle_{q}\right]\in\left(0,\infty\right),\,\forall q>1.\label{eq:4.3.1}
\end{equation}
\\
In particolare il teorema di Donsker-Varadhan fornisce anche gli esponenti
$\delta$ per tutte le dimensioni del reticolo:
\begin{equation}
\delta=\left\{ \begin{array}{l}
\frac{2}{3}\\
\frac{d}{d+2}
\end{array}\,\,\,\begin{array}{r}
d=2\\
d\neq2
\end{array}\right..\label{eq:4.3.2}
\end{equation}
Questo risultato (in qualche modo sorprendente, dato che un valore
di $q$ molto vicino all'unità comporta comunque un cambiamento qualitativo
rispetto al SRW) segue immediatamente anche da una approssimazione
\emph{saddle point} su $P_{n}^{\left(q\right)}\left(R\right)$. Infatti,
dalle relazioni \prettyref{eq:FROV}, \prettyref{eq:prevision}, \prettyref{eq:3.2.4},
la distribuzione del range $P_{n}^{\left(q\right)}\left(R\right)$
per $n^{-1}R\sim1$ è data dall'espressione
\begin{equation}
\log\left(P_{n}^{\left(q\right)}\left(R\right)\right)\sim-\lambda R^{-\frac{2}{d}}n-\log\left(q\right)R,
\end{equation}
il cui massimo in $R$ è proprio per $R\propto n^{d/\left(d+2\right)}$. 

Per quanto riguarda il range $\langle R\,\rangle_{q}$ con $q\in\left[0,\,1\right)$,
studi sulla funzione generatrice della WS (\cite{Van=000020den=000020Berg-Toth})
e alcune congetture derivanti dal caso $d=1$ (\cite{Redner-Kang,Chan-Huges}),
indicano $\langle R\,\rangle_{q}\sim n$. Dunque il numero di autointersezioni
nel modello di Stanley per $d\geq3$ dovrebbe scalare come per il
SRW, con l'ovvia esclusione per il caso $q=0$, in cui le intersezioni
sono assenti per definizione. Vedremo a breve come è possibile derivare
questo risultato attraverso le informazioni raccolte in questo lavoro.

\section{Energia libera del modello di Stanley repulsivo}

Si consideri la funzione di partizione \prettyref{eq:3.2.3}, con
$q=\exp\left(-\beta\right)$: come operazione preliminare conviene
eliminare il fattore moltiplicativo $\left(2d\right)^{n}$, che determina
semplicemente un termine additivo, indipendente da $M$, nell'energia
libera. Ridefiniamo la scala delle energie in modo da porre lo zero
su $-n\log\left(2d\right)$: la nuova funzione di partizione sarà
\begin{equation}
Z_{n}\left(\beta\right)=\sum_{M=0}^{n-1}P_{n}\left(M\right)e^{-\beta M}.
\end{equation}
Definiamo la funzione di partizione ridotta $z\left(\beta\right)$,
e la $u\left(\beta\right)$, energia libera per monomero nel limite
$n\rightarrow\infty$: dalle definizioni si trova che
\begin{equation}
z\left(\beta\right)\equiv\lim_{n\rightarrow\infty}Z_{n}\left(\beta\right)^{\frac{1}{n}}.\label{eq:PARTRID}
\end{equation}
\begin{equation}
u\left(\beta\right)\equiv-\lim_{n\rightarrow\infty}n^{-1}\log\left(Z_{n}\left(\beta\right)\right).\label{eq:GRAK}
\end{equation}
Si ricorda che la $u\left(\beta\right)$ è definita a meno del termine
additivo $-\log\left(2d\right)$. Introduciamo inoltre le funzioni
termodinamiche
\begin{equation}
\langle m\rangle_{\beta}=-\lim_{n\rightarrow\infty}n^{-1}\partial_{\beta}\log\left(Z_{n}\left(\beta\right)\right),\label{eq:DEFFF1}
\end{equation}
\begin{equation}
c\left(\beta\right)=-\lim_{n\rightarrow\infty}n^{-1}\beta^{2}\partial_{\beta}^{2}\log\left(Z_{n}\left(\beta\right)\right),\label{eq:DEFFF2}
\end{equation}
corrispondenti rispettivamente a numero medio di intersezioni e calore
specifico per monomero ($\langle\cdot\rangle_{\beta}$ indica la media
termodinamica).

Abbiamo visto in precedenza come, nel limite di $n$ grandi, $P_{n}\left(M\right)$
mostri il seguente andamento asintotico: 
\begin{equation}
n^{-1}\log\left(P_{n}\left(M\right)\right)\rightarrow\varphi\left(n^{-1}M\right),
\end{equation}
La funzione $\varphi\left(m\right)$ è stata ottenuta, a partire da
$\pi\left(m\right)$, come soluzione continua in $m\in\left[0,\,1\right]$
dell'equazione differenziale \prettyref{eq:EQUAZIONEFONDAMENTALE}.
Per semplificare l'esposizione, nei calcoli che svolgeremo di seguito
assumeremo che, nell'intervallo chiuso $m\in\left[0,\,C_{d}\right]$,
$\varphi\left(m\right)$ sia una funzione invertibile di classe $C^{\infty}$.

Sostituiamo ora $\varphi\left(n^{-1}M\right)$ in \prettyref{eq:GRAK},
esprimendo $u\left(\beta\right)$ attraverso il limite 
\begin{equation}
u\left(\beta\right)=-\lim_{n\rightarrow\infty}n^{-1}\log\left(\sum_{M=0}^{t-1}e^{n\varphi\left(n^{-1}M\right)-\beta M}\right).
\end{equation}
Introducendo $m=n^{-1}M$, è possibile passare al limite continuo:
cambiamo dunque la somma in $M$ con un integrale in $m$, ottenendo
\begin{equation}
u\left(\beta\right)=-\lim_{n\rightarrow\infty}n^{-1}\log\left(\int_{0}^{1}e^{n\left(\varphi\left(m\right)-\beta m\right)}dm\right).
\end{equation}
Il limite dell'equazione precedente può essere risolto esattamente
con una integrazione \emph{saddle point}. Introduciamo la quantità
$m_{\beta}$, definita come 
\begin{equation}
m_{\beta}\,|\,\dot{\varphi}\left(m_{\beta}\right)=\beta:\label{eq:TERMALEQU}
\end{equation}
l'energia libera può essere espressa in termini di $m_{\beta}$ attraverso
la relazione 
\begin{equation}
u\left(\beta\right)=\beta m_{\beta}-\varphi\left(m_{\beta}\right).\label{eq:EQUFROMHELL}
\end{equation}
Combinando \prettyref{eq:DEFFF1} con \prettyref{eq:EQUFROMHELL}
si trova che $\langle m\rangle_{\beta}=m_{\beta}$, mentre il calore
specifico è ovviamente
\begin{equation}
c\left(\beta\right)=-\beta^{2}\partial_{\beta}u\left(\beta\right).
\end{equation}

Dalle relazioni precedenti risulta intuitivamente chiaro come la media
termodinamica selezioni, per $n$ grandi, cammini appartenenti in
prevalenza all'insieme $\Omega_{m_{\beta}}$. Da questo segue immediatamente
che tutte le quantità del modello di Stanley per una determinata $\beta$
devono scalare (nel limite di $n$ grande) come quelle dell'ensemble
$\Omega_{m}$, con $m=m_{\beta}$. 

E' ora esplicita l'analogia, anticipata nel secondo capitolo, tra
gli andamenti di $\Re_{n}^{2}\left(\beta\right)=\langle\,\,\left|S_{n}-S_{0}\right|^{2}\rangle_{\beta}$
del modello di Stanley e di $\Re_{n}^{2}\left(m\right)$ per le catene
con $m<C_{d}$: le due funzioni, posto $m_{\beta}=m$, dovrebbero
essere uguali nel limite di grandi $n$. 

Quest'ultimo fatto ci è utile nei casi $d=3,\,4$. Sempre nel secondo
capitolo avevamo infatti introdotto il parametro $\varrho_{d}\left(m\right)$,
definito dal limite \prettyref{eq:VARRO}: detto $\delta m_{\beta}=C_{3}-m_{\beta}$,
e stante l'uguaglianza 
\begin{equation}
\Re_{n}^{2}\left(\beta\right)\simeq\Re_{n}^{2}\left(m_{\beta}\right),
\end{equation}
è possibile determinare la natura della transizione di $\varrho_{3}\left(\delta m\right)$
in $\delta m=0$ a partire dalla espressione di $\Re_{n}^{2}\left(\beta\right)$
per piccoli valori di $\beta$ (chiaramente questo non è possibile
per $d\geq5$, in quanto il punto di transizione è oltre il range
di $m_{\beta}$).

Una stima approssimativa di $\Re_{n}^{2}\left(\beta\right)$ per $\beta$
piccoli si può ottenere dalla teoria di Fisher-Flory. Si parte dall'energia
di Flory (\cite{Flory}) per una catena debolmente interagente:
\begin{equation}
\mathcal{F}_{\beta}\left(r\right)\sim\beta n^{2}r^{-d}+\frac{r^{2}}{2n}-\left(d-1\right)\log\left(r\right)+\Lambda\left(n\right),
\end{equation}
dove $r=\left|S_{n}-S_{0}\right|$ è la distanza end-to-end della
catena. Vogliamo minimizzare l'energia sotto l'ipotesi $r\sim\beta^{\theta}n^{\nu}$:
derivando in $r$, operando la sostiuzione $r\rightarrow\beta^{\theta}n^{\nu}$
ed annullando l'espressione si ottiene la relazione 
\begin{equation}
d\beta^{1-\theta\left(1+d\right)}n^{2-\nu\left(1+d\right)}+\left(d-1\right)\beta^{-\theta}n^{-\nu}=\beta^{\theta}n^{\nu-1}.
\end{equation}
Se $\nu d<2$, per grandi $n$ il termine $\mathcal{O}\left(n^{-\nu}\right)$
può essere trascurato rispetto a quello $\mathcal{O}\left(n^{2-\nu\left(1+d\right)}\right)$:
eguagliando gi esponenti si avrà $\nu=3/\left(d+2\right)$, $\theta=1/\left(d+2\right)$.
Se invece $\nu d\geq2$ si troverà $\nu=1/2$, $\theta=0$. Dunque
per $d=3$ la teoria di Flory prevede $\theta=1/5$.

Una stima, matematicamente più fondata, ma di precisione paragonabile,
è dedotta in \cite{Kleinert} utilizzando gli integrali di cammino:
l'espressione fornita per il raggio quadratico medio è identica a
quella ottenibile con il metodo di Flory, e vale 
\begin{equation}
\Re_{n}^{2}\left(\beta\right)\sim C_{0}\beta^{\frac{2}{d+2}}n^{\frac{6}{d+2}}.
\end{equation}
Questa relazione, certamente errata in quanto $\nu_{3}\neq3/5$, ha
il pregio di mostrare l'andamento in $\beta$, consentendoci di stimare
$\varrho_{3}\left(m_{\beta}\right)\propto\beta^{2/5}$, per piccoli
$\beta$. 

Abbiamo visto nel capitolo terzo che se $\pi\left(m_{0}\right)=m_{0}$,
$\dot{\pi}\left(m_{0}\right)=1/2$ (come accade appunto per $d=3$)
non è possibile definire $\varphi\left(m\right)$ in $m_{0}$: il
limite di $\log\left(P_{t}\left(m_{t}\right)\right)$ per $t\rightarrow\infty$,
$m_{t}\rightarrow m_{0}+\delta m$ è comunque compreso fra 
\begin{equation}
\lim_{t\rightarrow\infty}\log\left(P_{t}\left(m_{t}\right)\right)\in\left[c_{1}\left|\delta m\right|^{2},\,c_{2}\left|\delta m\right|^{2+\epsilon}\right],
\end{equation}
 con $\epsilon\rightarrow0^{+}$: segue dunque che l'andamento di
$\delta m_{\beta}$ per piccoli $\beta$ deve essere compreso tra
$\mathcal{O}\left(\beta^{1-\epsilon}\right)$ e $\mathcal{O}\left(\beta\right)$,
con $\epsilon\rightarrow0^{+}$. Questo ci dà che 
\begin{equation}
\varrho_{3}\left(\delta m\right)\in\left[\kappa_{1}\,\delta m^{\frac{2}{5}},\,\kappa_{2}\,\delta m^{\frac{2}{5}\left(1-\epsilon\right)}\right],
\end{equation}
con $\epsilon\rightarrow0^{+}$. Dunque, in $d=3$ l'analogia con
il modello di Stanley prevede una transizione continua in $C_{d}$:
l'esponente $\theta=1/5$ è quasi certamente errato, ma l'andamento
a potenza $\varrho_{3}\left(m_{\beta}\right)\propto\delta m^{2\theta}$,
con $\theta<1/2$, (e una qualche correzione di ordine logaritmico)
è del tutto plausibile. 

Per il caso $d=4$, la teoria di Flory non è applicabile (non può
rilevare la correzione logaritmica), e non è stato possibile reperire
altre stime di $\Re_{n}^{2}\left(\beta\right)$, dunque non vi sono
dati sufficienti per discuterne. 

Si noti che la teoria di Flory prevede, correttamente, $\theta=0$
per $d\geq5$: in questa situazione infatti $m=C_{d}$ è un punto
analitico di $\varrho_{d}\left(m\right)$, con $\varrho_{d}\left(C_{d}\right)>0$
(la transizione è prevista per $m_{c}>C_{d}$). 

Per quanto riguarda il range, dalla definizione di $\langle m\rangle_{\beta}=m_{\beta}$
segue che, se $m_{\beta}\in\left(0,\,1\right)$, il range del modello
di Stanley deve scalare come $\langle R\,\rangle_{\beta}\sim\left(1-m_{\beta}\right)n$
nel limite di $n$ grande.

Vedremo ora come è possibile determinare $u\left(\beta\right)$ unicamente
a partire dalla funzione ambiente $\pi\left(m\right)$. Sappiamo che
la funzione $\varphi\left(m\right)$ soddisfa l'equazione \prettyref{eq:EQUAZIONEFONDAMENTALE},
dunque si deve avere 
\begin{equation}
e^{\varphi\left(m_{\beta}\right)-m_{\beta}\,\dot{\varphi}\left(m_{\beta}\right)}=\pi\left(m_{\beta}\right)e^{-\dot{\varphi}\left(m_{\beta}\right)}+1-\pi\left(m_{\beta}\right).
\end{equation}
Sostituendo nell'equazione precedente le relazioni \prettyref{eq:TERMALEQU},
\prettyref{eq:EQUFROMHELL}, e ricordando che $m_{\beta}=\partial_{\beta}u\left(\beta\right)$,
otterremo la seguente equazione differenziale implicita in $u\left(\beta\right)$:
\begin{equation}
\pi\left(\partial_{\beta}u\left(\beta\right)\right)=\frac{1-e^{-u\left(\beta\right)}}{1-e^{-\beta}}.\label{eq:IMPLICITA}
\end{equation}
In conformità con le informazioni raccolte, si considera $\pi\left(m\right)$
invertibile; segue dunque che la precedente equazione può essere formalmente
riscritta come
\begin{equation}
\partial_{\beta}u\left(\beta\right)=\pi^{-1}\left(\frac{1-e^{-u\left(\beta\right)}}{1-e^{-\beta}}\right)
\end{equation}
dove $\pi^{-1}\left(\cdot\right)$ rappresenta la funzione inversa
di $\pi\left(m\right)$. 

L'equazione trovata è completamente equivalente alla \prettyref{eq:EQUAZIONEFONDAMENTALE}:
è infatti possibile determinare in forma parametrica $\varphi\left(m\right)$,
esprimendo $\varphi\left(m_{\beta}\right)=-u\left(\beta\right)+\beta m_{\beta}$
in funzione di $m_{\beta}$ al variare del parametro $\beta$.

La \prettyref{eq:IMPLICITA} sarà difficilmente solubile per una generica
$\pi\left(m\right)$, ma certamente più semplice di \prettyref{eq:EQUAZIONEFONDAMENTALE}
(almeno per quanto riguarda l'esplicitazione di $\dot{\varphi}\left(m\right)$).
Per esplicitare la \prettyref{eq:EQUAZIONEFONDAMENTALE} nella derivata
è infatti necessaria la soluzione generale $y\left(\kappa_{0},\kappa_{1},\kappa_{2}\right)$
di $y^{\kappa_{0}}+\kappa_{1}y+\kappa_{2}=0$, per cui certamente
non esiste alcuna espressione in forma chiusa. L'esplicitabilità di
\prettyref{eq:IMPLICITA} dipende invece dall'esistenza dell'inversa
di $\pi\left(m\right)$, non escludibile a priori.

\section{Approssimazione per $d\protect\geq5$}

Nel capitolo quarto abbiamo derivato una approssimazione lineare per
$\pi\left(m\right)$, piuttosto accurata quando $d\geq5$: vedremo
ora come, in questo caso, sia possibile risolvere esattamente la \prettyref{eq:IMPLICITA},
trovando l'espressione esatta di $u\left(\beta\right)$ in termini
di funzioni ipergeometriche. Partiamo dunque dalla forma lineare di
$\pi\left(m\right)$ nei parametri generici $a$, $b$, con $a>0$,
$a+b<1$:
\begin{equation}
\pi\left(m\right)=a+bm.
\end{equation}
Chiaramente, per ottenere il modello di Stanley vanno imposti i parametri
$a=C_{d}\left(1-B_{d}\right)$, $b=B_{d}$, che contengono la fisica
del sistema (si vedano le tabelle nei capitoli primo e quarto).

Sostituendo nella \prettyref{eq:IMPLICITA}, si ottiene la seguente
equazione differenziale in $\beta$, esplicitabile nella derivata:
\begin{equation}
a+b\,\partial_{\beta}u\left(\beta\right)=\frac{1-e^{-u\left(\beta\right)}}{1-e^{-\beta}}.\label{eq:LINER}
\end{equation}
Operiamo ora le seguenti sostituzioni:
\begin{equation}
\xi=\left(1-e^{-\beta}\right),
\end{equation}
\begin{equation}
z\left(\xi\right)=e^{-u\left(\beta\left(\xi\right)\right)},
\end{equation}
dove $\xi\in\left[0,\,1\right]$ equivale a $1-q$ del modello di
Stanley con $q<1$, e $z\left(\xi\right)$ è la funzione di partizione
ridotta \prettyref{eq:PARTRID} espressa in $\xi$. Dopo alcune semplificazioni,
la \prettyref{eq:LINER} nelle nuove variabili assume la forma 
\begin{equation}
\partial_{\xi}z\left(\xi\right)=\frac{z\left(\xi\right)^{2}+\left(a\,\xi-1\right)z\left(\xi\right)}{b\,\xi\left(1-\xi\right)}.\label{eq:RICCATI}
\end{equation}
Si tratta di una equazione di Riccati, che attraverso una ulteriore
serie di ridefinizioni (non riportate), è riconducibile ad una equazione
differenziale ipergeometrica. Alla fine dei calcoli, si trova la soluzione
della \prettyref{eq:RICCATI} nella forma 
\begin{equation}
\frac{z\left(\xi\right)}{1-a}=\left[1+\frac{a\,\xi^{\frac{1}{b}}\Psi_{+}}{\left(1-\xi\right)^{\frac{1-a}{b}}}-a\left(1-\xi\right){}_{2}F_{1}\left(1,1-\frac{a}{b};1-\frac{1}{b};\xi\right)\right]^{-1},\label{eq:SUPERZ}
\end{equation}
dove $_{2}F_{1}\left(\kappa_{0},\kappa_{1};\kappa_{2};\xi\right)$
è la funzione ipergeometrica di Gauss, e $\Psi_{+}$ è il parametro
libero. Imponendo la condizione $z\left(1\right)=1-a$, si ottiene
in fine il parametro $\Psi_{+}$, calcolabile attraverso il limite
\begin{equation}
\Psi_{+}=\lim_{\xi\rightarrow1}\left[\frac{_{2}F_{1}\left(1,1-\frac{a}{b};1-\frac{1}{b};\xi\right)}{\left(1-\xi\right)^{\frac{a-1}{b}-1}\xi^{\frac{1}{b}}}\right]=\left(\frac{1-a}{a\,b}\right)\frac{\Gamma\left(\frac{1-a}{b}\right)\Gamma\left(-\frac{1}{b}\right)}{\Gamma\left(-\frac{a}{b}\right)}.\label{eq:SUPERZ2}
\end{equation}
Si noti che la funzione $\Gamma\left(\kappa\right)$ diverge per $\kappa\in\mathbb{Z}^{-}$,
così come $_{2}F_{1}\left(1,\kappa_{1};\kappa_{2};\xi\right)$ diverge
per $\kappa_{1},\,\kappa_{2}\in\mathbb{Z}^{-}$. Dunque, quando $1/b$,
$\left(a-1\right)/b$ sono numeri interi positivi, la funzione $z\left(\xi\right)$
deve essere determinata attraverso un limite di parametri reali che
tendono a numeri interi.

A questo punto, possiamo scrivere l'energia libera 
\begin{equation}
u\left(\xi\right)=-\log\left(z\left(\xi\right)\right),
\end{equation}
e da questa le funzioni termodinamiche $m_{\beta}$ e $c\left(\beta\right)$
espresse nella nuova variabile:
\begin{equation}
m\left(\xi\right)=\left(1-\xi\right)\partial_{\xi}u\left(\xi\right),
\end{equation}
\begin{equation}
c\left(\xi\right)=-\left(1-\xi\right)\log\left(1-\xi\right)^{2}\left[\left(1-\xi\right)\partial_{\xi}^{2}u\left(\xi\right)-\partial_{\xi}u\left(\xi\right)\right].
\end{equation}
Per brevità, non si riportano le espressioni complete, che possono
essere facilmente calcolate con un CAS. 

\begin{figure}
\begin{centering}
\caption{}
\includegraphics[scale=0.5]{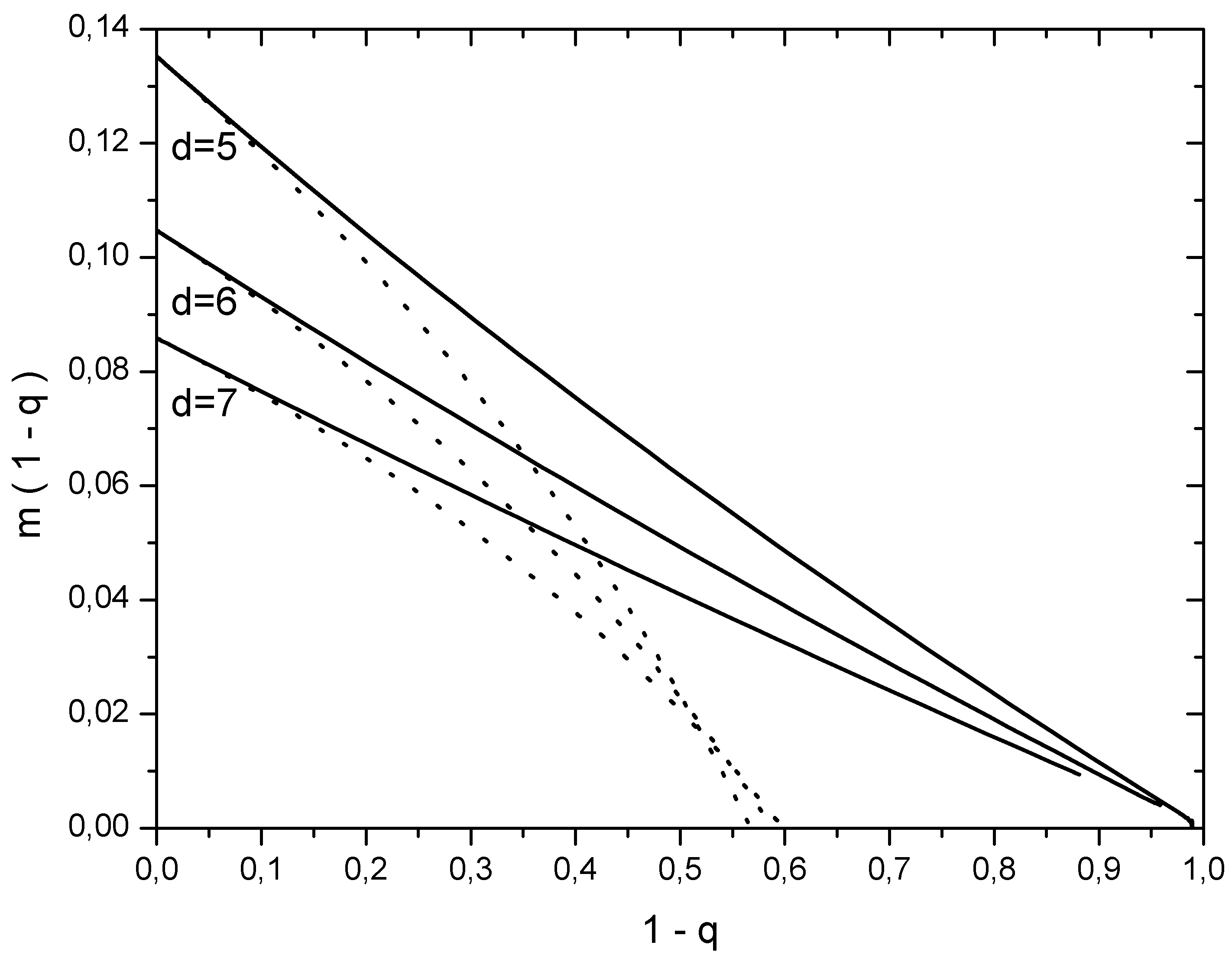}\label{Flo:GGR1}
\par\end{centering}
\emph{La figura mostra il numero medio di intersezioni per monomero
$m\left(1-q\right)$ per $d=5,\,6,\,7$, calcolato con l'approssimazione
\prettyref{eq:SUPERZ}. Le linee di punti indicano l'espansione al
primo ordine in $\beta=-\log\left(q\right)$: $m\left(1-q\right)\sim C_{d}+\sigma_{d}\log\left(q\right)$,
con $\sigma_{d}=C_{d}\bar{C}_{d}/\left(1-2B_{d}\right)$.}
\end{figure}

\begin{figure}
\begin{centering}
\caption{}
\includegraphics[scale=0.5]{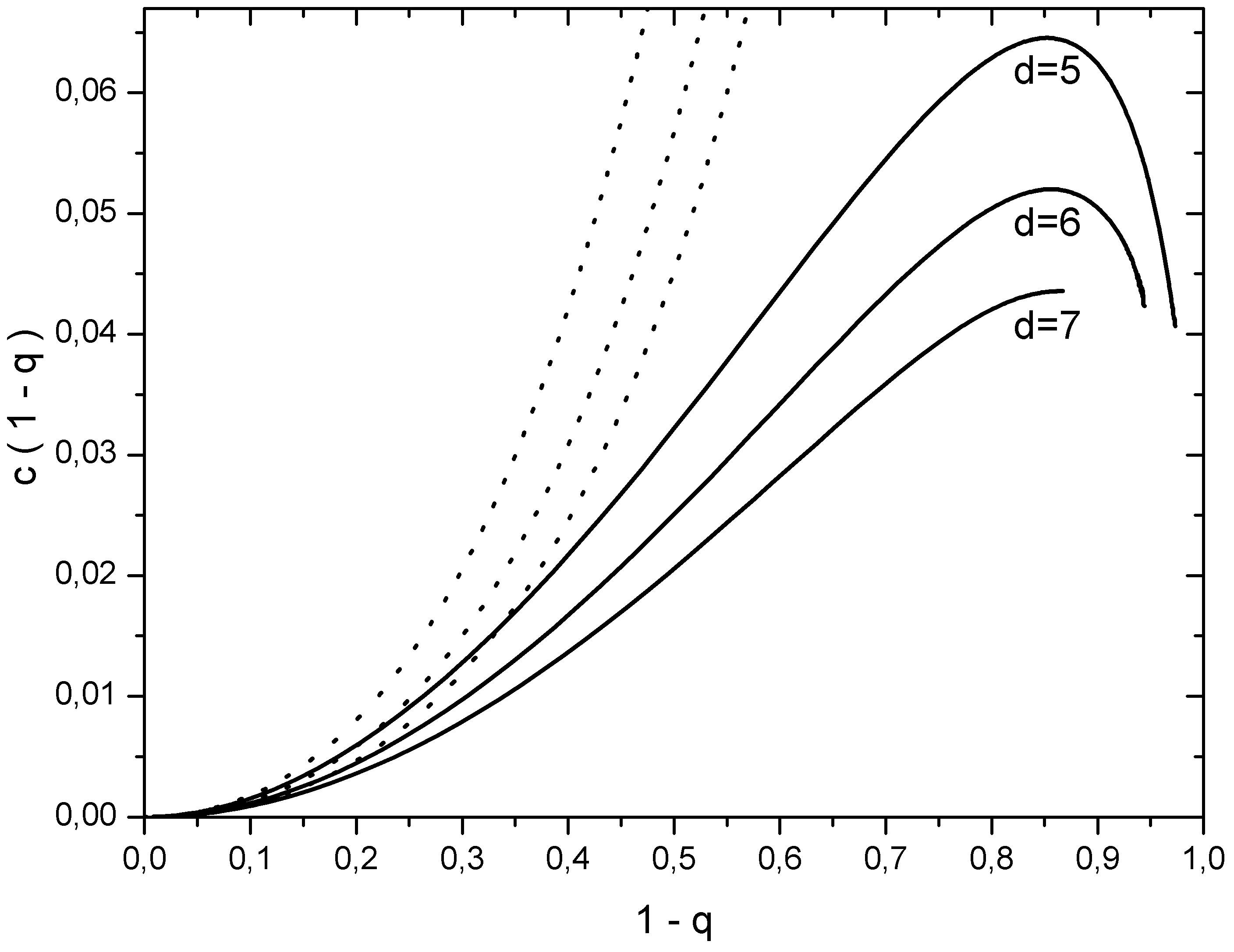}\label{Flo:GGR2}
\par\end{centering}
\emph{La figura mostra il calore specifico per monomero $c\left(1-q\right)$
per $d=5,\,6,\,7$, calcolato con l'approssimazione \prettyref{eq:SUPERZ}.
Le linee di punti indicano l'espansione al primo ordine in $\beta=-\log\left(q\right)$:
$c\left(1-q\right)\sim\sigma_{d}\log\left(q\right)^{2}$, con $\sigma_{d}=C_{d}\bar{C}_{d}/\left(1-2B_{d}\right)$.
I massimi per $d=4,\,5,\,6$ sono rispettivamente in $q_{5}=0.1478$,
$q_{6}=0.1440$, $q_{7}=0.1437$.}
\end{figure}

Imponiamo ora i valori relativi al modello reale per $d\geq5$: per
$a=C_{d}\left(1-B_{d}\right)$, $b=B_{d}$ non si osservano punti
di non analiticità nelle funzioni termodinamiche. In particolare,
come previsto il calore specifico non ha discontinuità (il modello
di Stanley non presenta transizioni di fase a temperatura finita),
e $m\left(\xi\right)\in\left(0,C_{d}\right)$ per $\xi\in\left(0,\,1\right)$
(Figure \prettyref{Flo:GGR1}, \prettyref{Flo:GGR2} per $d=5,\,6,\,7$).

Come ultimo argomento, presentiamo una piccola digressione sulla rappresentazione
parametrica di $\varphi\left(m\right)$ nel caso lineare. Abbiamo
già sottolineato come sia possibile descrivere $\varphi\left(m\left(\xi\right)\right)$
variando il parametro $\xi$, ma è chiaro che, prendendo $\xi\in\left[0,\,1\right]$,
siamo in grado di rappresentare la funzione solo nel range $m\left(\xi\right)\in\left[0,\,m_{0}\right]$
(con $m_{0}=\pi\left(m_{0}\right)$).

Per ottenere anche il ramo con $m\in\left(m_{0},\,1\right)$ occorre
prendere $\exp\left(\beta\right)$ al posto di $\exp\left(-\beta\right)$
nelle definizioni delle quantità termodinamiche: in particolare si
avrà che $m_{\beta}=-\partial_{\beta}u\left(\beta\right)$. Questo
ci porterà all'equazione

\begin{equation}
\pi\left(-\partial_{\beta}u\left(\beta\right)\right)=\frac{e^{-u\left(\beta\right)}-1}{e^{\beta}-1}.
\end{equation}
La sostituzione opportuna questa volta è
\begin{equation}
\vartheta=\left(e^{\beta}-1\right)\in\left[0,\,\infty\right),
\end{equation}
\begin{equation}
z\left(\vartheta\right)=e^{-u\left(\beta\left(\vartheta\right)\right)},
\end{equation}
da cui semplificando si ottiene l'equazione differenziale
\begin{equation}
\partial_{\vartheta}z\left(\vartheta\right)=\frac{z\left(\vartheta\right)^{2}-\left(a\,\vartheta+1\right)z\left(\vartheta\right)}{b\,\vartheta\left(1+\vartheta\right)}.\label{eq:GKO}
\end{equation}
Si tratta di una equazione di Riccati, dello stesso tipo della precedente
\prettyref{eq:RICCATI}: anche la sua soluzione è concettualmente
simile, e porta alla funzione 
\begin{equation}
z\left(\vartheta\right)=\left(1+\vartheta\right)^{\frac{1-a}{b}}\left[\vartheta^{\frac{1}{b}}\Psi_{-}+{}_{2}F_{1}\left(-\frac{1}{b},1+\frac{a-1}{b};1-\frac{1}{b};-\vartheta\right)\right]^{-1}.
\end{equation}
Determinare il parametro libero è più complesso del caso precedente,
essendo questo individuato dall'unica soluzione reale dell'equazione
\begin{equation}
\Psi_{-}\,|\,\lim_{\vartheta\rightarrow\infty}\partial_{\vartheta}z\left(\vartheta\right)\in\left(0,\infty\right):
\end{equation}
dimostrare l'unicità di $\Psi_{-}$ (e determinarne simbolicamente
il valore) è tecnicamente possibile a partire dalla \prettyref{eq:GKO},
ma comporta una mole di lavoro aggiuntivo che esula dagli scopi principali
di questo lavoro. 

Dall'espressione di $z\left(\vartheta\right)$ ricaviamo 
\begin{equation}
u\left(\vartheta\right)=-\log\left(z\left(\vartheta\right)\right),
\end{equation}
\begin{equation}
m\left(\vartheta\right)=-\left(1+\vartheta\right)\partial_{\vartheta}u\left(\vartheta\right),
\end{equation}
dove il range di $m\left(\vartheta\right)$ al variare di $\vartheta$
è $\left[m_{0},\,1\right]$: abbiamo dunque ottenuto anche il secondo
ramo di $\varphi\left(m\right)$. 

Si noti che questa seconda funzione $z\left(\vartheta\right)$ non
rappresenta il modello di Stanley per $q>1$, in quanto oltre il punto
$m=C_{d}$ la $\varphi\left(m\right)$ reale si annulla, dando origine
a comportamenti completamente differenti.

\pagebreak{}

\chapter{Conclusioni}

Prima di discutere le conclusioni, presentiamo un breve riassunto
dei principali risultati (originali) ottenuti nei vari capitoli. 

Il primo capitolo è introduttivo al problema delle intersezioni, e
riassume essenzialmente alcuni aspetti noti della teoria dei Random
Walks. Sono tuttavia presenti due risultati minori, come il calcolo
dei valori numerici di $\sigma_{d}$ per $d\in\left[4,\,8\right]$,
e la descrizione della distribuzione $P_{n}\left(M\right)$ in termini
della funzione \emph{ambiente} $\pi_{n}\left(M\right)$. 

Il secondo introduce invece l'ensemble $\Omega_{m}$, ed è dedicato
allo studio numerico delle sue proprietà al variare di $m$: in particolare,
vengono osservati gli andamenti della distanza media end-to-end $\Re_{n}^{2}\left(m\right)$,
e della funzione $\pi\left(m\right)$. 

Per quanto riguarda $\Re_{n}^{2}\left(m\right)$, le simulazioni indicano
la sussistenza della relazione $\Re_{n}^{2}\left(m\right)\propto n^{2\nu\left(m\right)}$,
con $\nu\left(m\right)$ esponente critico dipendente da $m$ e $d$.
Per $d\geq3$ si è rilevata l'esistenza di una transizione di fase
per un valore critico $0<m_{c}<1$ del parametro $m$: prima di questo
valore, le catene si comportano come dei Self-Avoiding Walks, con
$\nu\left(m<m_{c}\right)=\nu_{d}$, mentre per $m>m_{c}$ collassano
in una configurazione semi-compatta con $\nu\left(m>m_{c}\right)=1/d$
(range proporzionale al volume). Da argomentazioni euristiche, e per
analogia con i comportamenti della Wiener Sausage, si predice che
$m_{c}=C_{d}$ per $d=3,4$, $m_{c}>C_{d}$ per $d\geq5$. Dai risultati
delle simulazioni si predice inoltre che, al punto critico per $d=3,4$,
l'esponente assuma un valore $\nu\left(m_{c}\right)=1/2$.

I dati confermano l'esistenza della funzione $\pi\left(m\right)\in\left(0,1\right)$,
definita come limite di $\pi_{n}\left(M/n\right)$ per $n,\,M\rightarrow\infty$,
$M/n\rightarrow m$. Anche in $\pi\left(m\right)$ si osserva un punto
di non analiticità in $m_{c}$: a causa di limitazioni tecniche sul
metodo di simulazione, non è stato possibile indagare in maniera conclusiva
l'andamento asintotico nella zona clusterizzata $m>m_{c}$. 

Il terzo capitolo introduce un nuovo processo stocastico, esplicitamente
dipendente dal tempo, individuato dalla funzione $\pi\left(m\right)\in\left(0,1\right)$.
Viene introdotto un metodo per calcolare i momenti, e una equazione
differenziale implicita per la distribuzione stazionaria. I risultati
vengono successivamente applicati al calcolo della $P_{n}\left(m\right)$
del SRW.

In fine, il quarto capitolo è principalmente dedicato alla derivazione
microscopica di una approssimazione lineare per $\pi\left(m\right)$,
mentre il quinto utilizza l'analogia con il modello di Stanley per
estrarre ulteriori informazioni circa l'ensemble $P_{n}\left(m\right)$.
Sempre nel capitolo quinto vi é una riformulazione del processo stocastico
(definito nel capitolo terzo) in termini di funzioni generatrici,
che ha consentito di risolvere esattamente il caso lineare con $\pi\left(m\right)=a+bm$.

\section{Grandi deviazioni di $P_{n}\left(M\right)$ nella direzione negativa}

I risultati ottenuti sul processo stocastico \prettyref{eq:3.1.1-4},
\prettyref{eq:3.1.1-5} permettono di dimostrare alcune proprietà
della funzione $P_{n}\left(M\right)$ per deviazioni moderate $M=\mathcal{O}\left(\langle M_{n}\rangle\right)$
nella direzione $M<\langle M_{n}\rangle$ (direzione negativa). 

Benché il problema delle grandi deviazioni nella direzione $M>\langle M_{n}\rangle$
sia stato affrontato da molti autori (soprattutto nel caso della Wiener
Sausage, si veda \cite{Den=000020Hollander}), quello di deviazioni
di $P_{n}\left(M\right)$ nella direzione negativa è stato preso in
considerazione in soli due lavori: il \cite{Van=000020den=000020Berg-Toth}
riguarda la Wiener Sausage, mentre il \cite{Hamana} tratta del modello
discreto (SRW). In particolare, il lavoro \cite{Hamana}, in cui vengono
implicitamente provate esistenza e convessità della funzione $\varphi\left(m\right)$
(definita in \prettyref{eq:VARFIFUNCT}), sembra rappresentare l'attuale
stato dell'arte.

Il lavoro svolto ci permette di dedurre direttamente alcune proprietà
del limite 
\begin{equation}
\psi\left(\beta\right)=\lim_{n\rightarrow\infty}n^{-1}\log\left(\langle e^{-\beta M}\rangle\right),
\end{equation}
per SRW su reticolo $\mathbb{Z}^{d}$. In particolare, se $\pi\left(m\right)\in\left(\eta_{d},\,C_{d}\right)$
per $m\in\left(0,C_{d}\right)$, dalla \prettyref{eq:IMPLICITA} segue
immediatamente che 
\begin{equation}
\log\left(1+\eta_{d}\left(1-e^{-\beta}\right)\right)\leq\psi\left(\beta\right)\leq\log\left(1+C_{d}\left(1-e^{-\beta}\right)\right)
\end{equation}
per qualunque $\beta>0$. Inoltre,dai risultati del capitolo quinto
si ricava che, per $d\geq3$, sussiste il limite 
\begin{equation}
\lim_{\beta\rightarrow0}z\left(1-e^{-\beta}\right)e^{-\psi\left(\beta\right)}=1,
\end{equation}
con $z\left(\cdot\right)$ definita dalle equazioni \prettyref{eq:SUPERZ},
\prettyref{eq:SUPERZ2} (e $a=C_{d}\left(1-B_{d}\right)$, $b=B_{d}$).
Sempre dall'equazione \prettyref{eq:IMPLICITA} è possibile dimostrare
che, per $d=3$ ($B_{d}\rightarrow1/2$), 
\begin{equation}
\lim_{\delta m\rightarrow0^{-}}\frac{\varphi\left(\delta m+C_{3}\right)^{-1}\delta m^{2}}{W_{-1}\left(2\delta m\right)^{2}}\log\left(\frac{2\delta m}{W_{-1}\left(2\delta m\right)}\right)=1,
\end{equation}
dove $W_{-1}\left(\cdot\right)$ è il ramo inferiore della funzione
di Lambert, e $\delta m=m-C_{3}$.

Questi risultati costituiscono un significativo complemento ai lavori
\cite{Van=000020den=000020Berg-Toth} e \cite{Hamana}. In ultimo,
dalle considerazioni espresse nel capitolo quarto segue che l'equazione
\prettyref{eq:SUPERZ} costituisce un'ottima approssimazione non perturbativa
di $\exp\left(-\psi\left(\beta\right)\right)$ per $d\geq4$.

\section{Transizione Coil-Globule}

La transizione Coil-Globule (CG, \cite{de=000020Gennes,des=000020Cloiseaux-Jannink}),
in cui una catena flessibile collassa da una configurazione estesa
(\emph{random-coil}) a quella di un cluster semi-compatto, è un fenomeno
di grande interesse per la fisica dei polimeri, essenzialmente per
la somiglianza qualitativa con il \emph{protein folding} nelle macromolecole
biologiche.

La maggior parte dei modelli termodinamici, che mostrano una transizione
CG, considerano in genere una competizione tra interazioni di natura
geometrica differente: ad esempio, si può considerare il caso di una
catena ideale su reticolo, con interazioni on-site repulsive e link
attrattivi fra primi vicini. La transizione nasce dalla compensazione
di queste interazioni. Vale la pena di riportare che la transizione
CG non possiede ancora una teoria quantitativa affidabile: benché
siano stati ottenuti alcuni notevoli risultati attraverso la teoria
tricritica di ferromagneti con spin zero-dimensionali (\cite{De=000020Gennes=000020CG,Duplantier,Grassberger}),
le teorie RG perturbative sono probabilmente inadeguate per calcolare
le quantità di interesse nel di caso catene finite (si veda \cite{Douglas}
e referenze). 

Si consideri il random walk di $n$ passi $\omega=\left\{ S_{0},\,S_{1},\,...\,,\,S_{n}\right\} $:
definiamo il numero di visite per ogni sito come $\psi_{x}=\sum_{i}\delta\left(x-S_{i}\right)$.
Un modello tipico che incorpora sia il volume escluso che la transizione
CG è definito dall'hamiltoniana

\begin{equation}
E\left[\psi\right]=a\left(n\right)+\epsilon_{0}\sum_{x}\psi_{x}^{2}-\epsilon_{1}\sum_{\langle x,y\rangle}\psi_{x}\psi_{y},\label{eq:ENERGY}
\end{equation}
dove con $\langle x,y\rangle$ si sono indicate le coppie di siti
primi vicini.

Prendendo $\epsilon_{1}=0$ nella \prettyref{eq:ENERGY}, si avrà
il modello di Domb-Joyce (\cite{Domb-Joyce}), che permette di modellizzare
correttamente l'effetto di volume escluso, ma non mostra alcuna transizione
CG nel parametro di Boltzmann: in generale, tutti i modelli termodinamici
con un solo tipo (geometrico) di interazione non permettono di ottenere
una transizione CG in temperatura.

Benché sia possibile indurre una transizione CG nei modelli di tipo
Domb-Joyce cambiando il segno dell'accoppiamento $\epsilon_{0}$,
questo porterà a configurazioni estremamente collassate, e decisamente
poco realistiche (analogamente al caso del modello di Stanley attrattivo).

Nel caso $d=3$ il modello proposto realizza volume escluso, transizione
CG e cluster quasi-compatti con un singolo vincolo geometrico, mostrando
le stesse caratteristiche della transizione descritta dall'hamiltoniana
\prettyref{eq:ENERGY}. 

Se i due modelli dovessero trovarsi nella stessa classe di universalità,
questo sarebbe un risultato interessante, in quanto nel modello esaminato
la transizione nasce da un cambio radicale nella strategia ottima
per ottenere uno specifico rapporto di intersezione: la transizione
è quindi guidata dalla richiesta di massimizzare l'entropia configurazionale
dato il vincolo $m$ (ovvero, una correlazione a lungo raggio), senza
l'aggiunta esplicita di alcuna competizione tra interazioni.

\chapter*{Appendice}

\section*{Teorema di Jain-Pruitt per $d\protect\geq4$}

Presentiamo ora alcune parti rilevanti della dimostrazione per la
varianza (\cite{Jain-Pruitt}) in $d\geq4$. A questo scopo sarà necessario
introdurre le seguenti variabili (omettiamo per chiarezza le dipendenza
funzionale dalla realizzazione $\omega$):
\begin{equation}
Z_{i}^{n}=\mathbb{I}_{\omega}\left[S_{i}\neq S_{i+1}S_{i+2},\,...\,,\,S_{i}\neq S_{n}\right]\label{eq:1.2.6.0}
\end{equation}
\begin{equation}
Z_{i}=\mathbb{I}_{\omega}\left[S_{i}\neq S_{i+1},S_{i}\neq S_{i+2},\,...\right]\label{eq:1.2.6.1}
\end{equation}
\begin{equation}
W_{i}^{n}=Z_{i}^{n}-Z_{i}=\mathbb{I}_{\omega}\left[S_{i}\neq S_{i+1},\,...\,,\,S_{i}\neq S_{n};\,\exists S_{k}=S_{i},\,k>n\right]\label{eq:1.2.6.3}
\end{equation}
dove $\mathfrak{\mathbb{I}}_{\omega}\left[\Phi\right]$, definita
su $\Omega$, rappresenta la funzione caratteristica (\emph{indicatore})
del set $\Phi\in\Omega$ nella variabile funzionale $\omega\in\Omega$:
\begin{equation}
\mathfrak{\mathbb{I}}_{\omega}\left[\Phi\right]=\left\{ \begin{array}{c}
1\\
0
\end{array}\,\,\begin{array}{c}
\omega\in\Phi\\
\omega\notin\Phi
\end{array}\right..\label{eq:1.2.6.4}
\end{equation}
In termini pratici, $Z_{i}^{n}$ vale $1$ se nel cammino $\omega$
il sito $i-$esimo non viene rivisitato nei passi $i+1$, $i+2$,
... ,$n$ (ma può esserlo nei successivi passi) ed è nulla altrimenti.
$Z_{i}$ è il limite $Z_{i}^{n}$ per $n\rightarrow\infty$, in cui
il sito $i-$esimo non viene più visitato. In fine, $W_{i}^{n}$ vale
$1$ per quelle configurazioni in cui il sito $i$ non viene visitato
nei passi $i+1,\,i+2,\,...\,,\,n$, ed in cui $\exists k>n$ in cui
il cammino è tornato sul sito $i-$esimo. La variabile $R_{n}$ può
essere espressa in termini di queste variabili, secondo la relazione
\begin{equation}
R_{n}=\sum_{i=0}^{n}Z_{i}^{n}=\sum_{i=0}^{n-1}Z_{i}+\sum_{i=0}^{n-1}W_{i}^{n}+1.\label{eq:1.2.6.5}
\end{equation}
Definendo ancora due variabili $Y_{n}=\sum_{i=0}^{n-1}Z_{i}$, $W_{n}=\sum_{i=0}^{n-1}W_{i}^{n}$,
in modo da avere $R_{n}=Y_{n}+W_{n}+1$, con un po' di algebra possiamo
convenire che 
\begin{equation}
\langle\Delta R_{n}^{2}\rangle=\langle\Delta Y_{n}^{2}\rangle+\langle\Delta W_{n}^{2}\rangle+2\,\langle\Delta Y_{n}\Delta W_{n}\rangle.\label{eq:1.2.6.6}
\end{equation}
Dalla diseguaglianza di Schwartz sull'ultimo termine segue immediatamente
che
\begin{equation}
\langle\Delta Y_{n}\Delta W_{n}\rangle^{2}\leq\langle\Delta Y_{n}^{2}\rangle\langle\Delta W_{n}^{2}\rangle;\label{eq:1.2.6.7}
\end{equation}
è chiaro a questo punto che se $\langle\Delta W_{n}^{2}\rangle=o\left(\langle\Delta Y_{n}^{2}\rangle\right)$,
questo porta all'andamento $\langle\Delta R_{n}^{2}\rangle\sim\langle\Delta Y_{n}^{2}\rangle$
nel limite. La dimostrazione si riduce dunque a provare che $\langle\Delta W_{n}^{2}\rangle=o\left(\langle\Delta Y_{n}^{2}\rangle\right)$,
e a stimare l'andamento asintotico della $\langle\Delta Y_{n}^{2}\rangle$.
Omettiamo la dimostrazione della prima affermazione e vediamo come
ricavare l'andamento asintotico: dalla definizione di $Y_{n}$ troviamo
\begin{equation}
\langle\Delta Y_{n}^{2}\rangle=\sum_{j=1}^{n-1}\langle\Delta Z_{i}^{2}\rangle+2\sum_{j=1}^{n-1}\sum_{i=0}^{j-1}\langle\Delta Z_{i}Z_{j}\rangle,\label{eq:eq:1.2.6.8}
\end{equation}
dove $Z_{i}=1$ se il sito è occupato, e $0$ altrimenti. Dal fatto
che $\langle Z_{i}^{2}\rangle=\langle Z_{i}\rangle=1-F_{i}$ segue
che $\langle\Delta Z_{n}^{2}\rangle\sim F\left(1-F\right)n$. Avremo
dunque che 
\begin{equation}
\langle\Delta Y_{n}^{2}\rangle\sim F\left(1-F\right)n+2\sum_{j=1}^{n-1}a_{j}
\end{equation}
dove $a_{j}$ è la somma dalla matrice di covarianza $a_{j}=\sum_{i=0}^{j-1}\langle\Delta Z_{i}\Delta Z_{j}\rangle$
nella variabile $Z_{i}$. 

Dobbiamo ora stimare la $a_{j}$, e consideriamo a tal scopo il prodotto
$Z_{i}Z_{j}$: perché sia non-nullo è necessario che nel cammino avvengano
in sequenza i seguenti eventi: il cammino muove dall'origine al punto
$x$ in $j-i$ passi, senza essere tornato nell'origine, questo corrisponde
ad una probabilità $F_{j-1}\left(0,x\right)$ (si veda la sezione
precedente per le definizioni). Successivamente il camminatore non
rivisita più ne l'origine, ne il sito $x$. Questo comporta:
\begin{equation}
\langle Z_{i}Z_{j}\rangle=\sum_{x\neq0}F_{j-i}\left(x\right)\,\mbox{P}\left[\nexists k>\left(i-j\right)\,|\,S_{k}=\left\{ 0,x\right\} \right]\label{eq:1.2.7.1}
\end{equation}
\begin{equation}
\langle Z_{i}\rangle=\sum_{x\neq0}F_{j-i}\left(x\right)\,\mbox{P}\left[\nexists k>\left(i-j\right)\,|\,S_{k}=0\right]\label{eq:1.2.7.2}
\end{equation}
\begin{equation}
\langle Z_{j}\rangle=\mbox{P}\left[\nexists k>\left(i-j\right)\,|\,S_{k}=x\right].\label{eq:1.2.7.3}
\end{equation}
Introduciamo la funzione di supporto
\begin{equation}
\begin{array}{l}
b\left(x\right)=\mbox{P}\left[\nexists k>\left(i-j\right)\,|\,S_{k}=\left\{ 0,x\right\} \right]-\\
\\\,\,\,\,\,\,\,\,\,\,\,\,\,\,\mbox{+P}\left[\nexists k>\left(i-j\right)\,|\,S_{k}=0\right]\mbox{P}\left[\nexists k>\left(i-j\right)\,|\,S_{k}=x\right].
\end{array}\label{eq:1.2.7.4}
\end{equation}
La funzione $b\left(0,x\right)$ può essere riscritta per cammini
transienti (e dopo alcune manipolazioni algebriche) come
\begin{equation}
b\left(x\right)=\left(1-G\right)G\left(x\right)^{2}\left[1-G\left(x\right)^{2}\right]^{-1}.\label{eq:1.2.7.5}
\end{equation}
A questo punto possiamo scrivere $a_{j}$ come 
\begin{equation}
a_{j}=\sum_{x\neq0}\sum_{i=0}^{j-1}F_{j-i}\left(x\right)\,b\left(x\right)=\left(1-G\right)\sum_{x\neq0}\sum_{l=1}^{j}\frac{F_{l}\left(x\right)G\left(x\right)^{2}}{1-G\left(x\right)^{2}}.\label{eq:1.2.7.2-1}
\end{equation}
Consideriamo in fine la quantità $a$, limite $j\rightarrow\infty$
della precedente espressione:
\begin{equation}
a=\left(1-G\right)\sum_{x\neq0}\frac{F\left(x\right)G\left(x\right)^{2}}{1-G\left(x\right)^{2}}<\infty.\label{eq:1.2.7.7}
\end{equation}
E' facile notare come per $\sum_{x}F\left(x\right)G\left(x\right)^{2}<\infty$
si abbia $a=\mathcal{O}\left(1\right)$, e dunque $\langle\Delta R_{n}^{2}\rangle=\mathcal{O}\left(n\right)$.
Quest'ultima condizione è valida solo per $d\geq4$: se $d=3$ la
quantità $\sum_{x}F\left(x\right)G\left(x\right)^{2}$ è divergente,
e l'andamento $\langle\Delta R_{n}^{2}\rangle=\mathcal{O}\left(n\log\left(n\right)\right)$
necessita di lavoro supplementare per essere esplicitato (si veda
\cite{Jain-Pruitt}).

\section*{Self-Avoiding Walk}

Definiamo $\omega$, Self-Avoiding Walk (SAW) di $n$ passi su $\mathbb{Z}^{d}$
con estremi $x$ e $y$, come una sequenza di siti $\omega\equiv\left\{ S_{0},\,S_{1},...S_{n}\right\} $
con $S_{0}=x$, $S_{n}=y$. La sequenza rappresenta un SAW se valgono
le relazioni $\left|S_{j+1}-S_{j}\right|=1$, $S_{i}\neq S_{j}$ per
$\forall i\neq j$. Da qui in poi indicheremo come $\left|\omega\right|=n$
la lunghezza del cammino e come $\left(S_{j}\right)_{i}$ la componente
$i-$esima del vettore $S_{j}$ (con $i=1,\,...\,d$).

La nostra prima quantità di interesse sarà il numero $c_{n}$ di SAW
che hanno un estremo nell'origine e la cui lunghezza è $\left|\omega\right|=n$.
Benché sia facile calcolare esattamente $c_{n}$ per i primi valori
di $n$ (ad esempio $c_{1}=2d$, $c_{2}=2d\left(d-1\right)$, etc
...), l'approccio combinatorio esatto diventa presto impossibile all'aumentare
di $n$. I risultati di maggiore interesse circa $c_{n}$ riguardano
infatti l'andamento asintotico per $n\rightarrow\infty$ . 

Prima di introdurre altre quantità è opportuno fare alcune semplici
considerazioni su $c_{n}$. In primo, luogo è facile notare come il
numero $c_{n}$ dei SAW sia maggiorato dal numero di Random Walks
Semplici (SRW) di $n$ passi (che vale $w_{n}=\left(2d\right)^{n}$);
questo limite superiore può essere immediatamente migliorato considerando
il numero di Non-Reversal Walks (NRW: random walk semplice in cui
ad ogni passo è vietata l'inversione immediata del cammino) che è
$2d\left(2d-1\right)^{n-1}$. Per quanto riguarda il limite inferiore,
possiamo contare il numero di cammini in cui ogni passo è in direzione
di una delle coordinate positive (ad ogni passo una coordinata viene
aumentata e le altre rimangono invariate), che vale $d^{n}$. Otteniamo
dunque che 
\begin{equation}
d^{n}\leq c_{n}\leq2d\left(2d-1\right)^{n-1}.\label{eq:2.1.0}
\end{equation}

Per continuare dobbiamo definire una misura di probabilità per il
set di SAW di $n$ passi ($\left|\omega\right|=n$). La misura che
utilizzeremo è la misura uniforme, dove per ogni cammino viene assegnato
lo stesso peso $c_{n}^{-1}$ (in quest'ottica il problema del SAW
è quello della Self-Repelling Chain, in quanto la misura uniforme
è una misura su cammini completi di lunghezza $n$, e non definisce
alcun processo stocastico). 

Indicando il valore di aspettazione rispetto alla misura uniforme
con $\left\langle \cdot\right\rangle $ , il valore quadratico atteso
della distanza end-to-end $(\left|S_{0}-S_{n}\right|=\left|S_{n}\right|)$
sarà
\begin{equation}
\Re_{n}^{2}=\langle S_{n}^{2}\rangle=\frac{1}{c_{n}}\sum_{\omega}S_{n}^{2},\label{eq:2.1.1}
\end{equation}
dove la somma sui SAW $\omega$ è intesa su tutti i cammini con $\left|\omega\right|=n$.
Come per $c_{n}$, anche per $\Re_{n}^{2}$ l'approccio combinatorio
diventa troppo complicato all'aumentare di $n$.

Per il SAW si suppone che $c_{n}$ presenti, per $d\leq3$, un andamento
esponenziale con correzioni a potenza, invece dell'andamento puramente
esponenziale del SRW. Si ritiene anche che la distanza quadratica
end-to-end non sia lineare in $N$, come invece avviene per SRW. Queste
proposte, consistenti con le proprietà di altri modelli di meccanica
statistica, sono supportate da simulazioni numeriche e alcuni calcoli
non rigorosi di teoria dei campi. 

Gli andamenti congetturati per $c_{n}$ e $\Re_{n}^{2}$ sono (\cite{Madras-Slade})
\begin{equation}
c_{n}\sim A_{d}\,\mu_{d}^{n}\,n^{\gamma_{d}-1},\label{eq:2.1.2}
\end{equation}
\begin{equation}
\Re_{n}^{2}\sim D_{d}\,n^{2\nu_{d}},\label{eq:2.1.3}
\end{equation}
dove $A_{d}$, $D_{d}$, $\mu_{d}$ , $\gamma_{d}$ e $\nu_{d}$ sono
costanti positive dipendenti dalla dimensionalità $d$ del reticolo.
I coefficienti $\gamma_{d}$ e $\nu_{d}$ sono esempi di esponenti
critici, mentre $\mu_{d}$ è detta \emph{costante connettiva}. Come
vedremo si ritiene anche che, in $d=4$, per \prettyref{eq:2.1.2},
\prettyref{eq:2.1.3} siano necessarie delle correzioni logaritmiche,
mentre per $d\geq5$ si ha $\gamma_{d}=1$. Notiamo come nel SRW le
quantità analoghe abbiano $\gamma_{d}=1$, $\nu_{d}=1/2$, così come
per i NRW e altri cammini generalizzati il cui processo di generazione
è ancora markoviano.

La costante connettiva $\mu_{d}$ della formula \prettyref{eq:2.1.2}
viene definita attraverso il limite 
\begin{equation}
\mu_{d}=\lim_{n\rightarrow\infty}c_{n}^{1/n}\label{eq:2.1.4}
\end{equation}
la cui esistenza può essere facilmente provata in maniera rigorosa
. E' dimostrato anche che $c_{n}\geq\mu_{d}^{n}$ . Le semplici restrizioni
dell'equazione \prettyref{eq:2.1.0} implicano immediatamente che
\begin{equation}
d\leq\mu_{d}\leq2d-1.\label{eq:2.1.5}
\end{equation}
Benché esistano buone stime del valore di $\mu_{d}$ , per reticoli
ipercubici $\mathbb{Z}^{d}$ il valore esatto è tuttora ignoto in
$d>1$ (uno dei pochi risultati riguarda il reticolo esagonale in
$d=2$, per cui si congettura un valore $\mu_{2}=\sqrt{2+\sqrt{2}}$)
(\cite{Ninhenius}). E' nota una espansione asintotica di $\mu_{d}$
per alti valori della dimensionalità (\cite{Clisby-Slade}): per $d\rightarrow\infty$
si ha
\begin{equation}
\begin{array}{c}
\mu_{d}=2d-1-\frac{1}{2d}-\frac{3}{\left(2d\right)^{2}}-\frac{16}{\left(2d\right)^{3}}-\frac{102}{\left(2d\right)^{4}}-\frac{729}{\left(2d\right)^{5}}-\frac{5533}{\left(2d\right)^{6}}-\frac{42229}{\left(2d\right)^{7}}+\\
-\frac{288761}{\left(2d\right)^{8}}-\frac{288761}{\left(2d\right)^{8}}-\frac{1026328}{\left(2d\right)^{9}}+\frac{21070667}{\left(2d\right)^{10}}+\frac{780280468}{\left(2d\right)^{11}}+\mathcal{O}\left(\frac{1}{\left(2d\right)^{12}}\right).
\end{array}\label{eq:2.1.6}
\end{equation}
Intuitivamente si nota dalla \prettyref{eq:2.1.6} che, per $d$ grandi,
l'effetto principale della self-avoidance è di eliminare le inversioni
immediate (come nel NRW).

Vediamo ora alcune proprietà dell'esponente $\gamma_{d}$ . Dato che
$c_{n}\geq\mu_{d}^{n}$, allora si ha che $\gamma_{d}\geq1$ per qualunque
dimensione $d$. Tuttavia non esiste ancora alcuna prova rigorosa
del fatto che $\gamma_{d}$ sia finito per $d=2,\,3,\,4$ dove le
migliori restrizioni superiori note sono (\cite{Madras-Slade})
\begin{equation}
c_{n}\leq\left\{ \begin{array}{l}
\mu_{d}^{n}\exp\left(a_{d}n^{1/2}\right)\\
\mu_{d}^{n}\exp\left(a_{d}n^{\frac{2}{d+2}}\log n\right)
\end{array}\right.\begin{array}{l}
d=2\\
d=3,4
\end{array},\label{eq:2.1.7}
\end{equation}
con $a>0$. Per $d\geq5$ si può invece dimostrare che $\gamma_{d}=1$.

L'esponente $\gamma_{d}$ , oltre a caratterizzare le correzioni a
potenza dell'andamento di $c_{n}$, dà la misura di probabilità che
due SAW, $\omega^{\left(1\right)}$ e $\omega^{\left(2\right)}$,
di $n$ passi e con un estremo in comune, non siano intersecati. Infatti
questa probabilità vale (considerando la \prettyref{eq:2.1.2})
\begin{equation}
\Pi_{n}=\mbox{P}\left[\omega^{\left(1\right)}\cap\omega^{\left(2\right)}=S_{0}\right]=\frac{c_{2n}}{c_{n}^{2}}\sim\left(\frac{2^{\gamma_{d}-1}}{A}\right)n^{1-\gamma_{d}}.\label{eq:2.1.8}
\end{equation}
 Si nota immediatamente che per $\gamma_{d}>1$ la probabilità di
intersezione si annulla per $n\rightarrow\infty$ , mentre per $\gamma_{d}=1$
rimane costante e positiva. Per i SRW la stessa quantità è sempre
positiva per $d>4$, per $d=4$ decresce come $\left(\log n\right)^{-1/2}$
mentre in $d=2,3$ decresce come $n^{-\kappa}$.

La regola con cui abbiamo definito il SAW può essere interpretata
intuitivamente come una sorta di interazione repulsiva. In quest'ottica
la distanza end-to-end quadratica dovrebbe crescere più velocemente
nel SAW rispetto al SRW, ovvero l'esponente critico $\nu_{d}$ dovrebbe
essere maggiore o al più uguale a quello del SRW ($\nu_{d}\geq1/2$).
Ad ogni modo, una prova rigorosa della diseguaglianza $\Re_{n}^{2}\geq Cn$
per $\forall d$ è ancora un problema aperto. 

Anche per quel che riguarda il limite superiore, non ci sono risultati
rigorosi migliori di quello immediatamente ricavabile dalla definizione
\prettyref{eq:2.1.0}: $\Re_{n}^{2}\leq n^{2}$, $\nu_{d}\leq1$.
Tuttavia in $d\geq5$ è stato provato che $\nu_{d}=1/2$, da cui si
osserva che gli effetti self-avoiding in reticoli ad alta dimensionalità
non cambiano la legge di scala del SRW (\cite{Hara-Slade}).

Sempre per alte dimensioni, è dimostrato anche che il coefficiente
di diffusione $D_{d}$ della \prettyref{eq:2.1.4} è sempre $D_{d}>1$
(quindi strettamente maggiore di quello del SRW); ne deriva che il
SAW in alte dimensioni continua ad essere più esteso del SRW, ma solo
per quanto riguarda il coefficiente diffusivo. La tendenza a essere
più esteso diventa inoltre meno pronunciata all'aumentare della dimensione,
dunque si suppone che $\nu_{d}$ sia una funzione non crescente in
$d$.

Si ritiene (\cite{Hara-Slade}) che gli esponenti critici $\nu_{d}$
e $\gamma_{d}$ siano dipendenti dalla dimensionalità del reticolo
$d$, ma non dal tipo di passo compiuto (finché i passi ammessi sono
simmetrici e finiti) o dal tipo di reticolo su cui ci troviamo (ipercubico,
triangolare, esagonale, etc ...). Questa indipendenza dai dettagli
del sistema è chiamata \emph{universalità}: diremo dunque che i modelli
che hanno gli stessi esponenti critici appartengono alla stessa \emph{classe
di universalità}. Ad esempio la costante connettiva $\mu_{d}$, che
rappresenta il numero di coordinazione effettivo del reticolo, non
è universale, in quanto dipende dal tipo di reticolo, oltre che dalla
dimensione $d$.

E' chiaro che il SAW dovrebbe essere (in senso di classe di universalità)
più vicino al SRW in alte dimensioni, dato che il SRW tende ad auto-intersecarsi
sempre meno all'aumentare della $d$. Il caso $d=4$ è un caso speciale,
dato che il tempo $\tau_{d}$ di primo ritorno all'origine, condizionato
dal fatto che ciò accada, è una quantità finita per $d>4$; questo
suggerisce che, per alte dimensioni, l'effetto di self-avoidance sia
un effetto \emph{a corto raggio}, e dunque non abbia effetti sugli
esponenti critici.

\begin{table}
\begin{centering}
\caption{}
\ \ \label{Flo:tab2}
\par\end{centering}
\begin{centering}
\begin{tabular}{|c|c|c|c|c|}
\hline 
\noalign{\vskip\doublerulesep}
$d$ & $\mu_{d}$ & $D_{d}$ & $\gamma_{d}$ & $\nu_{d}$\tabularnewline[\doublerulesep]
\hline 
\hline 
$2$ & $2.63815\left(8\right)$ & $-$ & $43/32$ & $3/4$\tabularnewline
\hline 
$3$ & $4.683\left(9\right)$ & $-$ & $1.16\left(2\right)$ & $0.587597(7)$\tabularnewline
\hline 
$4$ & $6.774043$$\left(5\right)$ & $-$ & $\,\,\,\,\,\,\,1\,_{l.c.}$ & $\,\,\,\,\,\,\,\,1/2\,_{l.c.}$\tabularnewline
\hline 
$5$ & $8.838544\left(3\right)$ & $1.4767\left(1\right)$ & $1$ & $1/2$\tabularnewline
\hline 
$6$ & $10.878094\left(4\right)$ & $1.2940\left(6\right)$ & $1$ & $1/2$\tabularnewline
\hline 
$7$ & $12.902817\left(3\right)$ & $1.2187\left(3\right)$ & $1$ & $1/2$\tabularnewline
\hline 
$8$ & $14.919257\left(2\right)$ & $1.1760\left(2\right)$ & $1$ & $1/2$\tabularnewline
\hline 
\end{tabular}\\
$\,$
\par\end{centering}
\centering{}{\small\emph{l.c. indica la correzione logaritmica \prettyref{eq:2.1.11}}}{\small\par}
\end{table}

Per ora gli unici risultati rigorosi, che confermano gli andamenti
a potenza e i valori congetturati per $\nu_{d}$ e $\gamma_{d}$ ,
sono per $d\geq5$. Le correzioni logaritmiche in $d=4$ sono state
calcolate con il gruppo di rinormalizzazione e valgono (\cite{Le=000020Guillou-Zinn-Justin})
\begin{equation}
\begin{array}{c}
c_{n}\sim A_{4}\,\mu_{4}^{n}\,\log\left(n\right)^{\frac{1}{4}},\\
\Re_{n}^{2}\sim D_{4}\,n\,\log\left(n\right)^{\frac{1}{4}}.
\end{array}\label{eq:2.1.11}
\end{equation}

E' comune, in meccanica statistica, considerare l'esistenza di una
\emph{dimensione critica} $d_{c}$ oltre la quale gli esponenti critici
non dipendono più dalla dimensione, e sono gli stessi della versione
in campo medio ($d\rightarrow\infty$) del modello in esame. Dati
i risultati fin qui esposti, la dimensione critica per il SAW sembrerebbe
$d_{c}=4$. Per $d>4$ gli esponenti critici del SAW sono gli stessi
del SRW, possiamo quindi considerare il SRW come il modello in campo
medio del SAW, e ci riferiremo agli esponenti critici del SRW come
esponenti di campo medio.

Per quanto riguarda i valori numerici delle costanti citate in questa
sezione, lo stato dell'arte è riportato in tabella \prettyref{Flo:tab2}:
i valori riportati sono estrapolati dai lavori \cite{Ninhenius,Hara-Slade,Madras-Sokal,Clisby,Owczarek-Prellberg,Le=000020Guillou-Zinn-Justin}.

Abbiamo già sottolineato la carenza di risultati esatti sugli esponenti
per $d=2,3,4$. I valori riportati per $d=2$ derivano dalla soluzione
esatta, non rigorosa, del modello di Ising con spin $\mathrm{N}-$dimensionali
(il cosiddetto modello $\mathrm{O}\left(\mathrm{N}\right)$) che contiene
il SAW come caso particolare $\mathrm{N}\rightarrow0$ (\cite{de=000020Gennes}).
Questa straordinaria connessione tra sistemi di spin e SAW si basa
sul noto teorema $\mathrm{N}\rightarrow0$ di De Gennes. Per quanto
riguarda $d=3$, non esistono risultati analoghi e i valori presentati
degli esponenti vengono da simulazioni numeriche ad altissima precisione
(\cite{Clisby}).

Una semplice ma sorprendentemente accurata analisi (di tipo mean-field)
sul valore di $\nu$ è stata proposta da Flory (\cite{Flory}). Gli
esponenti di Flory valgono $\nu_{F}=3/\left(2+d\right)$ per $d\leq4$
e $\nu_{F}=1/2$ per $d>4$; questi esponenti sono in accordo con
quelli attualmente accettati per $d=2$, $d>4$, $d=4$ (a parte la
correzione logaritmica) e si avvicinano notevolmente al valore di
$0.587597(7)$ per $d=3$.

\section*{Volume della Wiener Sausage}

Prima di discutere l'oggetto principale di questa appendice, introduciamo
brevemente la Wiener Sausage. Detto $\beta\left(t\right)$, $t\in\mathbb{R}^{+}$
il moto Browniano standard in $\mathbb{R}^{d}$, definiamo la Wiener
Sausage (WS) $W^{a}\left(t\right)$ di raggio $a$ come 
\begin{equation}
W^{a}\left(t\right)=\bigcup_{0\leq s\leq t}B_{a}\left[\beta\left(s\right)\right],\,\,t\geq0,
\end{equation}
 dove $B_{a}\left[x\right]$ è una ipersfera $d-$dimensionale di
raggio $a$, centrata in $x\in\mathbb{R}^{d}$. Consideriamo il volume
$\left|W^{a}\left(t\right)\right|$ dell'insieme, equivalente del
Range nei SRW: come per il SRW gli andamenti asintotici dei primi
due momenti del volume sono noti: per $d\geq3$ si ha che (\cite{Den=000020Hollander}):
\begin{equation}
\langle\left|W^{a}\left(t\right)\right|\rangle\sim\kappa_{a}t,
\end{equation}
mentre per la varianza $\langle\Delta\left|W^{a}\left(t\right)\right|^{2}\rangle$
valgono le relazioni (\cite{Den=000020Hollander}) 
\begin{equation}
\langle\Delta\left|W^{a}\left(t\right)\right|^{2}\rangle\asymp\left\{ \begin{array}{l}
t\,\ln\left(t\right)+\mathcal{O}\left(t\right)\\
t+o\left(t\right)
\end{array}\,\,\,\begin{array}{r}
d=3\\
d\geq4
\end{array}\right..\label{eq:1.2.5.k}
\end{equation}
Si noti che sono gli stessi andamenti che si trovano per il range
del SRW. Inoltre, anche il volume della WS ha distribuzione gaussiana
attorno al valor medio per $d\geq3$.

In questa appendice vengono riportati alcuni risultati noti riguardo
alle deviazioni di $\left|W^{a}\left(t\right)\right|$ nella direzione
negativa (valori più piccoli di $\langle\left|W^{a}\left(t\right)\right|\rangle$),
in cui $\left|W^{a}\left(t\right)\right|=\mathcal{O}\left(t\right)$:
questo caso, dove le deviazioni sono dell'ordine di $\langle\left|W^{a}\left(t\right)\right|\rangle$
è detto anche di \emph{deviazione moderata}.

Consideriamo brevemente il caso delle \emph{grandi deviazioni}, definite
dal vincolo $\left|W^{a}\left(t\right)\right|\leq f\left(t\right)$,
con $f\left(t\right)=o\left(t\right)$, $\lim_{t\rightarrow\infty}f\left(t\right)=\infty$.
In una serie di studi condotti da Donsker e Varadhan (\cite{Donsker-Varadhan}),
Bolthausen (\cite{Bolthausen}) e Sznitman (\cite{Sznitman}) si è
arrivati a definire la seguente proprietà asintotica:
\begin{equation}
\lim_{t\rightarrow\infty}t^{-1}f\left(t\right)^{\frac{2}{d}}\ln\left(\mbox{P}\left[\left|W^{a}\left(t\right)\right|\leq f\left(t\right)\right]\right)=-\frac{1}{2}\lambda_{d},\label{eq:GDEV}
\end{equation}
con $\lambda_{d}$ minimo autovalore per l'operatore laplaciano $-\nabla^{2}$
sulla sfera di volume unitario.

Da questa relazione, segue che la strategia ottimale per realizzare
situazioni del tipo $\left|W^{a}\left(t\right)\right|\leq f\left(t\right)$
è quella di confinare la WS in una ipersfera di volume $f\left(t\right)$:
la WS copre per intero l'interno della ipersfera (il range non ha
buchi).

In un notevole lavoro esatto di Van den Berg, Bolthausen, den Hollander
(\cite{Den=000020Hollander}), la relazione \prettyref{eq:GDEV} è
stata estesa al caso delle deviazioni moderate: in particolare, per
$d\geq3$ viene dimostrata la relazione 
\begin{equation}
\lim_{t\rightarrow\infty}t^{1-\frac{2}{d}}\ln\left(\mbox{P}\left[\left|W^{a}\left(t\right)\right|\leq\bar{m}t\right]\right)=I^{\kappa_{a}}\left(\bar{m}\right)\in\left(0,\infty\right).
\end{equation}

Questa relazione suggerisce che la strategia ottimale per ottenere
delle WS con $\left|W^{a}\left(t\right)\right|\leq\bar{m}t$ sia quella
di arrangiarle in cluster semi-compatti, ovvero cluster connessi di
taglia $\mathcal{O}\left(t\right)$, contenenti buchi di dimensione
$\mathcal{O}\left(1\right)$ distribuiti casualmente all'interno (configurazione
\emph{a gruviera}). 

Definiamo ora la variabile $u=\bar{m}/\kappa_{a}$, e la funzione
$\chi\left(\bar{m}/\kappa_{a}\right)=2\kappa_{a}^{2/d}I^{\kappa_{a}}\left(\bar{m}\right)$:
nel lavoro \cite{Den=000020Hollander} viene dimostrato che la funzione
$\chi\left(u\right)$ è strettamente decrescente in $u$, e gode delle
proprietà 
\begin{equation}
\lim_{u\rightarrow1^{-}}\left(1-u\right)^{-\frac{2}{d}}\chi\left(u\right)=2^{\frac{2}{d}}\omega_{d},
\end{equation}
\begin{equation}
\lim_{u\rightarrow0^{+}}u^{-\frac{2}{d}}\chi\left(u\right)=\lambda_{d},
\end{equation}
dove $\omega_{d}\in\left(0,\infty\right)$, e $\lambda_{d}$ è lo
stesso valore definito nel caso di grandi deviazioni. Un ulteriore
risultato di notevole interesse è che, per $d\geq5$, la funzione
$\chi\left(u\right)$ presenta un punto di non analiticità per un
particolare valore $u_{c}\in\left(0,1\right)$: in questo punto si
ha una discontinuità finita nella funzione $\partial_{u}\chi\left(u\right)$. 

L'esistenza di $u_{c}$ trova la sua interpretazione nelle configurazioni
ottimali per ottenere una WS con deviazioni moderate (\cite{Den=000020Hollander}).
Se per $u\in\left(0,u_{c}\right)$ la configurazione ottimale è quella
compatta durante tutta la generazione della WS, per $u\in\left(u_{c},1\right)$
la strategia migliore potrebbe essere temporalmente inomogenea: in
particolare, una strategia in cui la WS cresce come cluster semi-compatto
fino al tempo $\alpha\left(\bar{m}\right)t$ (coprendo un volume $\bar{m}t-\upsilon\left(\bar{m}\right)t$),
e poi continua fino a $t$ come un moto browniano libero (su scala
$\sqrt{t}$), coprendo un ulteriore volume $\upsilon\left(\bar{m}\right)t$.

Per $d=3,4$, la funzione $\chi\left(u\right)$ risulta analitica
in tutto l'intervallo chiuso $u\in\left(0,1\right)$, suggerendo che
in questi due casi la strategia sia temporalmente omogenea (\cite{Den=000020Hollander}).

\section*{Pruned-Enriched Rosenbluth Metod e simulazione dell'ensemble $\Omega_{n}\left(M\right)$}

Di seguito verrà esposto il metodo computazionale utilizzato per studiare
l'ensemble $\Omega_{n}\left(M\right)$. Si tratta del Pruned-Enriched
Rosenbluth Metod (\cite{Grassberger,Prellberg-Krawczyk}), una evoluzione
sostanziale del metodo di Rosenbluth (\cite{Rosenbluth}) che si è
rivelata estremamente efficiente per simulare catene interagenti e
altri ensemble di interesse nella fisica dei polimeri. Dato che, per
definire il PERM, risulta essenziale introdurre il RM, andremo subito
a riassumerne le caratteristiche fondamentali.

Il metodo di Rosenbluth (RM) è stato uno dei primi algoritmi cinetici:
essenzialmente, un cammino è accresciuto con passi successivi, scelti
volta per volta fra le evoluzioni permesse. Dunque, se ad un dato
passo vi sono $k$ possibili prosecuzioni del cammino (vincolate all'ottenimento
delle proprietà richieste), ognuna di queste avrà una probabilità
$p=1/k$ di verificarsi. Dato che, in generale, $p$ è soggetto a
variazioni durante la crescita del cammino, per avere una distribuzione
piatta è necessario ripesare ogni cammino ottenuto con la sua probabilità
di generazione.

Si consideri la funzione \emph{atmosfera} $a_{n}=a\left[\omega_{n}\right]$,
indicante il numero di continuazioni possibili per il cammino $\omega_{n}$
(di $n$ passi): il peso $W$ associato a $\omega_{n}$ sarà il prodotto
delle atmosfere $a_{k}$, $k\in\left[0,\,n-1\right]$ incontrate lungo
la crescita, 
\begin{equation}
W_{n}=\prod_{k=0}^{n-1}a_{k}.
\end{equation}
Dopo aver creato $S$ cammini, possiamo definire uno stimatore $C_{n}^{*}$
del numero totale di configurazioni possibili $C_{n}$, mediando tutti
i cammini $\omega_{n}^{\left(i\right)}$ ($i\in\left[1,\,S\right]$)
generati con i rispettivi pesi $W_{n}^{\left(i\right)}$:
\begin{equation}
C_{n}^{*}=\langle W_{n}\rangle=S^{-1}\sum_{i}W_{n}^{\left(i\right)}.
\end{equation}
Si noti che la media è su tutti i cammini costruiti, anche quelli
impossibili rispetto alla regola imposta dalla funzione atmosfera:
in effetti, le configurazioni impossibili sono eliminate automaticamente,
in quanto avranno peso $W_{n}^{\left(i\right)}$ nullo. 

Quello che si assume è essenzialmente di voler misurare una variabile
$A$, fluttuante ma scorrelata da $W_{n}$: definiamo dunque media
e varianza delle $A$ sui sample:

\begin{equation}
\bar{A}_{n}=S^{-1}\sum_{i}A_{\omega_{n}^{\left(i\right)}},
\end{equation}
\begin{equation}
\overline{\Delta A_{n}^{2}}=S^{-1}\sum_{i}\left(A_{\omega_{n}^{\left(i\right)}}-\bar{A}_{n}\right)^{2}.
\end{equation}
Questa media è presa sulla misura non uniforme (definita da $a_{k}$);
quella con sample di eguale peso statistico sarà invece:
\begin{equation}
\langle A_{n}\rangle=\langle W_{n}\rangle^{-1}\sum_{i}A_{\omega_{n}^{\left(i\right)}}W_{n}^{\left(i\right)},
\end{equation}
\begin{equation}
\langle\Delta A_{n}^{2}\rangle=\langle W_{n}\rangle^{-1}\sum_{i}\left(A_{\omega_{n}^{\left(i\right)}}-\langle A_{n}\rangle\right)^{2}W_{n}^{\left(i\right)}.
\end{equation}
Risulta piuttosto semplice dimostrare che la varianza $\langle\Delta A_{n}^{2}\rangle$
sulla misura uniforme è legata a quella sulla distribuzione correlata
($\overline{\Delta A_{n}^{2}}$) attraverso la relazione 
\begin{equation}
\langle\Delta A_{n}^{2}\rangle=S^{-1}\left(1+\frac{\langle\Delta W_{n}^{2}\rangle}{\langle W_{n}\rangle^{2}}\right)\overline{\Delta A_{n}^{2}}:
\end{equation}
si noti da quest'ultima espressione come l'efficienza del RM sia legata
all'ordine della quantità $\langle\Delta W_{n}^{2}\rangle/\langle W_{n}\rangle^{2}$,
che, in generale, diverge molto velocemente all'aumentare di $n$.

In pratica, i problemi con il metodo RM sono essenzialmente due. In
primo luogo, i pesi $W_{n}^{\left(i\right)}$ possono variare su molti
ordini di grandezza, portando ad avere una media falsata dalla presenza
di pochi cammini con pesi molto grandi ($\langle\Delta W_{n}^{2}\rangle/\langle W_{n}\rangle^{2}$
molto grande). In secondo luogo, se la funzione ambiente esclude (probabilisticamente)
una certa frazione di continuazioni ad ogni passo, allora si avrà
una soppressione dei cammini che aumenta esponenzialmente in funzione
della loro lunghezza.

Il Pruned-Enriched Rosenbluth Method (PERM) permette di aggirare entrambe
queste limitazioni, inserendo nell'RM una procedura che seleziona
le configurazioni da continuare. L'idea di base è quella di sopprimere
artificialmente le fluttuazioni di $W_{n}^{\left(i\right)}$: dunque,
se il peso di una configurazione è troppo grande rispetto allo stimatore
$C_{n}^{*}$, questa viene duplicata in varie copie, riducendo di
conseguenza $W_{n}^{\left(i\right)}$. Se invece il peso è troppo
basso, la configurazione viene eliminata con una certa probabilità:
nel caso la configurazione non sia eliminata, questa viene continuata
con un peso statistico aumentato. Si noti che entrambe le operazioni
(che chiameremo rispettivamente di \emph{arricchimento} e \emph{potatura}),
non cambiano l'espressione di $C_{n}^{*}$.

Chiaramente, la scelta dei criteri di arricchimento e potatura è arbitraria,
ma solo una scelta accurata (basata sul tipo di modello) permetterà
di avere un algoritmo efficiente: il criterio utilizzato in questo
lavoro (che ha il pregio di essere \emph{parameter-free}) è stato
proposto alcuni anni fa in \cite{Prellberg-Krawczyk}. Il metodo propone
di arricchire e potare ad ogni passo, in modo da esplorare porzioni
più estese dello spazio delle configurazioni.

Sia $r=W_{n}^{\left(i\right)}/C_{n}^{*}$ il rapporto fra il peso
del cammino $i-$esimo e lo stimatore: se $r>1$ allora vengono costruite
$c=\min\left(\left\lfloor r\right\rfloor ,a_{n}\right)$ copie di
$\omega_{n}^{\left(i\right)}$, ognuna con peso $W_{n}^{\left(i\right)}/r$.
Se invece $r\leq1$, il cammino viene continuato con probabilità $r$
e peso $C_{n}^{*}$ (ovvero eliminato con probabilità $1-r$). Si
noti che che l'operazione di arricchimento/potatura va effettuata
solo dopo che la configurazione è stata inclusa nel calcolo di $C_{n}^{*}$,
per poi essere utilizzata nel passo successivo.

Un altro punto cruciale per l'efficienza dell'algoritmo è dato dalla
scelta dell'atmosfera $a_{k}$: data la natura dell'ensemble che ci
si propone di studiare, si è resa necessaria l'introduzione di un
parametro libero $q$, che permette di controllare la forma geometrica
dei cammini.

Sia $R_{x}\left[\omega\right]$ la funzione caratteristica del cammino
$\omega$, il cui valore è $1$ se $x\in\omega$, e $0$ altrimenti:
la crescita dei nostri cammini viene definita dalla seguente probabilità
di transizione:

\begin{equation}
\pi_{\omega_{n}^{\left(i\right)}}\left(q,\delta x\right)=\frac{q^{R_{S_{n}^{\left(i\right)}+\delta x}\left[\omega_{n}^{\left(i\right)}\right]}}{\sum_{\left|\delta x'\right|=1}q^{R_{S_{n}^{\left(i\right)}+\delta x'}\left[\omega_{n}^{\left(i\right)}\right]}},\label{eq:regola}
\end{equation}
dove $\pi_{\omega_{n}^{\left(i\right)}}\left(q,\delta x\right)$ è
la probabilità saltare nel sito $S_{n}^{\left(i\right)}+\delta x$
(trovandosi in $S_{n}^{\left(i\right)}$) al passo $\left(n+1\right)-$esimo.
Essenzialmente, si tratta di un Myopic Self-Avoiding Walk (\cite{Amit-Parisi})
controllato dal parametro $q$: variando $q$ possiamo aumentare ($q>1$)
o diminuire ($q<1$) il numero medio di intersezioni eseguite dal
cammino generato, ottenendo quindi una esplorazione parziale della
distribuzione delle intersezioni $P_{n}\left(M\right)$ in zone lontane
dal suo valore medio.

Il peso statistico legato al cammino $\omega_{n}^{\left(i\right)}$,
e il relativo stimatore per al funzione di partizione sono
\begin{equation}
W_{n}^{\left(i\right)}\left(q\right)=\prod_{k=0}^{n-1}\pi_{\omega_{k}^{\left(i\right)}}\left(q,\delta x_{k}\right),
\end{equation}
\begin{equation}
\langle W_{n}\rangle_{q}=S^{-1}\sum_{i}W_{n}^{\left(i\right)}\left(q\right).
\end{equation}
Ora, detto $\Omega$ l'insieme dei cammini possibili, la media sul
sottoinsieme $\Phi\in\Omega$ sarà data dalla relazione 
\begin{equation}
\langle A\rangle_{\Phi}=\langle W_{n}\rangle_{q}^{-1}\sum_{i}A_{\omega_{n}^{\left(i\right)}}W_{n}^{\left(i\right)}\left(q\right)\mathbb{I}_{\omega_{n}^{\left(i\right)}}\left[\Phi\right],\label{eq:STAT}
\end{equation}
con $\mathbb{I}_{\omega}\left[\Phi\right]=1$ se $\omega\in\Phi$,
e $0$ altrimenti. Siccome siamo interessati a studiare i cammini
con un rapporto fissato di intersezioni per monomero, nel nostro caso
avremo $\Phi=\Omega_{n}\left(M\right)$.

Viene riportato di seguito lo pseudo-codice della subroutine principale:
per semplicità si omettono tutte le costanti esterne, le variabili
di implementazione e gli array di supporto (ad esempio l'array relativo
allo stadio di \emph{learning} e il meccanismo di limitazione delle
ramificazioni).\\

$\mathsf{Ciclo}(x,\,n)\,\{$\\
\\
E' la subroutine che simula i cammini con il metodo PERM descritto
in precedenza: nel codice reale le variabili passate sono molto più
numerose di $x$ ed $n$ (variabili di supporto, puntatori, registro
delle configurazioni, etc ...), ma si è preferito ometterle, in favore
di una esposizione più astratta\\

~~~$\mathsf{Esegui\,il\,passo\,}x'=x+\delta x_{n}\,\mathsf{con\,distribuzione\,}\pi_{\omega_{n-1}^{\left(i\right)}}\left(q,\delta x_{n}\right);$\\
\\
La prima operazione è l'esecuzione del passo in accordo con la \prettyref{eq:regola}:
si omettono le variabili necessarie per l'esecuzione del passo, calcolate
nel ciclo precedente.\\

~~~$w_{n}^{\left(i\right)}=w_{n-1}^{\left(i\right)}+\log\left[(q-1)\pi_{\omega_{n-1}^{\left(i\right)}}\left(q,\delta x_{n}\right)+1\right];$

~~~$\mathsf{Calcola}\,\pi_{\omega_{n}^{\left(i\right)}}\left(q,\delta x_{n}\right);$

~~~$\mathsf{Aggiorna\,i\,registri};$\\
\\
Viene calcolata la $w_{n}^{\left(i\right)}=\log(W_{n}^{\left(i\right)})$
relativa al cammino $\omega_{n}^{\left(i\right)}$ in esecuzione,
e la $\pi_{\omega_{n}^{\left(i\right)}}\left(q,\delta x_{n}\right)$
da usare nell'eventuale ciclo successivo. Subito dopo, viene aggiornato
l'array che registra la configurazione dei cammini.\\

~~~$Z_{n}=Z_{n}+\exp\left(w_{n}^{\left(i\right)}\right);$

~~~$S_{n}=S_{n}+1;$

~~~$\mathsf{Registra\,la\,statistica};$\\
\\
Aggiornamento delle variabili globali $Z_{n}$, $S_{n}$ ($\langle W_{n}\rangle_{q}=Z_{n}/S_{n}$),
e registrazione delle quantità di interesse secondo la \prettyref{eq:STAT},
con $\Phi=\Omega_{n}\left(M\right)_{m}$ (nel codice reale è presente
un array esterno di \emph{learning}, compilato in uno stadio precedente
del programma, necessario per evitare eventuali \emph{overflow}).
Di seguito inizia invece la procedura di potatura/arricchimento:\\

~~~$\mathsf{if}\,\left(n<N\right)\,\{$

~~~~~~$r=S_{n}Z_{n}^{-1}\exp\left(w_{n}^{\left(i\right)}\right);$
\\
\\
Dopo aver verificato di non essere già arrivati alla lunghezza $N$
desiderata, si procede con il calcolo di $r=W_{n}^{\left(i\right)}\left(q\right)/\langle W_{n}\rangle_{q}$:
a questo punto\\

~~~~~~$\mathsf{if}\,\left(r>1\right)\,\{$

~~~~~~~~~$\mathsf{if}\,(r>2d)\,\left\{ r=2d;\right\} \,\mathsf{else}\,\left\{ r=\left\lfloor r\right\rfloor ;\right\} $\\
\\
Se $r>1$, $r<2d$ si procede trasformando $r$ nella sua parte intera
(ovviamente, si possono fare solo un numero intero di copie del cammino),
altrimenti, se $r>2d$, si impone $r=2d$, in quanto $2d$ è il massimo
numero di prosecuzioni effettivamente diverse realizzabili, e averne
di più non porterebbe alcun beneficio apprezzabile\\

~~~~~~~~~$w_{n}^{\left(i\right)}=w_{n}^{\left(i\right)}-\log\left(r\right);$

~~~~~~~~~$\mathsf{for}\,(k=0;\,k<r;\,k++)\,\{\mathsf{Ciclo}(x',\,n+1);\}$\\
\\
Si procede con l'abbassamento del peso $W_{n}^{\left(i\right)}\left(q\right)$
di un fattore $1/r$, e la successiva duplicazione dei cammini in
$r$ copie distinte, che verranno continuate indipendentemente (nel
codice reale è presente una variabile che limita il numero massimo
globale di biforcazioni, per evitare \emph{overflow} di memoria)\\

~~~~~~$\}\,\mathsf{else}\,\{$ 

~~~~~~~~~$w_{n}^{\left(i\right)}=w_{n}^{\left(i\right)}-\log\left(r\right);$

~~~~~~~~~$\mathsf{if}\,(\mathsf{rand}()<\mathsf{RANDMAX}\cdot r)\,\{\mathsf{Ciclo}(x',\,n+1);\}$

~~~~~~$\}$\\
\\
Se invece $r<1$ si aumenta il peso $W_{n}^{\left(i\right)}\left(q\right)$
di un fattore $1/r$, per poi continuare il cammino con probabilità
$r$, o \emph{potarlo} con probabilità $1-r$.\\

~~~$\mathsf{return};$ 

$\}$\\
\\
Quando si raggiunge la lunghezza $N$, la subroutine si ferma. Si
sottolinea che la subroutine reale è stata implementata in linguaggio
$\mathsf{C}$, su macchina $\mathsf{Intel\,Xeon\,E5520}$, con un
utilizzo di memoria sempre al di sotto della Cache di III livello
($8Mb$).

L'algoritmo presentato è risultato estremamente efficiente nel simulare
cammini in fase estesa ($m<m_{c}$), consentendo di calcolare le quantità
di interesse per catene di lunghezza $\mathcal{O}\left(10^{3}\right)$,
con errori $\sim1\%$ e in tempi relativamente brevi (alcune ore).
Per quanto riguarda la fase collassata ($m>m_{c}$), le simulazioni
hanno evidenziato un aumento esponenziale di $\langle\Delta W_{n}^{2}\rangle_{q}/\langle W_{n}\rangle_{q}$,
dovuto all'insorgenza di correlazioni a lungo raggio tra i monomeri
dei cluster: questa caratteristica (intrinseca al modello) ha reso
le simulazioni molto dispendiose in termini computazionali, e ha permesso
di simulare catene di lunghezza $\mathcal{O}\left(10^{3}\right)$
solo in una parte relativamente piccola dell'intervallo $m\in\left(m_{c}\,,1\right)$. 

In generale, l'impossibilità di simulare catene in fase collassata
da parte degli algoritmi PERM è ben nota, e va ricercata nella definizione
stessa di crescita cinetica: la regola di accrescimento nel RM e nel
PERM è intrinsecamente locale, e dunque non permette di costruire
efficientemente strutture clusterizzate, che sono invece caratterizzate
da fluttuazioni su scala paragonabile alla loro taglia complessiva.
Secondo questa linea di pensiero, algoritmi basati su movimenti globali,
come il metodo dei Pivot, potrebbero risultare più adatti allo scopo. 

Benché il PERM rappresenti ancora il metodo più diffuso per la simulazione
dei polimeri, la situazione è destinata a cambiare entro breve. Infatti,
una recente (e straordinariamente potente) evoluzione del metodo dei
Pivot (\cite{Clisby,Clisby=000020Special}) ha premesso di raggiungere
risultati fino ad ora impensabili con le catene estese, e potrebbe
rappresentare, in un prossimo futuro, una efficiente soluzione anche
per la simulazione di polimeri clusterizzati.

\section*{Ringraziamenti}

Si ringrazia Frank den Hollander (Università di Leiden), per aver
suggerito la correzione $C_{3}/\langle M_{n}\rangle$, e Riccardo
Balzan (EPFL), per le notevoli intuizioni riguardo al processo stocastico
definito nel capitolo terzo. Si ringraziano inoltre Jack F. Douglas
(NIST) e Giorgio Parisi (Università di Roma ``La Sapienza'') per
utili discussioni e suggerimenti, e Nathan Clisby (Università di Melbourne),
Gregory F. Lawler (Università di Chicago), Gordon Slade (Università
della British Columbia) per la disponibilità nel rispondere ad alcune
importanti domande.

\chapter*{~}
\end{document}